\documentclass[useAMS,usenatbib]{mn2e}
\input psfig.sty

\def \aj {AJ}
\def \apj {ApJ}
\def \apjl {ApJL}
\def \mnras {MNRAS}
\def \apjs {ApJS}
\def \aap {A\&A}
\def \etal {et~al.~}

\def \spose#1{\hbox  to 0pt{#1\hss}}  
\def \lta{\mathrel{\spose{\lower 3pt\hbox{$\sim$}}\raise  2.0pt\hbox{$<$}}}
\def \gta{\mathrel{\spose{\lower  3pt\hbox{$\sim$}}\raise 2.0pt\hbox{$>$}}}
\def \ion#1#2{#1{\footnotesize{#2}}\relax}

\def \hi       {\ion{H}{I}}

\def \kms {\ifmmode  \,\rm km\,s^{-1} \else $\,\rm km\,s^{-1}  $ \fi }
\def \kpc {\ifmmode  {\rm kpc}  \else ${\rm  kpc}$ \fi  }  
\def \hkpc {\ifmmode  {h^{-1}\rm kpc}  \else ${h^{-1}\rm kpc}$ \fi  }  
\def \hMpc {\ifmmode  {h^{-1}\rm Mpc}  \else ${h^{-1}\rm Mpc}$ \fi  }  
\def \Msun {\ifmmode {\rm M}_{\odot} \else ${\rm M}_{\odot}$ \fi} 
\def \hMsun {\ifmmode h^{-1}\,\rm M_{\odot} \else $h^{-1}\,\rm M_{\odot}$ \fi}
\def \hhMsun {\ifmmode h^{-2}\,\rm M_{\odot}\else $h^{-2}\,\rm M_{\odot}$ \fi}
\def \Lsun {\ifmmode L_{\odot} \else $L_{\odot}$ \fi} 
\def \hhLsun {\ifmmode h^{-2}\,\rm L_{\odot} \else $h^{-2}\,\rm L_{\odot}$ \fi}
\def \Mstar {\ifmmode M_{\rm star} \else $M_{\rm star}$ \fi} 

\def\LCDM{$\Lambda$CDM }
\def \LCDM {\ifmmode \Lambda{\rm CDM} \else $\Lambda{\rm CDM}$ \fi}
\def \sig8 {\ifmmode \sigma_8 \else $\sigma_8$ \fi} 
\def \OmegaM {\ifmmode \Omega_{\rm M} \else $\Omega_{\rm M}$ \fi} 
\def \Omegab {\ifmmode \Omega_{\rm b} \else $\Omega_{\rm b}$ \fi} 
\def \OmegaL {\ifmmode \Omega_{\rm \Lambda} \else $\Omega_{\rm \Lambda}$\fi} 
\def \Deltavir {\ifmmode \Delta_{\rm vir} \else $\Delta_{\rm vir}$ \fi}
\def \rhocrit {\ifmmode \rho_{\rm crit} \else $\rho_{\rm crit}$ \fi}

\def \rs {\ifmmode r_{\rm s} \else $r_{\rm s}$ \fi} 
\def \rh {\ifmmode r_{\rm h} \else $r_{\rm h}$ \fi} 
\def \Rvir {\ifmmode R_{\rm vir} \else $R_{\rm vir}$ \fi}
\def \Vvir {\ifmmode V_{\rm  vir} \else  $V_{\rm vir}$  \fi} 
\def \Vmax {\ifmmode V_{\rm  max} \else  $V_{\rm max}$  \fi} 
\def \Mvir {\ifmmode M_{\rm  vir} \else $M_{\rm  vir}$ \fi}  
\def \Nvir {\ifmmode N_{\rm  vir} \else $N_{\rm  vir}$ \fi}  
\def \Jvir {\ifmmode J_{\rm vir} \else $J_{\rm vir}$ \fi} 
\def \Evir {\ifmmode E_{\rm vir} \else $E_{\rm vir}$ \fi} 
\def \lam {\ifmmode \lambda  \else $\lambda$ \fi} 
\def \lamp {\ifmmode \lambda^{\prime} \else $\lambda^{\prime}$  \fi} 
\def \lampc {\ifmmode \lambda^{\prime}_{\rm c} \else
  $\lambda^{\prime}_{\rm c}$  \fi} 

\def \xoff {\ifmmode x_{\rm off} \else $x_{\rm off}$ \fi}
\def \rhorms {\ifmmode \rho_{\rm rms} \else $\rho_{\rm rms}$ \fi}
\def \qbar {\ifmmode \bar{q} \else $\bar{q}$ \fi}

\def \Mb {\ifmmode M_{\rm b} \else $M_{\rm b}$ \fi} 
\def \Md {\ifmmode M_{\rm d} \else $M_{\rm d}$ \fi} 
\def \Mg {\ifmmode M_{\rm g} \else $M_{\rm g}$ \fi} 
\def \Rb {\ifmmode R_{\rm b} \else $R_{\rm b}$ \fi} 
\def \Rd {\ifmmode R_{\rm d} \else $R_{\rm d}$ \fi} 
\def \Rg {\ifmmode R_{\rm g} \else $R_{\rm g}$ \fi} 
\def \mgal {\ifmmode m_{\rm gal} \else $m_{\rm gal}$ \fi} 
\def \lamgal {\ifmmode \lambda_{\rm gal} \else $\lambda_{\rm gal}$ \fi} 
\def \Vcirc {\ifmmode V_{\rm circ} \else $V_{\rm circ}$ \fi} 
\def \Vobs {\ifmmode V_{\rm obs} \else $V_{\rm obs}$ \fi} 
\def \Vopt {\ifmmode V_{\rm opt} \else $V_{\rm opt}$ \fi} 
\def \fdev {\ifmmode f_{\rm deV} \else $f_{\rm deV}$ \fi} 

\def \DeltaIMF {\ifmmode \Delta_{\rm IMF} \else $\Delta_{\rm IMF}$ \fi}

\def \VV {\ifmmode V_{\rm 2.2}/V_{200} \else $V_{2.2}/V_{200}$ \fi} 

\title[Dark halo response and the stellar IMF]
{Dark halo response and the stellar initial mass function in
  early-type and late-type galaxies} 

\author[Dutton et al.]  {Aaron A.
  Dutton$^{1}$\thanks{dutton@uvic.ca}\thanks{CITA National Fellow},
  Charlie Conroy$^2$, Frank C. van den Bosch$^3$, Luc Simard$^4$,
  \newauthor {J. Trevor Mendel$^1$, St\'ephane Courteau$^5$, Avishai
    Dekel$^6$, Surhud More$^7$, }
  \newauthor {\& Francisco Prada$^8$}\\
  $^1$Dept. of Physics and Astronomy, University of Victoria, Victoria, BC, V8P 5C2, Canada.\\
  $^2$Harvard-Smithsonian Center for Astrophysics, Cambridge, MA, USA.\\
  $^3$Astronomy Department, Yale University, P.O. Box 208101, New Haven, CT 06520-8101, USA.\\
  $^4$Herzberg Institue of Astrophysics, National Research Council of Canada, 5071 West Saanich Road, Victoria, B.C., V9E 2E7, Canada.\\
  $^5$Department of Physics, Engineering Physics \& Astronomy, Queen's
  University, Kingston, Ontario, Canada.\\
  $^6$Racah Institute of Physics, The Hebrew University, Jerusalem 91904, Israel.\\
  $^7$Kavli Institute for Cosmological Physics, University of Chicago, 933 East 56th Street, Chicago, IL 60637, USA.\\
  $^8$Instituto de Astrofisica de Andalucia (CSIC), E18008 Granada, Spain.\\
}

\begin{document}
             
\date{accepted to MNRAS}
             
\pagerange{\pageref{firstpage}--\pageref{lastpage}}\pubyear{2011}

\maketitle           

\label{firstpage}
             
%%%%%%%%%%%%%%%%%%%%%%%%%%%%%%%%%%%%%%%%%%%%%%%%%%%%%%%%%%%%%%%%%%%%%%

\begin{abstract}
  We investigate the origin of the relations between stellar mass and
  optical circular velocity for early-type (ETG) and late-type (LTG)
  galaxies --- the Faber-Jackson (FJ) and Tully-Fisher (TF) relations.
  We combine measurements of dark halo masses (from satellite
  kinematics and weak lensing), and the distribution of baryons in
  galaxies (from a new compilation of galaxy scaling relations), with
  constraints on dark halo structure from cosmological
  simulations. The principle unknowns are the halo response to galaxy
  formation and the stellar initial mass function (IMF). The slopes of
  the TF and FJ relations are naturally reproduced for a wide range of
  halo response and IMFs. However, models with a universal IMF and
  universal halo response cannot {\it simultaneously} reproduce the
  zero points of both the TF and FJ relations. For a model with a
  universal Chabrier IMF, LTGs require halo expansion, while ETGs
  require halo contraction. A Salpeter IMF is permitted for high mass
  ($\sigma \gta 180 \kms$) ETGs, but is inconsistent for intermediate
  masses, unless $V_{\rm circ}(R_{\rm e})/\sigma_{\rm e} \gta 1.6$. If
  the IMF is universal and close to Chabrier, we speculate that the
  presence of a major merger may be responsible for the contraction in
  ETGs while clumpy accreting streams and/or feedback leads to
  expansion in LTGs. Alternatively, a recently proposed variation in
  the IMF disfavors halo contraction in both types of galaxies.
  Finally we show that our models naturally reproduce flat and
  featureless circular velocity profiles within the optical regions of
  galaxies without fine-tuning.
\end{abstract}

\begin{keywords}
dark matter 
-- galaxies: fundamental parameters 
-- galaxies: structure 
-- galaxies: haloes 
-- galaxies: spiral 
-- galaxies: elliptical lenticular, cD

\end{keywords}

\setcounter{footnote}{1}

%%%%%%%%%%%%%%%%%%%%%%%%%%%%%%%%%%%%%%%%%%%%%%%%%%%%%%%%%%%%%%%%%%%%%%
%% SECTION 1: INTRODUCTION
%%%%%%%%%%%%%%%%%%%%%%%%%%%%%%%%%%%%%%%%%%%%%%%%%%%%%%%%%%%%%%%%%%%%%%

\section{Introduction}
\label{sec:intro}

%% V-L relations
Galaxy properties obey several fundamental relations, which
have long been thought to hold important clues about the physical
processes that influenced their formation and evolution.  The
relations between rotation velocity and luminosity (for late-types)
and velocity dispersion and luminosity (for early-types) are
particularly interesting as they connect luminous mass with dynamical
mass (which includes not only baryons but also dark matter). These
relations are also known as the Tully-Fisher (Tully \& Fisher 1977)
and Faber-Jackson (Faber \& Jackson 1976) relations.

% small scatter
The scatter in these relations is small, 0.07 dex in velocity for FJ
(e.g., Gallazzi \etal 2006) and 0.05 dex in velocity for TF (e.g.,
Courteau \etal 2007b; Pizagno \etal 2007), which has enabled these
relations to be used as secondary distance indicators.  The smallness
and source of the scatter is also interesting from a galaxy formation
point of view. For early-type galaxies the scatter in velocity
dispersion correlates with galaxy size, or surface brightness.  This
leads to the so-called fundamental plane of early-type galaxies, a
correlation between size, surface brightness and velocity dispersion
(Dressler \etal 1987; Djorgovski \& Davis 1987). By contrast the
scatter in the TF relation is independent of size or surface
brightness (Zwaan \etal 1995; Courteau \& Rix 1999; Courteau \etal
2007b; Pizagno \etal 2007), suggesting that the TF relation is the edge
on projection of the fundamental plane for late-type galaxies.

%% origin of scaling relations
The origin of these scaling relations in \LCDM cosmologies is
typically thought to be the relation between halo virial velocity and
virial mass, which scale as $\Vvir \propto \Mvir^{1/3}$. Accounting
for the higher halo concentrations in lower mass dark matter haloes
results in a shallower slope for the relation between the maximum
circular velocity of dark matter haloes and the halo virial mass:
$V_{\rm max,h} \propto \Mvir^{0.29}$ (Bullock \etal 2001). This
scaling is similar to the observed stellar mass TF and FJ relations:
$\sigma_{\rm e} \propto \Mstar^{0.29}$ (Gallazzi \etal 2006) and
$V_{2.2}\propto \Mstar^{0.28}$ (Dutton \etal 2010b), where
$\sigma_{\rm e}$ is the velocity dispersion within the half-light
radius, and $V_{2.2}$ is the rotation velocity at 2.2 disk scale
lengths. In what follows we define $\Vopt=V_{2.2}$ for late-types, and
$\Vopt \propto \sigma_{\rm e}$ for early-types.  However, for the
$V_{\rm max,h}-\Mvir$ relation to be the direct origin of the TF and
FJ relations requires that $\Vopt/V_{\rm max,h}$, and $\Mstar/\Mvir$
are constants.  In Dutton \etal (2010b) we showed that this is at best
only approximately the case.
%% Vopt/Vvir
The relation between $\Vopt$ and $\Vvir$ depends on three factors:
1) The contribution of baryons to $\Vopt$; 2) The structure of the
``pristine'' dark matter halo (i.e., without the influence of
baryons); and 3) The response of the dark matter halo to galaxy
formation.

%% VV: baryons
For low mass star-forming galaxies gas dominates their baryonic
budget, but for high mass star-forming galaxies and most non
star-forming galaxies, stars dominate the baryon budget. A key
uncertainty in measuring stellar masses is the stellar initial mass
function (IMF). There is a factor of $\sim 2$ difference between the
masses derived assuming the traditional Salpeter (1955) IMF compared
with those derived assuming a Chabrier (2003) or Kroupa (2001)
IMF. These latter IMFs are based on more recent measurements in the
solar neighbourhood.

%% VV: DM
The structure of ``pristine'' dark matter haloes, has been extensively
studied using cosmological N-body simulations (e.g., Navarro, Frenk,
\& White 1996a; Navarro, Frenk, \& White 1997; Bullock \etal 2001;
Eke, Navarro \& Steinmetz 2001; Zhao \etal 2003; Navarro \etal 2004;
Diemand \etal 2005, 2007; Macci\`o \etal 2007; Neto \etal 2007;
Macci\`o \etal 2008; Zhao \etal 2009; Navarro \etal 2010; Klypin \etal
2010; Mu{\~n}oz-Cuartas \etal 2011). While the nature of the density
profile in the inner 0.1\% of the virial radius is still uncertain,
and the three parameter Einasto profile provides better fits than the
broken power law of Navarro, Frenk, \& White (1997, hereafter NFW),
(Merritt \etal 2006), on the scales relevant for modeling galaxy
kinematics \LCDM dark matter haloes are well described by the two
parameter NFW function.  These two parameters are correlated with
small scatter, such that the structure of dark matter haloes is almost
completely determined by their mass.

%% VV: halo response
The response of the halo to galaxy formation has traditionally been
modeled assuming galaxy formation is adiabatic, i.e., changes in
potential are slow compared to the dynamical time. Under this
assumption the halo is expected to contract, resulting in so-called
adiabatic contraction (Blumenthal \etal 1986). The standard model
assumes dark halo particles are on circular orbits. Using more
realistic orbits results in weaker halo contraction (Wilson 2003;
Gnedin \etal 2004; Sellwood \& McGaugh 2005). Recent hydrodynamical
simulations of galaxy formation yield even weaker halo contraction
(Abadi \etal 2010; Duffy \etal 2010; Pedrosa \etal 2010; Tissera
\etal 2010). If galaxy formation is adiabatic, then the halo response
should depend only on the final distribution of the baryons, and not
on the assembly history. In these simulations the assembly history
does matter, and thus this calls into question the basic assumption
that halo response to galaxy formation is adiabatic.

It is also possible for haloes to expand in response to galaxy
formation, via a number of processes: rapid mass loss from the galaxy,
e.g., driven by supernovae (Navarro, Eke, \& Frenk 1996b; Gnedin \&
Zhao 2002; Read \& Gilmore 2005; Governato \etal 2010); dynamical
friction operating on baryonic clumps (El-Zant, Shlosman \& Hoffman
2001; El-Zant \etal 2004; Elmegreen \etal 2008; Jardel \& Sellwood
2009) or galactic bars (Weinberg \& Katz 2002; Holley-Bockelmann \etal
2005; Sellwood 2008).  Thus while the underlying structure of dark
matter haloes is well understood, in order to understand the origin of
the TF and FJ relations, it is necessary to know the stellar IMF, and
how dark matter haloes respond to galaxy formation.

%What does current literature say about constraints on the IMF?
Upper limits to stellar mass-to-light ratios, and hence the IMF, can
be obtained from galaxy dynamics. For spiral galaxies, a Salpeter IMF
is ruled out from maximal disk fits to resolved rotation curves (Bell
\& de Jong 2001). An IMF with stellar masses 0.15 dex lower than a
Salpeter, the so-called diet-Salpeter IMF, is the upper limit.  For
elliptical galaxies a Salpeter IMF is also ruled out for some galaxies
(Cappellari \etal 2006).  But massive elliptical galaxies ($\sigma>200
\kms$) are consistent with a Salpeter IMF (Bernardi \etal 2010), which
may even be favored over lighter IMFs (Treu \etal 2010; Auger \etal
2010a). An alternative constraint on the stellar IMF comes from
choosing the stellar mass-to-light ratio normalization that minimizes
the scatter in the baryonic\footnote{The baryonic TF relation is the
  relation between rotation velocity and baryonic mass (i.e., stars
  plus cold gas).} TF relation (Stark, McGaugh, Swaters 2009). This
favours a diet-Salpeter IMF with an uncertainty of $\pm 0.1$ dex in
the stellar mass-to-light ratio.

%What does current literature say about constraints on halo contraction?
For a Chabrier IMF, elliptical galaxies require halo contraction to
explain the mass discrepancy between the observed dynamical masses and
those predicted when galaxies are embedded in NFW haloes (Schulz \etal
2010; Tollerud \etal 2011).  However, for late-type galaxies,
models with halo contraction over-predict the rotation velocities at
fixed luminosity or stellar mass (Dutton \etal 2007; Dutton \& van den
Bosch 2009; Trujillo-Gomez \etal 2010).

In this paper we construct bulge-disk-halo models of early-type and
late-type galaxies. These models are constrained to reproduce the
distribution of stars and gas in galaxies, the relation between
stellar mass and halo mass, and the structure of dark matter haloes in
cosmological simulations.  The key unknowns are the stellar IMF and
the halo response to galaxy formation. We use the observed TF and FJ
relations to place constraints on these two unknowns.  As a by-product
of this exercise, we measure the average dark matter fractions within
the optical regions of early-type and late-type galaxies as a function of
stellar mass.

This paper is organized as follows.  In \S 2 we describe the mass
models. In \S 3 we discuss the observational constraints. In \S 4 we
present model TF and FJ relations and compare to the observations.  A
discussion is given in \S5 and a summary in \S 6.

%%%%%%%%%%%%%%%%%%%%%%%%%%%%%%%%%%%%%%%%%%%%%%%%%%%%%%%%%%%%%%%%%%%%%%%%%%%%
%% SECTION 2 Mass Models
%%%%%%%%%%%%%%%%%%%%%%%%%%%%%%%%%%%%%%%%%%%%%%%%%%%%%%%%%%%%%%%%%%%%%%%%%%%%%%%
\section{Mass Models}
\label{sec:mm}
This section describes the mass models that we construct to compare to the
observed scaling relations. The mass models consist of 3 components for
early-types (stellar bulge, stellar disk, and dark matter halo), and 4
components for late-types (stellar bulge, stellar disk, gas disk, and
dark matter halo).  The circular velocity at radius, $r$, in the plane
of the disk is given by the quadratic sum of the circular velocities
of the various components:
\begin{equation}
V(r) = \sqrt{ V_{\rm b}^2(r) + V_{\rm d}^2(r) + V_{\rm g}^2(r) + V_{\rm h}^2(r)},
\end{equation}
where the subscripts, b, d, g, and h, refer to the stellar
bulge, stellar disk, gas disk, and dark matter halo, respectively.

\subsection{Baryons}
\label{sec:models}
We model the stellar bulge with a Hernquist profile (Hernquist 1990) 
\begin{equation}
\rho(r) = \frac{\Mb }{2 \pi} \frac{\rh}{r(r+\rh)^3},
\end{equation}
where $\Mb$ is the bulge mass, and $\rh$ is a scale radius.  We adopt
the Hernquist profile as it provides a convenient analytical
approximation to the de-projected de Vaucouleurs profile, which is the
assumed profile in our bulge plus disk fits (see \S~\ref{sec:sdss}).
The projected half mass radius is given by $R_{50\rm b}=1.815 \rh$.
The enclosed mass of a Hernquist sphere is given by
\begin{equation}
\Mb(r) =   \Mb \frac{r^2}{(r+\rh)^2},
\end{equation}
and the circular velocity is given by
\begin{equation}
V_{\rm b}(r) =   \sqrt { G \Mb \frac{r}{(r+\rh)^2}}.
\end{equation}

We model the stellar disk with an exponential surface density profile 
\begin{equation}
\Sigma_{\rm d}(R) = \frac{\Md}{2 \pi \Rd^2} \exp(-R/\Rd),
\end{equation}
which is specified by two parameters: the disk mass $\Md$ and disk
scale length $\Rd$. The half mass radius of the disk is given by
$R_{50\rm d} = 1.678 \Rd$.  For late-type galaxies we include a gas disk
which we also model with an exponential profile
\begin{equation}
\Sigma_{\rm g}(R) = \frac{\Mg}{2 \pi \Rg^2} \exp(-R/\Rg),
\end{equation}
which is specified by two parameters: the gas mass $\Mg$ and gas disk
scale length $\Rg$.

For late-type galaxies we assume the stellar disks are infinitesimally
thin.  The circular velocity at radius, $r$, (in the plane of the
disk) of a thin exponential disk of mass $\Md$ and scale length $\Rd$
is given by:
\begin{equation}
V_{\rm d}^2 (r) = \frac{G \Md}{\Rd} 2 y^2 [ I_0(y)K_0(y) - I_1(y)K_1(y)],
\end{equation}
where $y=r/(2 \Rd)$, and $I_n$ and $K_n$ are modified Bessel
functions (Freeman 1970). 

For early-type galaxies the ``disk'' (i.e., the exponential) component
that we fit to the observed photometry is usually not a true disk (see
\S~\ref{sec:sdss}). Thus in our dynamical models for early-types we
assume the exponential component is spherical.  In this case the 3D
density profile is given by (e.g., van den Bosch \& de Zeeuw 1996)
\begin{equation}
\label{eq:exps}
\rho(r)=\rho_{0,\rm d} K_0(r/R_{\rm d}),
\end{equation}
where $\rho_{0,\rm d}=M_{\rm d}/ (2 \pi^2 R_{\rm d}^3)$, and $K_0$ is
a modified Bessel function. We then obtain the circular velocity by
integrating Eq.~\ref{eq:exps}.

\subsection{Dark matter}
For the dark matter halo, we use a spherical NFW (Navarro, Frenk, \&
White 1997) profile
\begin{equation}
\rho(r) = \frac{ \delta_{\rm c} \rhocrit}{ (r/r_{-2}) ( 1 + r/r_{-2})^{2}}.
\end{equation}
Here $r_{-2}$ is the radius where the logarithmic slope of the density
profile is $-2$, $\rhocrit$ is the critical density of the universe,
and $\delta_{\rm c}$ is the characteristic halo density.  We
reparametrize the halo by the virial mass, $M_{200}$ and the
concentration $c= R_{200}/r_{-2}$, where $R_{200}$ is the virial
radius.  We adopt the definition of the virial mass such that the
average density inside the virial radius is equal to 200 times the
critical density of the universe. Thus, for a given $M_{200}$, the
virial radius and circular velocity at the virial radius, $V_{200}$
are related to each other at redshift $z=0$ by:
\begin{eqnarray}
\frac{R_{200}}{(h^{-1}{\rm kpc})} = \frac{V_{200}}{({\kms})} = \left[ G \,\frac{M_{200}}{(h^{-1}{\rm \Msun})} \right]^{1/3},  
\end{eqnarray}
where $G\simeq 4.301 \times 10^{-6} \,\rm km^2\,s^{-2}\,kpc\,\Msun^{-1}$.
The mass enclosed within a spherical radius, $r$, is given by
\begin{equation}
M(r) = M_{200} A(r/r_{-2})/A(c), 
\end{equation}
where $A(x) = \ln(1+x) - x/(1+x)$. And thus the circular velocity at
radius, $r$, is given by
\begin{equation}
  V_{\rm h}(r) =   V_{200} \sqrt{\frac{A(r/r_{-2})}{r/r_{-2}} \frac{c}{A(c)}}.
\end{equation}

\subsubsection{dark matter halo response to galaxy formation}
\label{sec:ac}
The process of galaxy formation is expected to modify the pristine
dark matter density profile.  If galaxy formation is a slow and smooth
process then the halo is expected to contract (e.g., Blumenthal \etal
1986; Gnedin \etal 2004; Sellwood \& McGaugh 2005).  However, if the
initial stages of galaxy formation involve the mergers of massive
clumps of gas, dynamical friction between the baryons and dark matter
may result in net halo expansion (e.g., El-Zant \etal 2001).  In
addition other processes such as rapid ejection of baryons from the
disk through supernova driven winds (e.g., Navarro \etal 1996b), and
galactic bars (Weinberg \& Katz 2002) may also contribute to expanding
the halo.

In the standard formalism the adiabatic invariant is $r M(r)$, and thus
\begin{equation}
\label{eq:ac1}
r_{\rm f}/r_{\rm i} = M_{\rm i}(r_{\rm i})/M_{\rm f}(r_{\rm f}),
\end{equation}
where $M_{\rm i}(r)$ and $M_{\rm f}(r)$ are the initial and final mass
distributions, and $r_{\rm i}$ and $r_{\rm f}$ are initial and final
radii. Assuming no shell crossing of the dark matter, $M_{\rm
  h,i}(r_{\rm i})=M_{\rm h,f}(r_{\rm f})$, and that baryons initially
follow the same mass profile as the dark matter, $M_{\rm b,i}(r_{\rm
  i})=m_{\rm gal}M_{\rm h,i}(r_{\rm i})$, then
\begin{equation}
\label{eq:ac2}
r_{\rm f}/r_{\rm i} = M_{\rm h,i}(r_{\rm i})/[M_{\rm b,f}(r_{\rm f}) + (1-m_{\rm gal})M_{\rm h,i}(r_{\rm i})],
\end{equation}
where $M_{\rm b,f}$ is the final mass distribution of the baryons, and
$m_{\rm gal}$ is the baryon mass fraction.  For our mass models we
know $M_{\rm b,f}(r)$, $M_{\rm i}(r)$, and $m_{\rm gal}$, and thus one can solve
Eq.~\ref{eq:ac2} for the mapping between $r_{\rm i}$ and $r_{\rm f}$.

Using cosmological hydrodynamical simulations of galaxy clusters,
Gnedin \etal (2004) advocate a modified adiabatic invariant
$rM(<\bar{r})$, where $\bar{r}/R_{200}= 0.85 (r/R_{200})^{0.8}$, and
thus
\begin{equation}
  r_{\rm f}/r_{\rm i} = M_{\rm i}(\bar{r}_{\rm i})/M_{\rm f}(\bar{r}_{\rm f}).
\end{equation}
This formula results in slightly less contraction than the standard formula
(Eq.~\ref{eq:ac1}).

More recent cosmological hydrodynamical simulations of galaxies (Abadi
\etal 2010; Pedrosa \etal 2010; Tissera \etal 2010) have found less
contraction than predicted by the Blumenthal \etal (1986) and Gnedin
\etal (2004) models. Abadi \etal (2010) advocate the following formula
\begin{equation}
r_{\rm f}/r_{\rm i} = 1 + 0.3[ (M_{\rm i}(r_{\rm i})/M_{\rm f}(r_{\rm f}))^2 -1].
\end{equation}

In order to explore the possibility of expansion we also consider the
generalized contraction formula from Dutton \etal (2007). A modified
contraction parameter $\Gamma$ can be defined as
\begin{equation}
\Gamma=(r_{\rm f}/r_{\rm i})^{\nu}.
\label{eq:nu}
\end{equation}
For $\nu=1$, we have the standard Blumenthal \etal (1986) contraction
formula (Eq.~\ref{eq:ac1}), if $\nu=0$ there is no contraction, and if
$\nu < 0$ there is expansion. The Gnedin \etal (2004) model can be
well approximated with $\nu =0.8$, while the Abadi \etal (2010) model
corresponds to $\nu \sim 0.4$. 

%%% FIGURE 1
\begin{figure*}
\centerline{
\psfig{figure=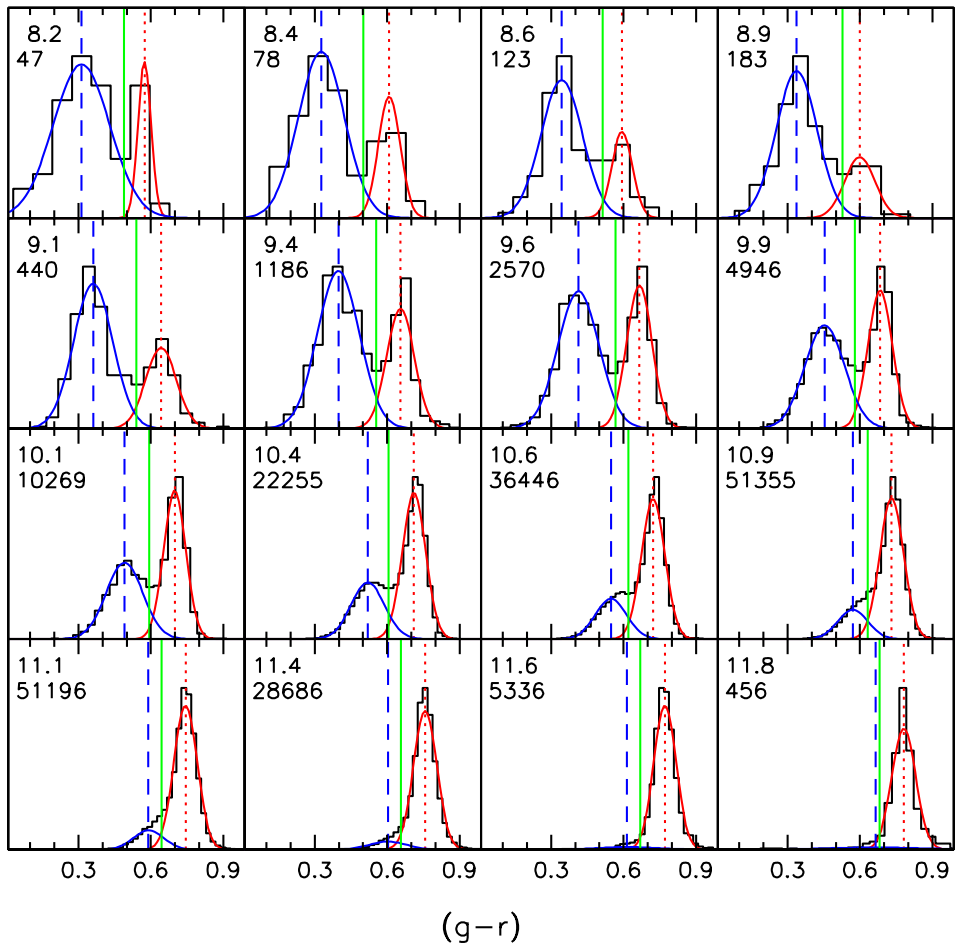,width=0.75\textwidth}}
\caption{Histograms of $(g-r)$ color in bins of stellar mass for
  galaxies that meet our redshift, minimum stellar mass and disk axis
  ratio cuts. The mean of $\log_{10}(\Mstar/\Msun)$ and the number of
  galaxies are indicated at the top left corner of each
  panel. Galaxies have been separated into red and blue using a Monte
  Carlo method that assigns color based on double Gaussian fits to the
  galaxy color distribution in each mass bin. The mean colors of the
  red and blue galaxies are shown as red dotted and blue dashed
  vertical lines. The separator between red and blue galaxies
  (Eq.~\ref{eq:gv}) that we use for the rest of this paper is given by
  the solid green vertical lines.}
\label{fig:cmhist}
\end{figure*}

%% FIGURE 2
\begin{figure*}
\centerline{
\psfig{figure=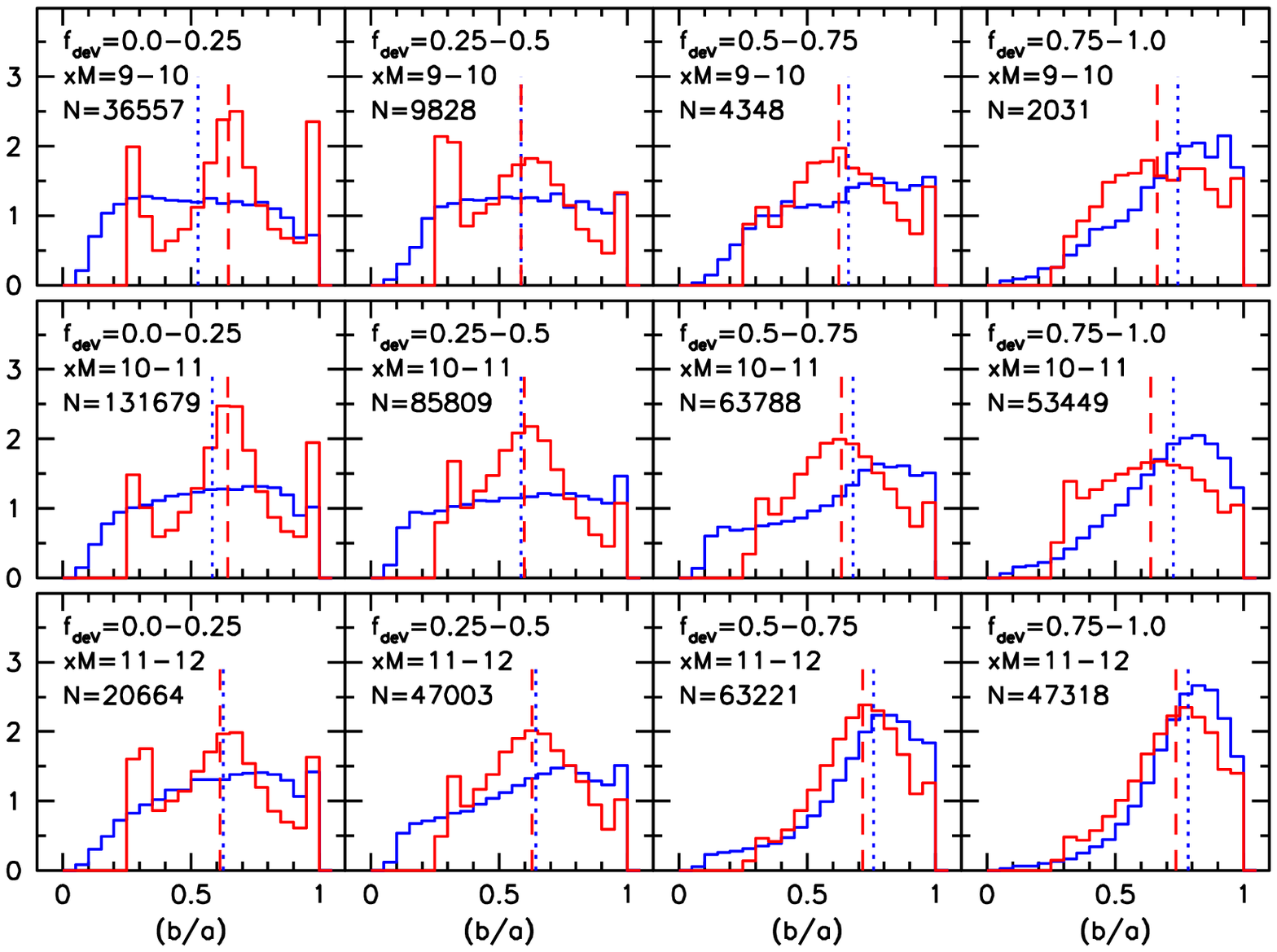,width=0.80\textwidth}}
\caption{Axis ratio distributions of $n=4$ components ``bulges'' (red)
  and $n=1$ components ``disks'' (blue) for different stellar masses,
  ${\rm xM}\equiv \log_{10}(\Mstar/\Msun)$, and de Vaucouleurs
  fractions, $f_{\rm deV}$. for galaxies that meet our redshift
  cuts and redshift dependent stellar mass limit (based on early-type
  galaxies). The vertical lines show the median axis ratio for $n=4$
  components (dashed) and $n=1$ components (dotted). For thin disks
  the distribution of axis ratios is uniform. This figure shows that
  the $n=1$ components in galaxies with low $\fdev$ are disks, whereas
  the $n=1$ components in galaxies with high $\fdev$ are
  spheroids. The value of $\fdev$ above which the $n=1$ components are
  predominantly spheroids decreases with increasing stellar mass.  }
\label{fig:bahist}
\end{figure*}

%%%%%%%%%%%%%%%%%%%%%%%%%%%%%%%%%%%%%%%%%%%%%%%%%%%%%%%%%%%%%%%%%%%%%%
%% SECTION 2: OBSERVATIONS
%%%%%%%%%%%%%%%%%%%%%%%%%%%%%%%%%%%%%%%%%%%%%%%%%%%%%%%%%%%%%%%%%%%%%%

%%%%%%%%%%%%%%%%%%%%%%%%%%%%%%%%%%%%%%%%%%%%%%%%%%%%%%%%%%%%%%%%%%%%%%%%%%%%

\section{Constraints}
\label{sec:obs}
In this section we discuss the observational and theoretical
constraints to our mass models.  Readers that are interested in the
results of our mass models, but not the details of the constraints,
may skip to \S~\ref{sec:overview}.

Here we present new determinations of the structural and dynamical scaling
relations of early-type and late-type galaxies. We use both new
measurements as well as data compiled from the literature.  We start
with a discussion of the scaling relations derived from Sloan Digital
Sky Survey (SDSS; York \etal 2000) data release seven (DR7) (Abazajian
\etal 2009), and then proceed to other observational constraints.

The main use for the observed structural scaling relations we present
is to enable a determination of the average mass of baryons (i.e., stars
and gas), within a specified radius, as a function of the stellar mass
of a galaxy.  These scaling relations are also of interest in their
own right as constraints to cosmological galaxy formation models.

\subsection{General properties of the SDSS sample}
\label{sec:sdss}
A number of the scaling relations we present here are based on a large
sample of galaxies from the SDSS/DR7. Structural properties are
derived from two component fits (nominally referred to as bulge and
disk) on SDSS $g$- and $r$-band images performed using \textsc{gim2d}
(Simard \etal 2002; Simard \etal, in prep). The two components are
special cases of the S\'ersic function
\begin{equation}
\Sigma(R)=\Sigma_0 \exp [-(R/R_0)^{1/n}],
\end{equation}
which is characterized by the S\'ersic index, $n$.  Bulges are modeled
as elliptical S\'ersic $n=4$ profiles (i.e., de Vaucouleurs) with axis
ratio $q_{\rm b}$, while disks are modeled as elliptical S\'ersic
$n=1$ profiles (i.e., exponential) with axis ratio $q_{\rm d}$.  The 2D
models are convolved with the point spread function due to seeing, and
fitted to the observed images using a Monte Carlo Markov Chain
technique.

Rest frame $g$ and $r$ magnitudes have been derived from
\textsc{gim2d} model magnitudes with a k-correction to $z=0$ based on
SDSS petrosian $ugriz$ magnitudes.  Stellar masses are from the
MPA/JHU group\footnote{ Available at
  http://www.mpa-garching.mpg.de/SDSS/DR7/}, who fit SDSS $ugriz$
model magnitudes with Bruzual \& Charlot (2003) stellar population
synthesis models, and adopting a Chabrier (2003) IMF. Our full sample
consists of $\sim 655\; 000$ galaxies with spectroscopic redshifts,
\textsc{gim2d} fits and stellar masses.  We prune this down to $\sim
170\; 000$ early-type and $\sim 100\; 000$ late-type galaxies using a
number of cuts as described below.

\begin{itemize}

\item Redshift Range: $0.005 \le z \le 0.2$. This leaves $\sim 614\;000$
  galaxies, or $\sim 94\%$ of the sample.

\item Color Cuts: We split galaxies into red (early-types) and blue
  (late-types) based on the bimodality in the ($g-r$) color-stellar
  mass plane (Fig.~\ref{fig:cmhist}). As a separator we adopt
\begin{equation}
\label{eq:gv}
  (g-r) = 0.59 +0.052 [\log_{10}(\Mstar/\Msun) -10],
\end{equation}
which is shown as solid green vertical lines in Fig.~\ref{fig:cmhist}.
This results in $\sim 246\;000$ late-type galaxies and $\sim 368\;000$
early-type galaxies.

\item Stellar Mass Limit: $M_{\rm min}= 10^{10.2}\Msun [(z/0.1)^2
  +0.2(z/0.1)^4]$ for late-type galaxies, and a factor of 2 higher for
  early-type galaxies. This removes the color bias towards bluer
  galaxies in the $(g-r)$- stellar mass plane caused by the $r$-band
  magnitude limit for SDSS spectroscopy (see Appendix A in van den
  Bosch \etal 2008 for a discussion of this bias).  This selection
  leaves $\sim 150\;000$ late-type galaxies ($61\%$ that passed the
  redshift cut) and $\sim 255\; 000$ early-type galaxies ($69\%$ that
  passed the redshift cut).

\item Disk Axis Ratio Cut: $b/a > 0.5$. For thin disks this
  corresponds to an inclination of less than 60 degrees. This
  reduces contamination of the red sequence with dust reddened
  late-types, and reduces extinction biases to the structural
  parameters. This selection leaves $\sim 106\;000$ late-type galaxies
  ($71\%$ that passed mass and redshift cuts) and $\sim 168\;000$ early-type
  galaxies ($66\%$ that passed mass and redshift cuts). 

\item Stellar Velocity Dispersion: For early-type galaxies we also
  require a stellar velocity dispersion from the SDSS/DR7
  pipeline. This requires that the spectrum has been classified as an
  early-type. This selection leaves $\sim 136\;000$ early-type
  galaxies ($81\%$ that passed previous cuts).

\end{itemize}

\subsubsection{Interpretation of bulge-disk fits}
\label{sec:bdfits}
How robust are our bulge-disk fits?  For late-type galaxies it is well
known that bulges are often better fit with S\'ersic indices of $n
\sim 1-2$ (e.g., Courteau \etal 1996), rather than the $n=4$ that we
adopt here. However, since the bulge half-light radii are typically
much smaller than the radii at which we measure the rotation velocity
(2.2 disk scale lengths), the structure of the bulge is not
critical. What matters for our purposes is that the bulge-disk fits
provide an accurate measurement of the {\it total} light within 2.2
disk scale lengths.  We also do not attempt to distinguish between
classical bulges, pseudo bulges and bars. Firstly, because this is an
unnecessary complication, and secondly because for typical galaxies
(at redshifts $z\gta 0.1$) in our sample the $\sim 1.5$ arcsec
resolution of the SDSS imaging is not capable of distinguishing
between these components.

The distribution of bulge and disk axis ratios gives an important
diagnostic to the physical interpretation of our bulge and disk
components, as these components are expected to have different
distributions. For disks, the observed distribution should be uniform,
with a cut off at low axis ratios corresponding to the finite
thickness of galaxy disks, and a deficit of perfectly round galaxies
due to spiral arms. For elliptical galaxies the distribution is
expected to be skewed towards high axis ratios.

Fig.~\ref{fig:bahist} shows distribution of axis ratios for $n=4$
components (``bulges'', red lines) and $n=1$ components (``disks'',
blue lines) as a function of stellar mass, and the fraction of light
in the $n=4$ component, hereafter the de Vaucouleurs fraction,
$\fdev$\footnote{Note that our parameter $\fdev$ should not be
  confused with the SDSS parameter fracDev. While both parameters
  measure the fraction of the light in an $n=4$ component, they are
  calculated using different methods, with our method being more
  robust than that used by SDSS.}.  We use galaxies that meet our
redshift cut and redshift dependent stellar mass limit (based on
early-type galaxies).  For galaxies with $\fdev < 0.25$ (far left
panels) the observed distribution of disk axis ratios is as
expected. But for galaxies with $\fdev > 0.75$ (far right panels) the
observed distribution of disk axis ratios is skewed towards high
values, indicating that these $n=1$ components are in fact spheroids.
The value of $\fdev$ above which the $n=1$ components are
predominantly spheroids decreases with increasing stellar mass. For
example, for low stellar mass galaxies, the transition is at $\fdev
\simeq 0.75$, whereas for high stellar mass galaxies the transition is
at $\fdev \simeq 0.5$.

These results should not be considered a surprise, as while the
average S\'ersic index of elliptical galaxies is $n \simeq 4$, it is
known that in general $n \ne 4$. The mean S\'ersic index of elliptical
galaxies increases with increasing luminosity (e.g., Prugiel \& Simien
1997; Graham \& Guzman 2003; Ferrarese \etal 2006; Kormendy \etal
2009). In particular high luminosity ellipticals have $n>4$, and thus
will be better fitted by a model with an $n=4$ plus an $n=1$ component
rather than just a single $n=4$ component.

Based on these results, we assume that the majority of the $n=1$
components in our fits to early-type galaxies are not true disks.
Thus in our dynamical models for early-type galaxies we assume that
the $n=1$ components are spherical. There are differences between the
circular velocity profiles of a thin disk and a sphere of the same
projected mass profile, with thin disks having 10-15\% higher maximum
circular velocities (Binney \& Tremaine 1987). However, for early-type
galaxies the $n=4$ component dominates the baryons within the
half-light radius, so the exact details for how one models the 3D
structure of the $n=1$ component is not critical to the conclusions of
this paper.

%% FIGURE 3
\begin{figure}
\centerline{
\psfig{figure=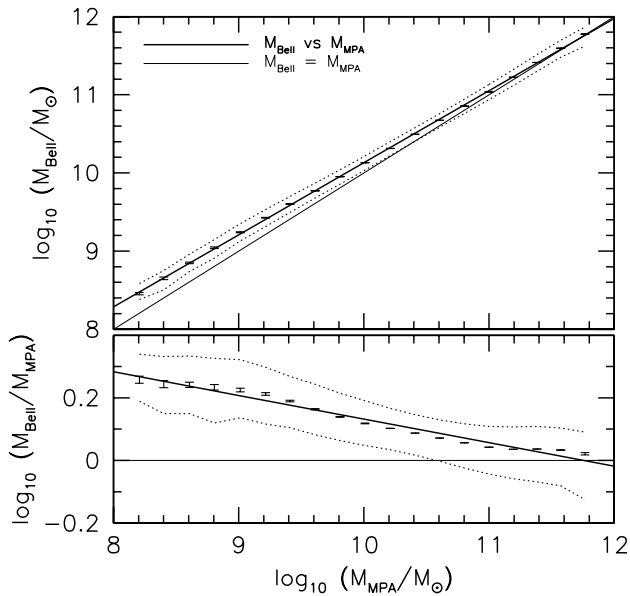,width=0.47\textwidth}
}
\caption{Comparison between stellar masses from the MPA/JHU group,
  $M_{\rm MPA}$, and those derived from the $(g-r)$ - $\Mstar/L_r$
  relation in Bell \etal (2003), $M_{\rm Bell}$. We have subtracted
  0.1 dex from the Bell \etal (2003) formula, so that $M_{\rm Bell}$
  and $M_{\rm MPA}$ both nominally correspond to a Chabrier (2003)
  IMF. The error bars show the error on the median, while the dotted
  lines show the 16th and 84th percentiles. The solid line gives our
  fitting formula Eq.~\ref{eq:mm}. }
\label{fig:mm}
\end{figure}

\subsection{Comparison between stellar mass estimates}
In this paper we use stellar masses estimated using two different
methods: The relations between optical color and stellar mass-to-light
ratio from Bell \etal (2003), $M_{\rm Bell}$; and from $ugriz$
spectral energy distribution (SED) fitting from the MPA/JHU group,
$M_{\rm MPA}$.  The MPA/JHU masses have been explicitly calculated
with a Chabrier IMF. We have subtracted 0.1 dex from the Bell \etal
(2003) masses, so that these correspond to those from a Salpeter IMF
-0.25 dex, and hence a Chabrier IMF. Thus both of the mass estimates
we use nominally correspond to a Chabrier (2003) IMF.

Here we measure the differences between these masses, so that scaling
relations based on either of these methods can be more directly
compared. Fig.~\ref{fig:mm} shows that the scatter between the two
different masses is small ($\lta 0.1$ dex), and at high masses the two
methods agree, but there is a systematic difference that increases
with decreasing mass.  We fit the relation between the two masses
using the following power law
\begin{equation}
\label{eq:mm}
\log_{10}\left(\frac{M_{\rm Bell}}{10^{10}\Msun} \right)
= 0.130 +0.922 \log_{10}\left(\frac{M_{\rm MPA}}{10^{10}\Msun} \right).
\end{equation}
Similar differences between masses from Bell \etal (2003) and masses
from Blanton \& Roweis (2007) were shown by Li \& White (2009). The
masses from Blanton \& Roweis (2007) use a similar methodology to
those from the MPA/JHU group, so the good correspondence is expected.

Both the MPA/JHU and Bell \etal (2003) masses are based on SED fits to
SDSS photometry, but there are a number of possible sources of the
systematic differences between the MPA/JHU and Bell \etal (2003)
masses, which we briefly list here. 1) Curvature in color-M/L relation
at blue colors (e.g., Portinari \etal 2004). 2) Different treatments
for extinction (MPA/JHU apply an extinction correction, Bell \etal
does not). 3) Different methods for deriving stellar masses from an
SED. 4) Different stellar population synthesis (SPS) codes: Bruzual \&
Charlot (2003) vs PEGASE (Fioc \& Rocca-Volmerange 1997).

Since the MPA/JHU masses are more direct measurements, throughout this
paper we adopt these as our fiducial masses. All scaling relations
presented or used in this paper use MPA/JHU masses directly, or Bell
\etal (2003) masses converted to MPA/JHU masses using Eq.~\ref{eq:mm}.

\begin{table}
 \centering
 \caption{Parameters of double power-law fitting formula (Eq.~\ref{eq:power2})
   to the $y=R_{50}/[\kpc]$ vs $x=\Mstar/[\Msun]$ relations in Fig.~\ref{fig:rm}.}
  \begin{tabular}{cccccc}
\hline
\hline  
Component  &  $\alpha$ & $\beta$ & $\log_{10} M_0$ & $\log_{10} R_{0}$ & $\gamma$\\
\hline
\multicolumn{6}{c}{Late-types}\\
  Disks  $R_{50}$ &0.20 & 0.46 & 10.39 & 0.75 & 1.95 \\
  Bulges $R_{50}$ &0.17 & 0.35 & 11.02 & 0.51 & 2.8 \\
  Total $R_{50}$ &0.21 & 0.47 & 10.86 & 0.84 & 2.2 \\
\hline
\multicolumn{6}{c}{Early-types}\\
  Disks  $R_{50\rm c}$ &0.27 & 0.59 & 9.97 & 0.32 & 1.5 \\
  Bulges $R_{50\rm c}$ &0.03 & 0.54 & 10.07 & -0.07 & 4.3 \\
  Total $R_{50\rm c}$ &0.03 & 0.64 & 10.09 & 0.16 & 1.3 \\
\hline
\hline
\label{tab:rmfit}
\end{tabular}
\end{table}

%% FIGURE 4
\begin{figure*}
\centerline{
\psfig{figure=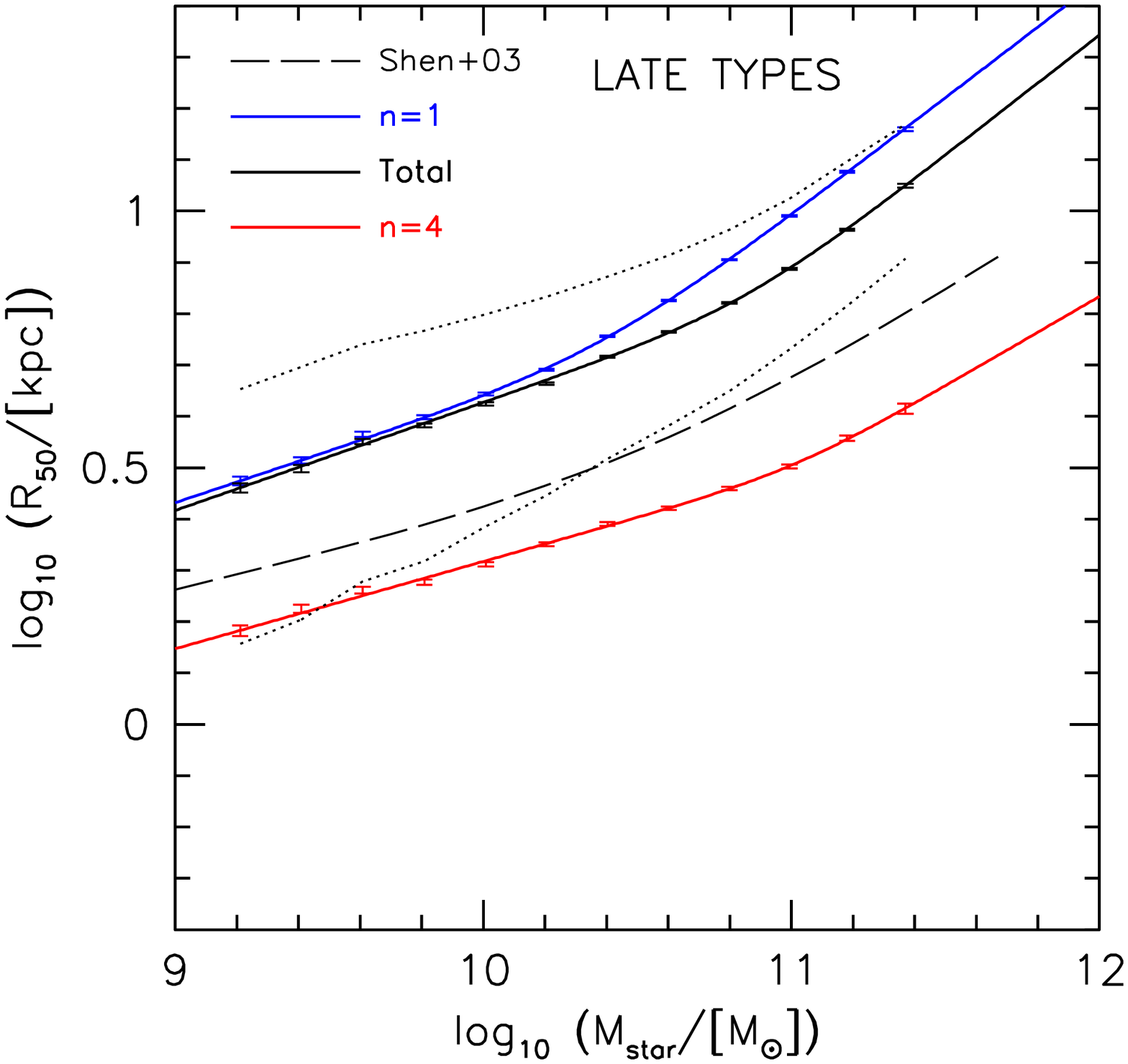,width=0.49\textwidth}
\psfig{figure=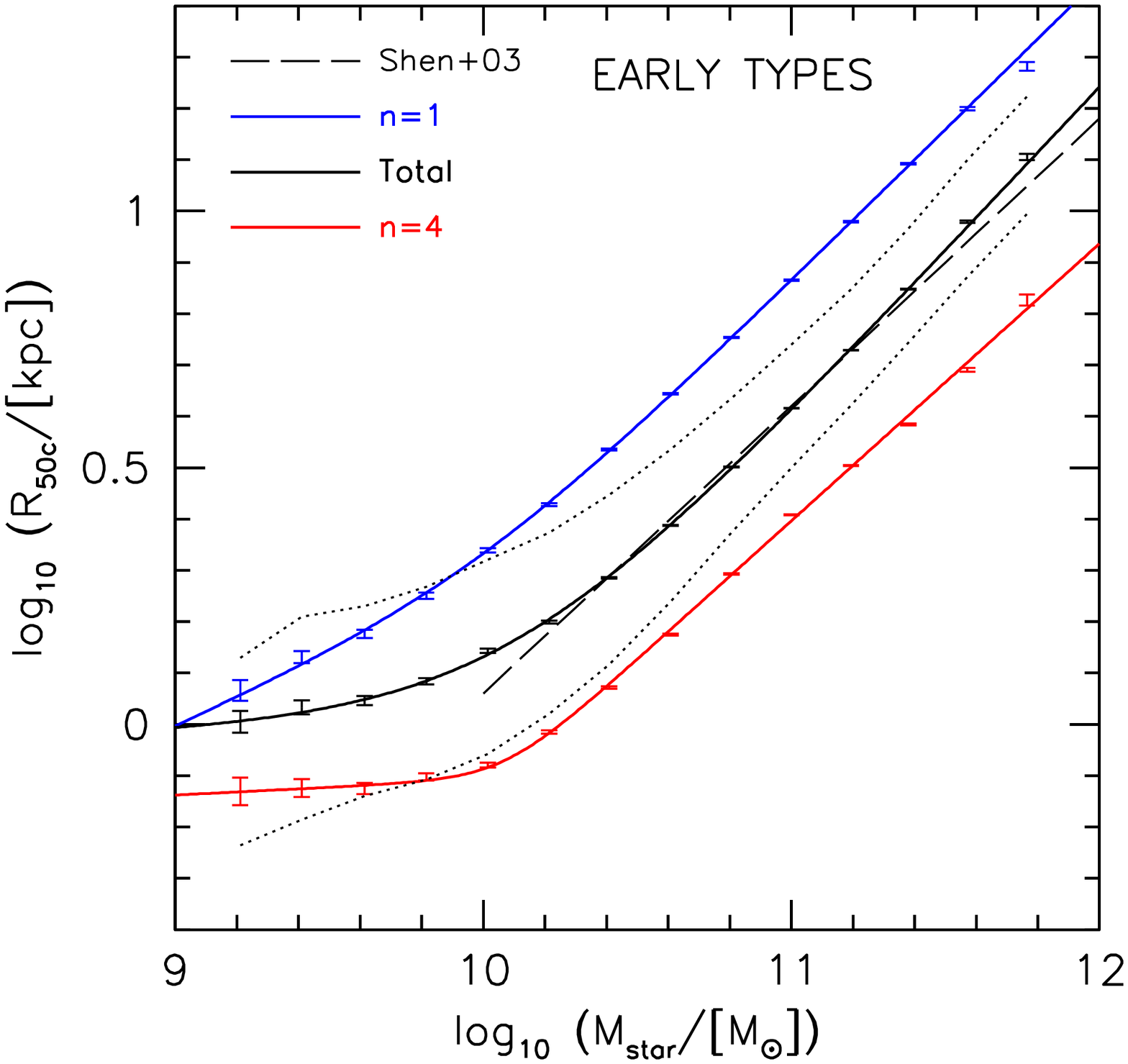,width=0.49\textwidth}}
\caption{Size-stellar mass relations for late-type (left panel) and
  early-type (right panel) galaxies. For the early-type galaxies we
  show circularized half-light radii, $R_{50\rm c}$, whereas for the
  late-type galaxies we show the major-axis half-light radii,
  $R_{50}$. In both panels the blue error bars show the median sizes
  of the disk (S\'ersic $n=1$ component), the red error bars show the
  median sizes of the bulge (S\'ersic $n=4$ component), and the black
  error bars show the median sizes of the whole galaxy. The solid
  lines show fits to these data using Eq.~\ref{eq:power2}. The
  parameters of these fits are given in Table.~\ref{tab:rmfit}.  For
  comparison we show the circularized half-light radii for early-type
  ($n>2.5$) and late-type ($n<2.5$) galaxies from Shen \etal
  (2003). For early-types there is good agreement, but for late-types
  the sizes from Shen \etal (2003) are smaller by $\simeq 0.15$ dex
  due to circularization. The dotted lines show the 16th and 84th
  percentiles for the total half-light sizes, which shows that the
  scatter in sizes decreases at higher masses. }
\label{fig:rm}
\end{figure*}

\subsection{Size - stellar mass relations}

Fig.~\ref{fig:rm} shows the size - stellar mass relations for
early-type (right panel) and late-type (left panel) galaxies.  The
red, blue and black lines show the relations for the $n=4$ components,
$n=1$ components, and total, respectively. The mass is always the
total stellar mass, but the sizes are those of the separate
components. For the late-type galaxies the sizes are the major-axis
$r$-band half-light sizes, $R_{50}$.  For the early-type galaxies the
sizes are the circularized $r$-band half-light sizes,
$R_{50c}=\sqrt{q} R_{50}$, where $q$ is the luminosity weighted
minor-to-major axis ratio of the galaxy ($q=q_{\rm d} f_{\rm d} +
q_{\rm b} f_{\rm b}$), and $R_{50}$ is the major axis half-light size.
Our use of major axis sizes for late-types and circularized sizes for
early-types is motivated by our desire to construct dynamical
models. For disk-dominated galaxies the major axis size is the only
sensible size to use. For bulge-dominated galaxies the situation is
more complicated due to the unknown intrinsic 3D shape. We use
circularized sizes (i.e., intermediate axis) for early-type galaxies
for two reasons: Firstly, in our dynamical models we assume that the
bulges are spherical; Secondly, for an ellipsoidal mass profile, the
potential is more spherical than the mass profile.

We fit the data using the following double power-law
\begin{equation}
  y= y_0\left(\frac{\Mstar}{M_0}\right)^{\alpha}\left[\frac{1}{2}+\frac{1}{2}\left(\frac{\Mstar}{M_0}\right)^\gamma\right]^{(\beta-\alpha)/\gamma},
\label{eq:power2}
\end{equation}
with $y=R_{50}$. These fits are shown with solid lines in
Fig.~\ref{fig:rm} and the parameters are given in
Table~\ref{tab:rmfit}.

For late-type galaxies the size-mass relation has a pronounced
curvature, with a slope of $0.47$ at high masses and $0.21$ at low
masses. Curvature in the half light radius - stellar mass relation for
late-types (defined by having $n<2.5$) was measured by Shen \etal
(2003). Their relation is shown by the long-dashed line. It has
similar slopes at low and high masses, but a significantly lower
zero-point.  This difference can mostly be attributed to the fact that
the sizes used by Shen \etal (2003) were performed using circular
apertures, which biases the sizes of disk dominated galaxies low, on
average by $\simeq 0.15$ dex

For early-type galaxies the half-light radius stellar mass relation
has a slope that ranges from $\simeq 0$ at low masses to $\simeq 0.6$
at high masses. Flattening of the slope at low luminosities has been
reported previously (e.g., Graham \& Guzman 2003; Graham \& Worley
2008).  At high masses our results are in good agreement with the
power-law scaling from Shen \etal (2003), who found $R_{50} \propto
\Mstar^{0.56}$ for early-types (defined to have S\'ersic index
$n>2.5$) with stellar masses greater than $\sim 10^{10}\Msun$. The
absolute differences between our size - mass relation and that from
Shen \etal (2003) are less than 10\%.

\subsubsection{Conversion from optical sizes to stellar mass sizes}
In theoretical models of disk galaxy formation, disks form inside-out
(e.g., Dutton \etal 2011). This results in color gradients, with
progressively larger scale lengths when going from stellar mass to
$K$-band light to $B$-band light. Observations of face-on disk
galaxies yield $B$-band sizes that are $\simeq 0.03$ dex larger than
$V$-band sizes, and that are $\simeq 0.06$ dex larger than $R$-band
sizes (MacArthur \etal 2003).

Assuming that the disks are exponential in each pass-band, and a
power-law relation between stellar mass-to-light ratio and color
(e.g., Bell \& de Jong 2001), then one can show that the stellar disk
will also have an exponential stellar mass density profile, with a
scale length, $R_{\rm d,*}$, given by
\begin{equation}
R_{\rm d,*}=R_{\rm d,R} / [ 1 + 2.5b(1-R_{\rm d,R}/R_{\rm d,B})],
\end{equation}
where $b$ is the slope of the relation between $M_*/L_R$ and $(B-R)$
color, $R_{\rm d,B}$ and $R_{\rm d,R}$ are the scale lengths of the
disk in $B$- and $R$-bands respectively. Adopting $b=0.683$ (Bell
\etal 2003), and $R_{\rm d,R}/R_{\rm d,B}=0.87$ implies $R_{\rm
  d,*}=0.82R_{\rm d,R}$, and thus $R_{\rm d,*}=0.76R_{\rm d,V}$.  The
rest frame disk scale lengths of galaxies in our SDSS sample are
roughly $V$-band, so we apply this correction to our late-type
galaxies.  We don't apply any corrections to the sizes of late-type
bulges, or to the sizes of early-type galaxies.

\subsection{De Vaucouleurs fraction - stellar mass relation}
\begin{figure}
\centerline{
\psfig{figure=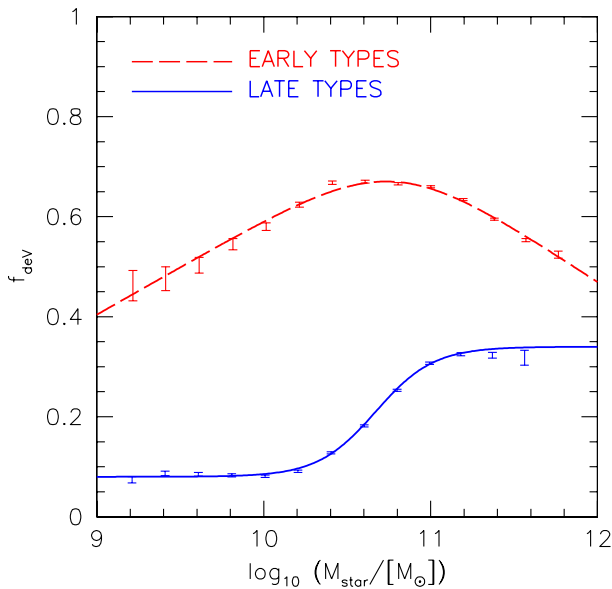,width=0.47\textwidth}}
\caption{De Vaucouleurs fraction, $f_{\rm deV}$, vs stellar mass,
  $\Mstar$, for early (red lines) and late-type (blue lines)
  galaxies. The de Vaucouleurs fraction is the fraction of $r$-band
  light in the S\'ersic $n=4$ component, derived from two component
  ($n=4$, $n=1$) fits to the galaxy images.  The error bars show
  the median and error on the median $r$-band bulge fraction in bins
  of width 0.2 dex in stellar mass.  The lines show fits to these data
  points, using Eq.~\ref{eq:bt} with parameters given in
  Table~\ref{tab:btfit}. }
\label{fig:bt}
\end{figure}

Using the \textsc{gim2d} two component fits as described above, we
compute median de Vaucouleurs fractions, $\fdev$, in
$r$-band light of early-type and late-type galaxies.  These relations are
shown in Fig.~\ref{fig:bt}. For late-type galaxies the relation is
fitted with the following equation:
\begin{equation}
\label{eq:bt}
f_{\rm deV}= f_{\rm deV2} + \frac{ f_{\rm deV1} - f_{\rm deV2}}{1 + (\Mstar/M_0)^{\gamma}}.   
\end{equation}
Here $f_{\rm deV1}$ is the asymptotic de Vaucouleurs fraction at low
masses, $f_{\rm deV2}$ is the asymptotic de Vaucouleurs fraction at high masses,
$M_{0}$ is the transition mass, and $\gamma$ controls the sharpness of
the transition.  For early-type galaxies, we use Eq.~\ref{eq:power2}
with $\log_{10} y = f_{\rm deV}$.  The parameters of the fits are given
in Table~\ref{tab:btfit}.

%% FIGURE 5

\begin{table}
 \centering
 \caption{Parameters of fits to the de Vaucouleurs fraction - stellar
   mass relations for early-type and late-types galaxies in
   Fig.~\ref{fig:bt}. For early-type galaxies we use Eq.~\ref{eq:power2} with
   $\log_{10}y=\fdev$. For late-type galaxies we use Eq.~\ref{eq:bt}.}
  \begin{tabular}{cccccc}
    \hline
    \hline  
    Late-type    &  $f_{\rm deV1}$ & $f_{\rm deV2}$ & $\log_{10} M_0$ & $\gamma$\\
    galaxies &0.08 & 0.34 & 10.67 & 2.5\\
    \hline
    Early-type  & $\alpha$ & $\beta$ & $\log_{10}M_0$ & $f_{\rm deV0}$ & $\gamma$\\
   galaxies  &0.19 & -0.22 & 10.77 & 0.67 & 1.75\\
\hline
\hline
\label{tab:btfit}
\end{tabular}
\end{table}

As discussed in \S\ref{sec:bdfits} for late-type galaxies, the $n=1$
component usually corresponds to a disk, while the $n=4$ component
usually corresponds to a spheroid. Thus for late-type we refer to the
de Vaucouleurs fractions as bulge fractions.  The median bulge
fractions are $\sim 10\%$ for stellar masses below $\Mstar \simeq
2\times 10^{10}\Msun$. Above this mass scale the median bulge
fractions increase rapidly with mass.  These trends are qualitatively
the same as those between global S\'ersic index and stellar mass from
Dutton (2009). In particular, our new results confirm the prevalence
of galaxies with low bulge fractions at stellar masses below $\Mstar =
3\times 10^{10}\Msun$.

As discussed in \S\ref{sec:bdfits} for early-type galaxies the $n=1$
component does not always correspond to a physical disk, and thus the
interpretation of $\fdev$ in terms of bulges and disks is less
straightforward. For our dynamical models our fiducial assumption is
that the $n=1$ component in early-type galaxies is spherical.

\subsection{The Tully-Fisher relation}
The Tully-Fisher (TF) relation is the relation between rotation
velocity (or linewidth) and luminosity (or stellar mass).  Here we
present a new determination of the stellar mass TF relation using data
from Verheijen (2001) and Pizagno \etal (2007).  As our velocity
measure we adopt $V_{2.2}$: the rotation velocity measured at 2.2 disk
scale lengths ($R$-band for the Verheijen sample and $i$-band for
Pizagno \etal sample). For the Verheijen (2001) galaxies we have
computed $V_{2.2}$ using disk scale lengths from McDonald, Courteau,
\& Tully (2009).  Stellar masses were first calculated using the
relations between color and stellar mass-to-light ratio from Bell
\etal (2003), with an offset of -0.1 dex, corresponding to a Chabrier
(2003) IMF, and then converted into the MPA/JHU masses using
Eq.~\ref{eq:mm}.

%% FIGURE 6
\begin{figure}
\centerline{
\psfig{figure=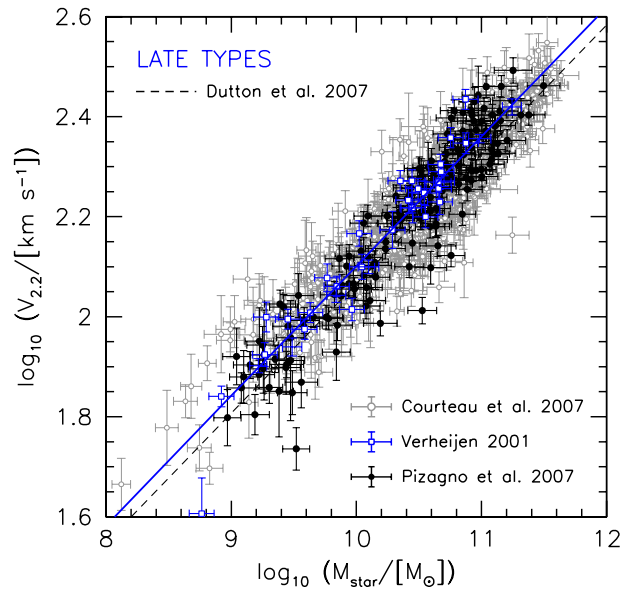,width=0.47\textwidth}
}
\caption{Tully-Fisher relation for spiral galaxies using data from
  Verheijen (2001); Pizagno \etal (2007) and Courteau \etal
  (2007b). The rotation velocity is measured at 2.2 disk scale
  lengths. The data is well fit by a power-law, as given by the solid
  blue line. This fit is in good agreement with that from Dutton \etal
  (2007) which was based on the Courteau \etal (2007b) sample. The
  small offset is due to a different stellar mass normalization.}
\label{fig:tf}
\end{figure}

The TF data are shown in Fig.~\ref{fig:tf}. The Verheijen (2001)
galaxies are shown with blue symbols, the Pizagno \etal (2007)
galaxies with black symbols. In grey we also show data from Courteau
\etal (2007b).  A fit to the Verheijen (2001) and Pizagno \etal (2007)
galaxies (solid blue line in Fig.~\ref{fig:tf}) is given by:
\begin{equation}
\label{eq:tf}
\log_{10} \left(\frac{V_{2.2}}{\rm \kms}\right) = 2.179+ 0.259 \log_{10} \left( \frac{M_{\rm star}}{10^{10.3}\Msun}\right),
\end{equation}
where the $1\sigma$ uncertainty on the zero point is 0.005 and the
$1\sigma$ uncertainty on the slope is 0.01. We do not include the
galaxies from Courteau \etal (2007) in this fit as this sample has a
heterogeneous selection function. However, the data set compiled by
Courteau \etal (2007b) has a TF relation (as calculated by Dutton
\etal 2007, dashed line in Fig.~\ref{fig:tf}) with a slope that is in
good agreement with Eq.~\ref{eq:tf}.  The difference in zero point is
explained by the different stellar mass normalizations used by Dutton
\etal (2007) and this paper.

\subsection{The Faber-Jackson relation}
The Faber-Jackson (FJ) relation is the relation between velocity
dispersion and luminosity (or stellar mass).  Here we present a new
determination of the FJ relation for early-type galaxies using data
from the SDSS.  This sample has the same selection criteria as used
for the early-type size-mass relation (Fig.~\ref{fig:rm}).

As is common practice, we correct SDSS fiber velocity dispersions to
velocity dispersions measured within the effective radius using the
empirical calibration from Jorgensen \etal (1995). Since we wish to
convert velocity dispersions to within $R_{\rm e}$ rather than $R_{\rm
  e}/8$ we use the quadratic formula from Jorgensen \etal (1995):
\begin{equation}
  \label{eq:apquad}
  \log_{10} \frac{\sigma_{\rm ap}}{\sigma_{\rm e}} = -0.065 \log_{10}\left(\frac{R_{\rm ap}}{R_{\rm e}}\right) 
  -0.013\left[\log_{10}\left(\frac{R_{\rm ap}}{R_{\rm e}} \right) \right]^2,
\end{equation}
where $\sigma_{\rm ap}$ is the velocity dispersion measured within the
aperture radius, $R_{\rm ap}$. For SDSS spectra $R_{\rm ap}=1.5$
arcsec. The quadratic formula is more accurate than the more commonly
used linear relation from Jorgensen \etal (1995):
\begin{equation}
\label{eq:aplin}
  \log_{10} \frac{\sigma_{\rm ap}}{\sigma_{\rm e}} = -0.04 \log_{10}\left(\frac{R_{\rm ap}}{R_{\rm e}}\right).
\end{equation}
Using Eq.~\ref{eq:apquad} results in average corrections that vary
between $0.00 \lta \log_{10}(\sigma_{\rm ap}/\sigma_{\rm e}) \lta
0.03$, i.e., the velocity dispersion within $R_{\rm e}$ is smaller
than the velocity dispersion with the SDSS aperture. The correction is
approximately zero for stellar masses of
$\log_{10}(\Mstar/\Msun)\simeq 10.4$, and increases to lower and
higher masses, reaching $\simeq 0.015$ at
$\log_{10}(\Mstar/\Msun)\simeq 9.5$ and $\simeq 0.03$ at
$\log_{10}(\Mstar/\Msun)\simeq 11.7$. Since the aperture corrections
to the velocity dispersion are small, they are not a significant
source of systematic uncertainty.
 
%% FIGURE 7
\begin{figure}
\centerline{
\psfig{figure=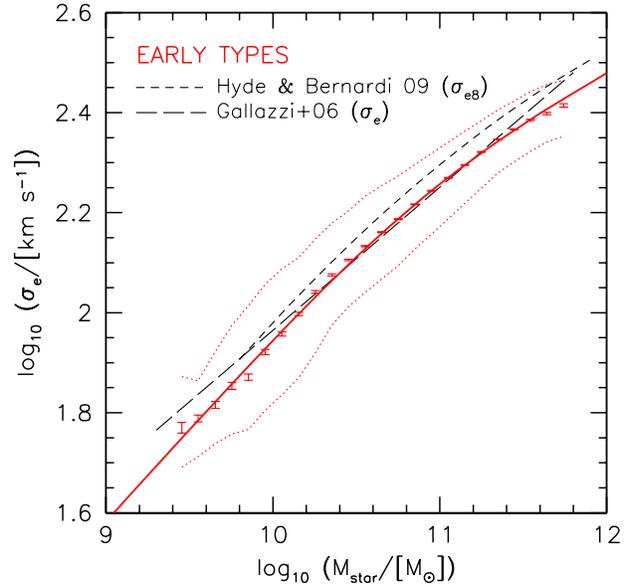,width=0.47\textwidth}
}
\caption{Velocity dispersion - stellar mass relation for early-type
  galaxies (also known as the Faber-Jackson relation). Velocity
  dispersions, $\sigma_{\rm e}$, have been corrected to the effective
  radius.  The dotted lines show the 16th and 84th percentiles of the
  distribution of $\sigma_{\rm e}$ at fixed mass.  The error bars show
  the median and error on the median velocity dispersion in bins of
  width 0.1 dex in stellar mass.  The solid line shows our double
  power-law fit to these data points, given in Eq.~\ref{eq:fj2}.  The
  long-dashed line shows the power-law relation from Gallazzi \etal
  (2006), Eq.~\ref{eq:fj1}, which over-predicts the velocity
  dispersions at low and high masses. The short-dashed line shows the
  quadratic relation from Hyde \& Bernardi (2009), which agrees well
  with our relation, except for a $\simeq 0.04$ dex offset due to a
  different definition of velocity dispersion. }
\label{fig:fj}
\end{figure}

The resulting FJ relation is shown in Fig.~\ref{fig:fj}. The error
bars show median velocity dispersions in bins of width 0.1 dex in
stellar mass, while the dotted lines show the 16th and 84th
percentiles of the distribution.  We fit the FJ relation using
Eq.~\ref{eq:power2} with $y=\sigma_{\rm e}$, finding
\begin{eqnarray}
\log_{10}\left(\frac{\sigma_{\rm e}}{\kms}\right)= 2.23 + 0.37\log_{10}\left(\frac{\Mstar}{10^{10.9}\Msun}\right) \nonumber \\
-0.19\log_{10}\left[\frac{1}{2}+\frac{1}{2}\left(\frac{\Mstar}{10^{10.9}\Msun}\right)\right].
\label{eq:fj2}
\end{eqnarray}

The long-dashed line shows the FJ relation based on SDSS data from
Gallazzi \etal (2006),
\begin{equation}
\label{eq:fj1}
\log_{10} \left( \frac{\sigma_{\rm e}}{\rm km s^{-1}}\right) = 2.051+ 0.286 \log_{10} \left( \frac{M_{\rm star}}{10^{10.3}\Msun}\right), 
\end{equation}
who use the same IMF and velocity dispersion definition as used here.
This power-law relation from Gallazzi \etal (2006) is in reasonable
agreement with our measurements.  There is, however, a curvature to
our relation. Curvature at high and low stellar masses has been
reported previously (e.g., Hopkins \etal 2008; Hyde \& Bernardi 2009;
Bernardi \etal 2011). In addition, curvature in the velocity
dispersion - luminosity relation was hinted at by early studies (Tonry
1981; Davies \etal 1983), and is now firmly established (Matkovi\'c \&
Guzman 2005). The short-dashed line shows the quadratic FJ relation
from Hyde \& Bernardi (2009). This agrees well with our measurements,
except for a $\simeq 0.04$ dex zero point offset which is due
different definitions of velocity dispersion ($\sigma_{\rm e}$ vs
$\sigma_{e8}$).

At lower masses $\Mstar \lta 10^{10}\Msun$ the velocity dispersions
are below $80\kms$ and are subject to larger systematic
uncertainties. However, none of our results are strongly sensitive to
whether we use the power-law FJ from Gallazzi \etal (2006;
Eq.~\ref{eq:fj1}) or our double power-law fit using Eq~\ref{eq:fj2} at
low masses.  We note that the deviations from power-laws of the
$\sigma_{\rm e}-\Mstar$ and $R_{\rm e}-\Mstar$ relations are
correlated: i.e., where $\sigma_{\rm e}$ is lower than the power-law
fit, $R_{\rm e}$ is higher. Such a correlation is expected from the
fundamental plane.

\subsection{Converting velocity dispersions into circular velocities}
\label{sec:vsigma}
In order to use the TF and FJ relations as dynamical constraints to
our mass models, we need to convert velocity dispersions and rotation
velocities into circular velocities (i.e., the rotation velocity of a
massless test particle moving in a circular orbit). For late-type
galaxies, we assume that the rotation velocity is equal to the
circular velocity. This assumption is expected to be correct for
galaxies with rotation velocities greater than about 50 $\kms$. For low
mass galaxies pressure support from turbulence is expected to result
in rotation velocities that are significantly lower than the
circular velocities (e.g., Dalcanton \& Stilp 2010).

Converting velocity dispersions into circular velocities is less
straightforward. Based on the spherical Jeans equation, there is
dependence on the density profile of the tracer population, the slope
of the velocity dispersion profile, and the anisotropy parameter.
Conversion factors motivated by the Jeans equation that are used in
the literature typically range from $\sqrt{2}$ to $\sqrt{3}$ (Courteau
\etal 2007a).  

In this section we use results and data from the literature to
motivate a conversion factor between the velocity dispersion within
the effective radius, $\sigma(<R_{\rm e})\equiv \sigma_{\rm e}$, and
circular velocity at the effective radius, $\Vcirc(R_{e})$, of
$f=V_{\rm circ}(R_{\rm e})/\sigma_{\rm e}= 1.54^{+0.11}_{-0.10}$,
i.e., roughly half way between $\sqrt{2}$ and $\sqrt{3}$.

We start our discussion with the result from Padmanabhan \etal (2004)
who used models of elliptical galaxies in NFW haloes to argue that
\begin{equation}
\label{eq:pad}
\Vcirc(R_{\rm e})=1.65\sigma_{\rm e},
\end{equation}
with a $\sim 10\%$ uncertainty in the conversion factor depending on
the anisotropy profile. Using similar arguments Schulz \etal (2010)
adopt
\begin{equation}
\label{eq:schulz}
  \Vcirc(R_{\rm e})=1.7 \sigma(<R_{\rm e}).
\end{equation}
Wolf \etal (2010) argue that $M_{1/2}\simeq 3\sigma^2_{\rm los}
r_{1/2}/G$, where $M_{1/2}$ is the spherical mass enclosed within the
3D half-light radius, $r_{1/2}$, and $\sigma_{\rm los}$ is the line of
sight velocity dispersion of the whole system. This is equivalent to
\begin{equation}
\label{eq:wolf}
\Vcirc(r_{1/2})=1.73\sigma_{\rm los}.
\end{equation}
For most galaxy light profiles $r_{1/2}=1.33 R_{\rm e}$ (Wolf \etal
2010), and since the circular velocity profiles of galaxies typically
are close to constant, we expect that $\Vcirc(r_{1/2})\simeq
\Vcirc(R_{\rm e})$ to within a few percent.  Since the integrated velocity
dispersion is expected to be slightly lower than the velocity
dispersion within the projected half-light
radius. Eqs. ~\ref{eq:pad}-\ref{eq:wolf} are expected to be roughly
equivalent.

Using Schwarzschild models of 25 early-type galaxies and the
assumption of a mass follows light model, Cappellari \etal (2006) find
that the dynamical mass is given by
\begin{equation}
\label{eq:mdyn}
M_{\rm dyn}=5.0 (\pm 0.1) R_{\rm e} \sigma^2_{\rm e} / G.
\end{equation}
We note that the dynamical mass so defined only makes physical sense
if the system has a finite mass (which is the case for a mass follows
light model of a galaxy). In all other cases, the radius at which the
dynamical mass is measured needs to be specified. The mass follows
light assumption is obviously incorrect for galaxies embedded in dark
matter haloes, but if the baryons dominate within the optical part of
the galaxy (for example the half light radius), then mass follows
light models may be a reasonable approximation.  Models with a dark
matter halo typically have flatter velocity dispersion profiles than
those without (e.g., Fig.~5 in Cappellari \etal 2006). Thus when one
fits a galaxy which contains stars and a dark halo using a model with
just stars, the model will tend to overestimate the total mass at
small radii and underestimate it at large radii. Since the SAURON
kinematics used by Cappellari \etal (2006) are typically confined to
less than $R_{\rm e}$, we expect that their mass follows light models
underestimate the total mass within $R_{\rm e}$.

The conventional way of writing the dynamical mass
(Eq.~\ref{eq:mdyn}), can be reinterpreted in terms of a more
dynamically meaningful mass, namely the mass enclosed within the half
light radius.  The half-light radius contains half the projected
light, by definition. But for a Hernquist sphere (which approximates a
de Vaucouleurs profile in projection) the half-light radius encloses
(in 3D space) 41.6\% of the light.  The mass follows light assumption
means that $M_{\rm dyn}=\Mstar$, and thus the 3D mass within the 2D
half-light radius is given by
\begin{equation}
  M(<R_{\rm e})=0.416 M_{\rm dyn}=2.08(\pm 0.04) R_{\rm e}\, \sigma^2_{\rm e}/G.
\end{equation}
 Writing this in terms of circular velocities gives
\begin{equation}
\Vcirc(R_{\rm e})=1.44(\pm0.01)\,\sigma_{\rm e}.
\end{equation}

We thus see that the dynamical masses derived by Cappellari \etal
(2006) are inconsistent with those derived by Padmanabhan \etal
(2004), Schulz \etal (2010), and Wolf \etal (2010). In particular, the
dynamical masses used by Schulz \etal (2010) are a factor of 1.39
times higher than those of Cappellari \etal (2006). This difference
helps to explain why the dark matter fractions are so much higher
($\sim 60\%$) for galaxies in Schulz \etal (2010) than in Cappellari
\etal (2006) ($\sim 30\%$), even though the IMFs are nominally the
same.

%% FIGURE 8
\begin{figure}
\centerline{
\psfig{figure=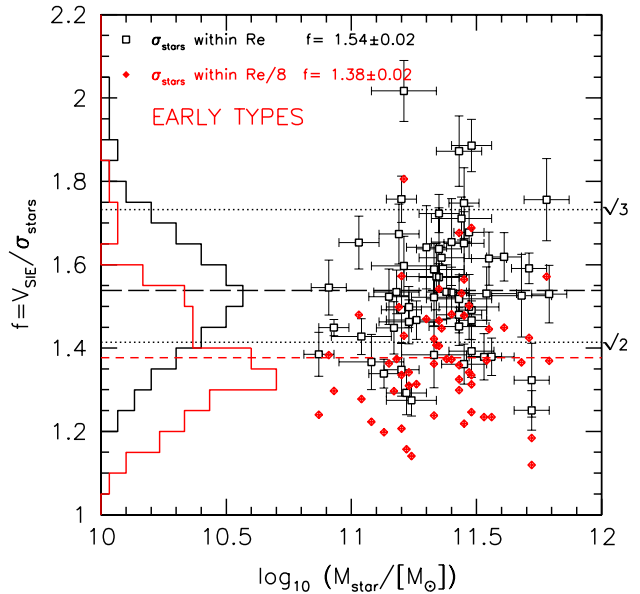,width=0.47\textwidth}
}
\caption{Relation between circular velocity from strong gravitational
  lensing to stellar velocity dispersion. The black squares show the
  ratio with velocity dispersion measured within the (rest-frame
  V-band) effective radius, $R_{\rm e}$, while the red diamonds show
  the ratio with velocity dispersion measured within $R_{\rm
    e}/8$. For velocity dispersion measured within $R_{\rm e}$ the
  median conversion factor between velocity dispersion and circular velocity
  is $f=1.54$.}
\label{fig:vv}
\end{figure}

Strong gravitational lensing gives robust measurements (typical
accuracy of $\sim 1\%$) of the total projected mass within the
Einstein radius.  Using lenses from the SLACS survey (Bolton \etal
2006, 2008a), Bolton \etal (2008b) find that the velocity dispersion
of the singular isothermal ellipsoid (SIE) lens model, $\sigma_{\rm
  SIE}$, is similar to the velocity dispersion of the stars. When
using the SDSS fiber velocity dispersions they find $\sigma_{\rm
  fiber}=0.948\pm0.008\sigma_{\rm SIE}$, and after correcting velocity
dispersions to $R_{e}/8$ using the empirical formula of Jorgensen
\etal (1995), they find $\sigma(R_{\rm e}/8)\equiv\sigma_{\rm
  e8}=1.019\pm0.008 \sigma_{\rm SIE}$.  Converting lens velocity
dispersions into circular velocities using $V_{\rm
  SIE}=\sqrt{2}\sigma_{\rm SIE}$ results in
\begin{equation}
V_{\rm SIE}=1.39(\pm0.01)\sigma_{\rm e8}.
\end{equation}
This motivates a conversion factor of $\simeq \sqrt{2}$ between
central velocity dispersion and circular velocity.  However, since the
velocity dispersion decreases with increasing radius, measuring the
velocity dispersion within a larger aperture will result in a larger
conversion factor between dispersion and circular velocity.
Furthermore, since the typical Einstein radii of SLACS lenses is
$\simeq 0.5 R_{\rm e}$, the lensing data more closely constrain the
circular velocity at radii significantly greater than $R_{\rm e}/8$.
 
In Fig.~\ref{fig:vv} we show results using data from SLACS survey as
presented in Auger \etal (2009, 2010b). The median stellar mass is
$2\times 10^{11}\Msun$, with a range of $\sim 1-4 \times 10^{11}$
(assuming a Chabrier IMF). We correct the SDSS fiber (3 arcsec diameter)
velocity dispersions into dispersions within $R_{\rm e}$ and $R_{\rm
  e}/8$ using the quadratic formula\footnote{Using the quadratic
  formula results in a correction of $\sigma_{\rm e8}/\sigma_{\rm
    e}=1.117$, compared to $\sigma_{\rm e8}/\sigma_{\rm e}=1.087$
  from the linear formula.}  from Jorgensen \etal (1995). We find
\begin{equation}
V_{\rm SIE}/\sigma_{\rm e8}=1.38\pm0.02,
\end{equation}
 and 
\begin{equation}
  V_{\rm SIE}/\sigma_{\rm e}=1.54\pm0.02,
\end{equation}
where the errors are the statistical uncertainty. The main systematic
uncertainty is from the conversion of fiber velocity dispersions to
other apertures. For $\sigma_{e}$ the mean correction is -0.01 dex,
and thus the uncertainty on the correction can be ignored.  For
$\sigma_{e8}$ the mean correction is 0.038 dex, and thus the
systematic uncertainty might be comparable or larger to the
statistical uncertainty.

To summarize, the strong gravitational lensing masses favor a
conversion factor between velocity dispersion within the effective
radius and circular velocity at the effective radius, $f=V_{\rm
  circ}(R_{\rm e})/\sigma(<R_{\rm e}) $, roughly half way between
$\sqrt{2}$ and $\sqrt{3}$. The dynamical modeling from Cappellari
\etal (2006) results in a slightly lower conversion factor, but we
expect that this is underestimated due to the assumption that mass
follows light.

In what follows we will adopt a conversion factor of
$f=1.54^{+0.11}_{-0.10}$, i.e., the uncertainty on $f$ is 0.03 dex.
This uncertainty encompasses the conversion factors of 1.44
from Cappellari \etal (2006) and 1.65 from Padmanabhan \etal (2004).

\subsection{Gas mass -  stellar mass relation}
\begin{figure}
\centerline{
\psfig{figure=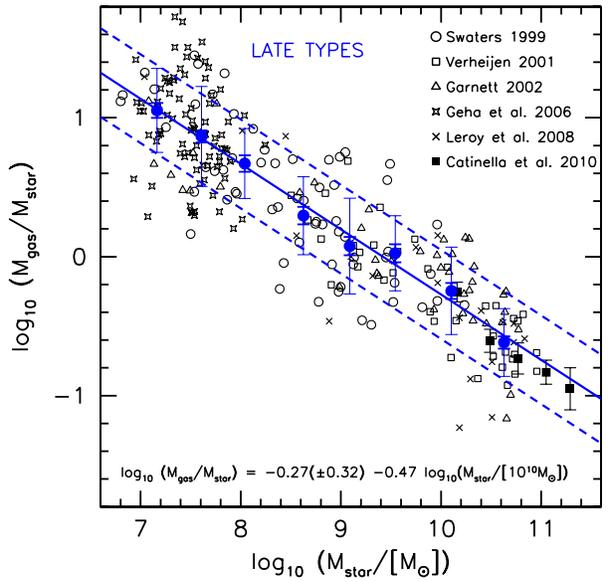,width=0.47\textwidth}}
\caption{Gas-to-stellar mass ratio vs stellar mass for late-type
  galaxies.  The gas is the cold atomic and molecular gas, including
  Helium. The relation is well fit with a power-law as given, with a
  scatter of $0.32$ dex, which is independent of stellar mass.}
\label{fig:gs}
\end{figure}

Late-type galaxies generally have gas fractions that increase with
decreasing stellar mass or luminosity (e.g., McGaugh \& de Blok 1997;
Kannappann 2004, Geha \etal 2006, Baldry \etal 2008, Catinella \etal
2010).  Fig.~\ref{fig:gs} shows the relation between gas-to-stellar
mass ratio and stellar mass for late-type galaxies using data from
Swaters (1999), Verheijen (2001), Garnett (2002), Geha \etal (2006),
Leroy \etal (2008) and Catinella \etal (2010). A factor of 1.36
correction for helium has been included in all gas masses either by us
or the original authors. The stellar masses have been measured using
relations between color ($B-V$ or $B-R$) and $M/L$ from Bell \etal
(2003), with 0.1 dex subtracted, and then converted to MPA/JHU masses
using Eq.~\ref{eq:mm}. All data sets include atomic hydrogen, Garnett
(2002) and Leroy \etal (2008) also include molecular hydrogen.  For
stellar masses above $10^{10}\Msun$ the mean ratio between molecular
hydrogen and stellar mass is 8\%. For the samples without molecular
gas we apply this factor to derive the mean molecular gas mass.
Theoretical support for this conversion comes from semi-analytic
models (Dutton, van den Bosch, \& Dekel 2010a), which find that the
ratio between molecular gas and stellar mass at redshift $z=0$ is
$\simeq 8\%$, independent of stellar mass, and with small scatter
$\simeq 0.1$ dex.

%% FIGURE 9

The gas mass - stellar mass relation in Fig.~\ref{fig:gs} is well fitted by
\begin{equation}
 \log_{10} \left(\frac{M_{\rm gas}}{M_{\rm star}}\right) = 
 -0.27 -0.47 \log_{10} \left(\frac{M_{\rm star}}{10^{10}\Msun}\right).
\end{equation}
As specific examples, for a stellar mass of $\Mstar=10^{11} \Msun$,
the mean gas fraction is 15\% (which is divided roughly equally
between atomic and molecular gas), while for a stellar mass
$\Mstar=10^{9}\Msun$, the cold gas fraction is 61\% (which is
dominated by atomic gas).  The observed scatter in gas-to-stars ratio
is a factor of $\simeq 2$, and is independent of stellar mass.

The majority of galaxies with non-detections in the massive galaxy
sample ($\Mstar >10^{10}\Msun$) of Catinella \etal (2010), and the
dwarf galaxy sample ($\Mstar \sim 10^{8}\Msun$) of Geha \etal (2006)
are on the red sequence. We thus assume that early-type galaxies have,
in general, insignificant amounts of gas to be dynamically important.

%% FIGURE 10
\begin{figure}
\centerline{
\psfig{figure=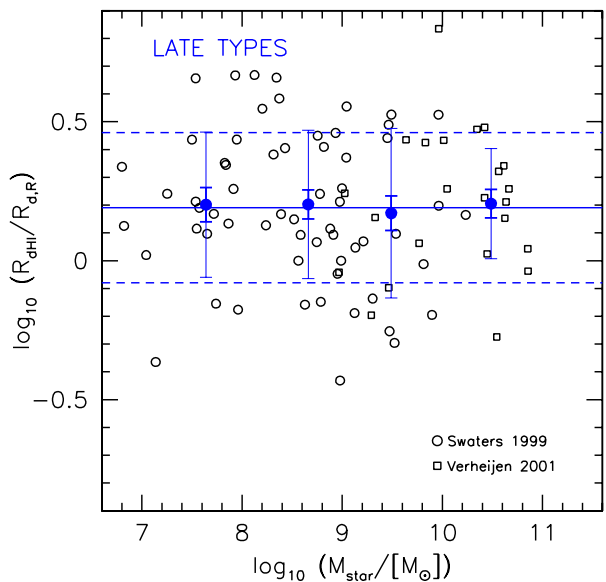,width=0.47\textwidth}}
\caption{Ratio between H{\scriptsize I} ($R_{\rm d,HI}$) and $R$-band
  ($R_{\rm d,R}$) disk scale lengths vs stellar mass for late-type
  galaxies, using data from Swaters (1999) and Verheijen (2001). The
  blue points with error bars show the mean and error on the mean in 4
  stellar mass bins of width 1 dex. The mean ratio between
  H{\scriptsize I} and R-band disk scale lengths is independent of
  stellar mass and is $\log_{10} R_{\rm d,HI}/R_{\rm
    d,R}=0.19\pm0.03$, with a scatter of 0.26 dex.}
\label{fig:RHI}
\end{figure}

\subsection{Gas size - optical size relation}
It is well known that the sizes of atomic gas disks are, on average,
larger than the sizes of the stellar disks (e.g., Swaters 1999;
Verheijen 2001).  Fig.~\ref{fig:RHI} shows the ratio between the disk
scale lengths of the atomic \hi\ gas and the $R$-band light, for
late-type galaxies from the samples of Swaters (1999) and Verheijen
(2001). For the Swaters (1999) sample we use the scale lengths
determined by the author. For the Verheijen (2001) sample we use disk
scale lengths from McDonald, Courteau, \& Tully (2009), and \hi\ disk
scale lengths measured by us. In both cases the \hi\ scale lengths are
determined from marked fits. That is, the exponential part of the \hi\
density profile is marked by hand, and an exponential profile is
fitted to this region.  We note that these \hi\ scale lengths are
typically measured from the outer part of the \hi\ density
profile. The inner part is often constant density or contains a
hole. Molecular gas typically dominates over atomic gas in these
regions, so that the total gas profile is approximately exponential.

These data span three orders of magnitude in stellar masses, but show
no evidence for a mass dependence to the size ratios of atomic and
stellar disks. The ratio between the scale lengths of the atomic gas
disk and the $R$-band light is approximately log-normally distributed,
with a mean
\begin{equation}
\log_{10} (R_{\rm d,\hi}/R_{\rm d,R})=0.19\pm0.03
\end{equation}
and observed scatter of $0.26$ dex.  A larger scale length of the
atomic gas disk compared to the stellar disk is a natural outcome of a
density dependent star formation law (Dutton 2009). 

%% FIGURE 11
\begin{figure}
\centerline{
\psfig{figure=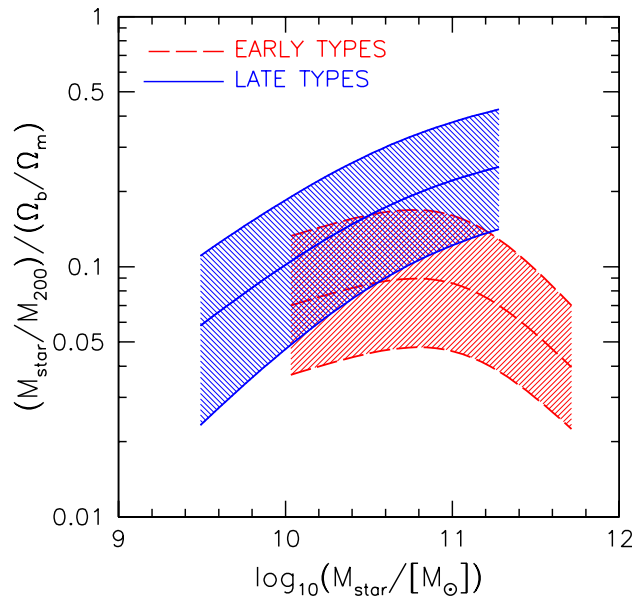,width=0.47\textwidth}
}
\caption{Halo mass vs Stellar mass relation for early-types (red) and
  late-types (blue) expressed in terms of the integrated star
  formation efficiency, assuming a Chabrier (2003) IMF. The shaded
  regions correspond to the systematic uncertainty in halo masses at
  fixed stellar mass. For early-type galaxies of all masses and most
  late-type galaxies the integrated star formation efficiency is less
  than 20\%.}
\label{fig:mmvir}
\end{figure}

\subsection{Halo mass - stellar mass relation}
\label{sec:mmvir}
We use the relations between the dark matter halo mass, $M_{200}$, and
stellar mass for central early-type and late-type galaxies from Dutton
\etal (2010b).  The virial radius of the dark matter haloes are
defined such that the mean density of the halo is 200 times the
critical density of the universe.  The halo mass is calculated as the
log of the mean halo mass at fixed stellar mass:
$\log_{10}<M_{200}>(\Mstar)$.  Dutton \etal (2010b) combines halo
masses measured from satellite kinematics (Conroy \etal 2007; More
\etal 2011), weak lensing (Mandelbaum \etal 2006; Schulz \etal 2010),
group catalogs (Yang \etal 2009) and halo abundance matching (Moster
\etal 2010; Guo \etal 2010; Behroozi \etal 2010), finding generally
good agreement between the different techniques. The results from
Dutton \etal (2010b) are shown in Fig.~\ref{fig:mmvir}, where we have
converted the Bell \etal (2003) masses to the MPA/JHU masses using
Eq.~\ref{eq:mm}. The shaded regions show the systematic uncertainty in
mean halo mass at fixed stellar mass, which is $\sim 0.25$ dex
($2\sigma$) for both early-types and late-types.

\subsection{Halo mass - halo concentration relation}
\label{sec:cm}
Dark matter only simulations in \LCDM cosmologies have shown there is a
tight correlation between halo concentration and halo mass (Navarro,
Frenk, \& White 1997; Bullock \etal 2001). We adopt the concentration
- mass relation for relaxed haloes in a WMAP 5th year cosmology (Dunkley
\etal 2009) from Macci\`o \etal (2008):
\begin{equation}
\label{eq:cm}
  \log_{10} c_{200} = 0.830 -0.098 \log_{10} \left( \frac{M_{200}}{10^{12}h^{-1}\Msun}\right). 
\end{equation}

The scatter in this relation for relaxed haloes is $\simeq 0.11$ dex
(Jing 2000; Wechsler \etal 2002; Macci\`o \etal 2007, 2008). The
concentration is correlated with the formation history of the halo
(Wechsler \etal 2002), with earlier forming haloes having higher
concentrations.

In this paper we are constructing mass models for the average
early-type and late-type galaxy of a given stellar mass. At low masses
late-types dominate, so they must form in typical haloes, likewise for
high mass early-types. However, low mass early-types, and high mass
late-types are minorities, and thus it is plausible that they could
form in a biased subset of haloes, and thus their concentrations may
be different than the mean. We refer to this bias as a formation bias.
The magnitude of the formation bias is constrained by the relatively
small scatter in the concentration mass relation.

An additional consideration is that Eq.~\ref{eq:cm} is for central
haloes. Sub-haloes have higher mean concentrations (e.g., Bullock
\etal 2001), due to their higher formation redshifts.  In the
simulation from Klypin \etal (2010) sub-haloes of masses between
$\Mvir = 10^{11}$ and $10^{12} \Msun$ have $\sim 30\%$ higher
concentrations than parent haloes of the same mass. This provides an
upper limit to the increase in halo concentrations. However, at fixed
stellar mass, sub-haloes are likely to have lower halo masses than
parent haloes (Neistein \etal 2011). This counteracts the effect of
higher concentrations in sub-haloes, and thus reduces the likelihood
that we are underestimating the effective concentrations of low mass
dark matter haloes.

Observations show that the satellite fraction is a strong function of
stellar mass, with lower satellite fractions at higher stellar masses
(Yang \etal 2008). For late-type galaxies the satellite fractions are
only $\sim 30\%$ for stellar masses of $\sim 10^{9}\Msun$ and drops to
below $\sim 10\%$ for stellar masses of $\sim 10^{11}\Msun$. Thus we
do not expect satellite galaxies to bias our results for
late-types. At a given stellar mass, the satellite fraction is higher
for early-type galaxies than late-type galaxies.  For stellar masses
of $\sim 10^{10}\Msun$ the satellite fraction for early-types is $\sim
50\%$. Thus it is possible that we are underestimating the
concentrations for low mass early-type galaxies by up to $\sim 30\%$.

%% FIGURE 12
\begin{figure*}
\centerline{
\psfig{figure=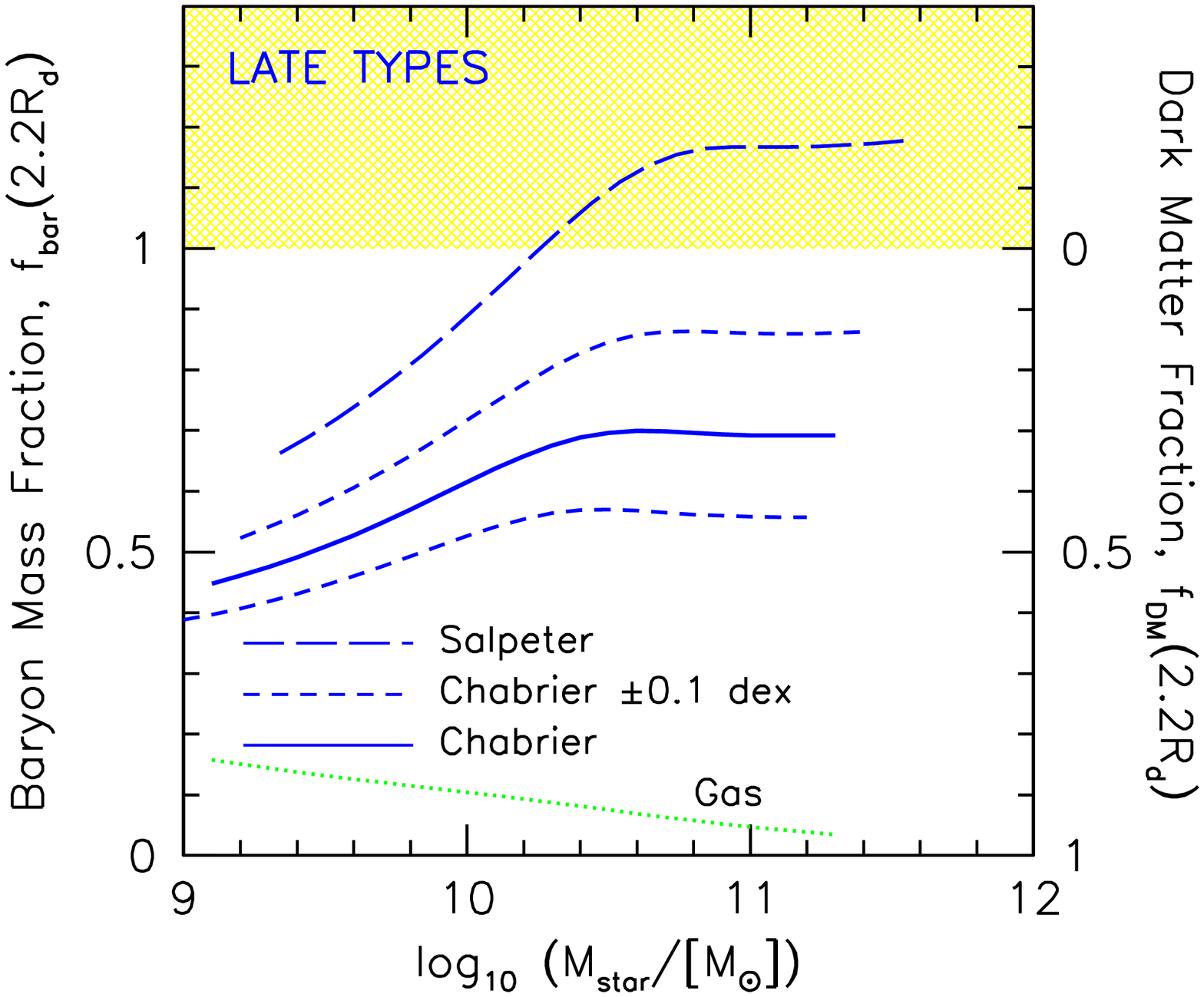,width=0.49\textwidth}
\psfig{figure=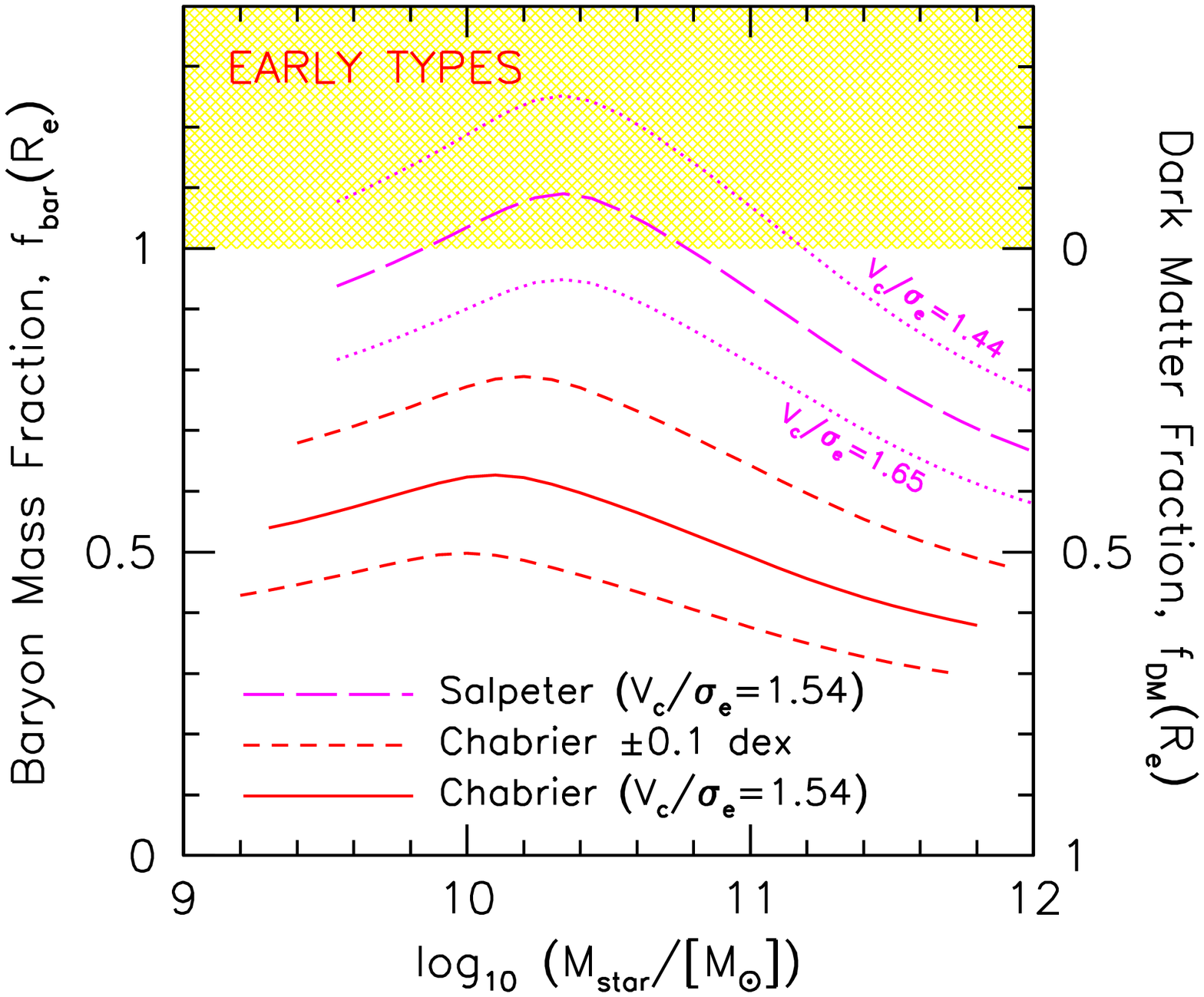,width=0.49\textwidth}}
\caption{Mean baryon mass fractions within the optical part of
  galaxies as a function of total galaxy stellar mass. For late-types
  (left panel) the radius is 2.2 stellar disk scale lengths, while for
  early-types (right panel) it is the projected circularized
  half-light radius. The solid lines show the results for a Chabrier
  (2003) IMF. The effect of changing the stellar mass normalization by
  $\pm 0.1$ dex is shown by the short-dashed lines, and by $+0.24$ dex
  (i.e., a Salpeter IMF) is shown by the long-dashed lines.  For
  late-type galaxies the baryon fractions increase with stellar mass
  and saturates at $f_{\rm bar}\sim 0.7$ at high masses. For
  early-types the baryon fractions reach a maximum near a stellar mass
  of $10^{10}\Msun$, and declines to higher and lower masses.  The
  yellow shading corresponds to the forbidden region where the baryons
  over-predict the dynamical mass. A Salpeter IMF over-predicts the
  dynamical masses for massive late-types, but is allowed for all
  early-type galaxies unless $V_{\rm c}/\sigma_{\rm e} \lta 1.6$.}
\label{fig:fdm}
\end{figure*}

\subsection{Overview of model parameters and constraints}
\label{sec:overview}
Our mass model has 9 parameters: 3 for the dark matter halo (mass
$M_{200}$, concentration $c$, and halo contraction model); and 6 for
the baryons (stellar mass, $\Mstar$, bulge fraction, $f_{\rm b}$,
bulge size, $R_{\rm b}$, stellar disk size, $R_{\rm d}$, cold gas
mass, $M_{\rm gas}$, gas disk size, $R_{\rm g}$).
To construct a model galaxy we then apply the following procedure:
\begin{enumerate}
\item Pick the type of galaxy (i.e., early or late). 
\item Pick a stellar mass, $\Mstar$, assuming a Chabrier IMF.
\item The six parameters for the baryons are then specified by the
  observational constraints (\S 3.3-3.9), up to the normalization of
  the stellar mass to light ratio, which we term $\Delta_{\rm IMF}$ to
  suggest this uncertainty is due to the unknown stellar IMF, although
  it also includes systematic uncertainties in measuring stellar
  masses for a given IMF.
\item Determine the halo mass from the halo mass - stellar mass
  relation from Dutton \etal (2010b) as discussed in \S
  \ref{sec:mmvir}, and shown in Fig.~\ref{fig:mmvir}.
\item Determine the halo concentration using Eq.~\ref{eq:cm} as
  discussed in \S\ref{sec:cm}.
\end{enumerate}

Thus our nine model parameters can be reduced to 2 primary unknowns:
the stellar IMF normalization, and the halo contraction model.  We
have one additional constraint, which is the model has to reproduce
the observed TF or FJ relation Eqs.~\ref{eq:tf} \& \ref{eq:fj2}.  We
construct model TF/FJ relations by computing the model at a range of
stellar masses.  Thus by comparing the observed and model TF/FJ
relations for a given IMF we can solve for the halo contraction
model. Alternatively, we can assume a halo contraction model and solve
for the stellar IMF.  Since the model is under constrained, we expect
there to be a degeneracy between the IMF and the halo contraction
model.

\subsection{Dark matter fractions}
\label{sec:fdm}
Before we compare our bulge-disk-halo models to the observed TF and FJ
relations, some useful insight into the dark matter content of
galaxies can be obtained by comparing the baryonic mass within a
sphere of some fiducial radius (2.2 disk scale lengths for late-types,
the half light radius for early-types) with the spherical mass derived
from the optical circular velocity.  The baryonic masses within the
fiducial radii, $M_{\rm bar}(<r)$ are determined using the mass models
as described in \S~\ref{sec:mm}, with the observational constraints
discussed earlier in this section. The total masses within the
fiducial radii are determined assuming spherical symmetry,
i.e., $M_{\rm tot}(<r) = r V^2_{\rm circ}(r)/G$. Thus the baryon
fraction is given by $f_{\rm bar}(<r)=[V_{\rm bar}(<r)/V_{\rm
  tot}(<r)]^2$, and the dark matter fraction is given by $f_{\rm
  DM}(<r)=1-f_{\rm bar}(<r)$. Note that for galaxies with a dominant
disk component the assumption of spherical symmetry will cause the
total mass to be over-estimated slightly.

The results of this calculation is shown in Fig.~\ref{fig:fdm}.  For
early-type galaxies (left panel) the dark matter fraction is lowest in
galaxies with stellar mass $\sim 10^{10}\Msun$, and increases for
higher and lower mass galaxies. The trend of increasing dark matter
fractions in higher mass (or luminosity) early-type galaxies, for a
universal IMF, is well established (e.g., Padmanabhan \etal 2004;
Gallazzi \etal 2006; Tortora \etal 2009; Auger \etal 2010b, Napolitano
\etal 2010).  The breaking of this trend at lower masses should not be
considered a surprise as dwarf spheroidal galaxies are known to be
dark matter dominated within their half-light radii (e.g., Tollerud
\etal 2011).

For early-type galaxies a Salpeter IMF is consistent with the
dynamical masses, as long as $V_{\rm circ}(R_{\rm e})/\sigma(<R_{\rm
  e})\equiv V_{\rm c}/\sigma_{\rm e} \gta 1.6$. As shown in
\S~\ref{sec:vsigma} the dynamical models of Cappellari \etal (2006)
imply that $V_{\rm c}/\sigma_{\rm e}=1.44\pm 0.01$, which, as noted by
these authors, disfavours a universal Salpeter IMF.  For late-type
galaxies (right panel) the dark matter fraction decreases with
increasing stellar mass in agreement with previous studies (e.g.,
McGaugh 2005; Pizagno \etal 2005; Dutton \etal 2007).  In late-type
galaxies the contribution of the cold gas (as shown by the green
dot-dashed line) is small, and for early-type galaxies we assume that
the contribution of the cold gas is negligible.

%%%%%%%%%%%%%%%%%%%%%%%%%%%%%%%%%%%%%%%%%%%%%%%%%%%%%%%%%%%%%%%%%%%%%%
%% SECTION 4: RESULTS
%%%%%%%%%%%%%%%%%%%%%%%%%%%%%%%%%%%%%%%%%%%%%%%%%%%%%%%%%%%%%%%%%%%%%%

\section{RESULTS}
\label{sec:results}
\begin{figure*}
\centerline{
 \psfig{figure=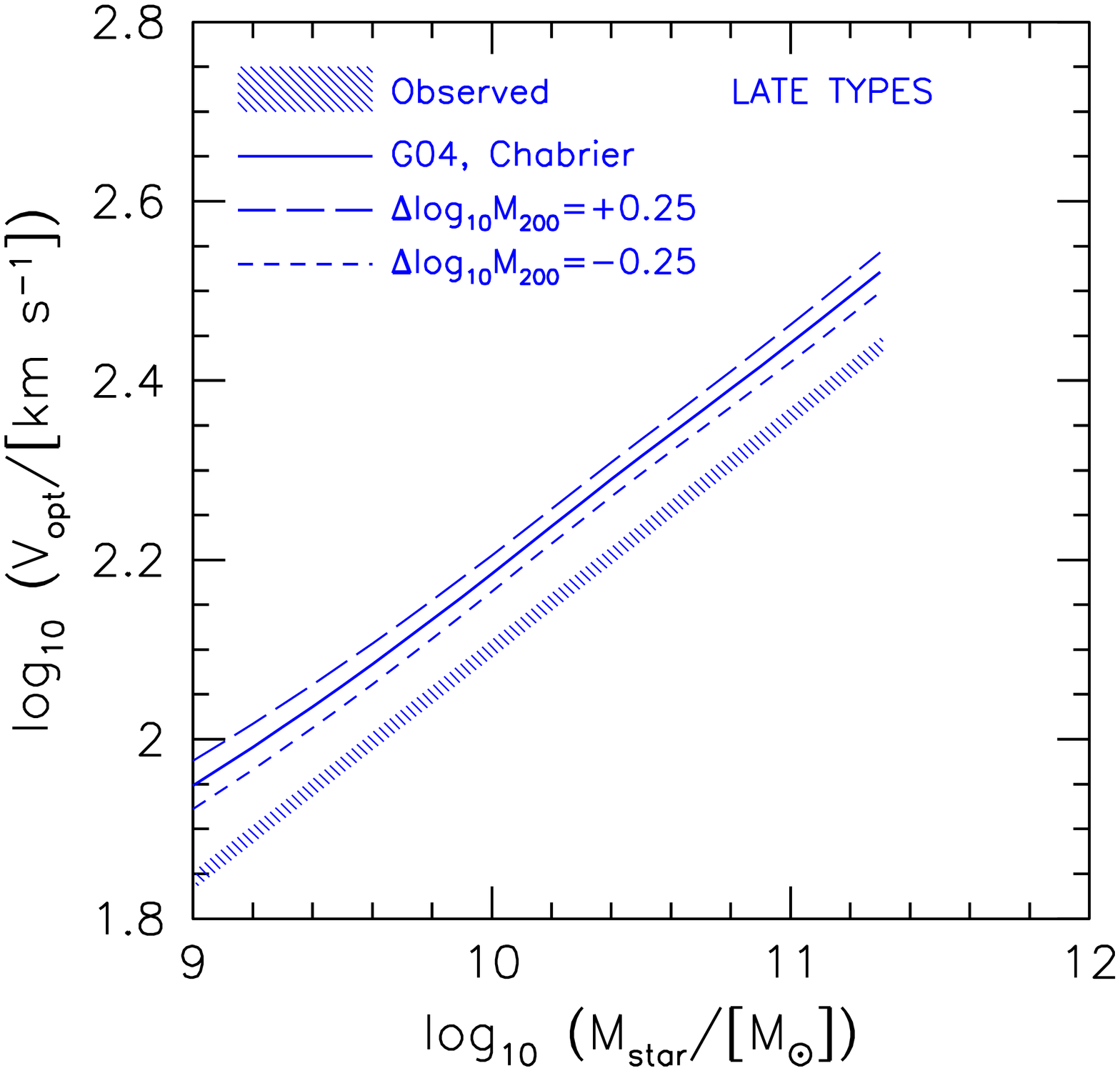,width=0.46\textwidth}
 \psfig{figure=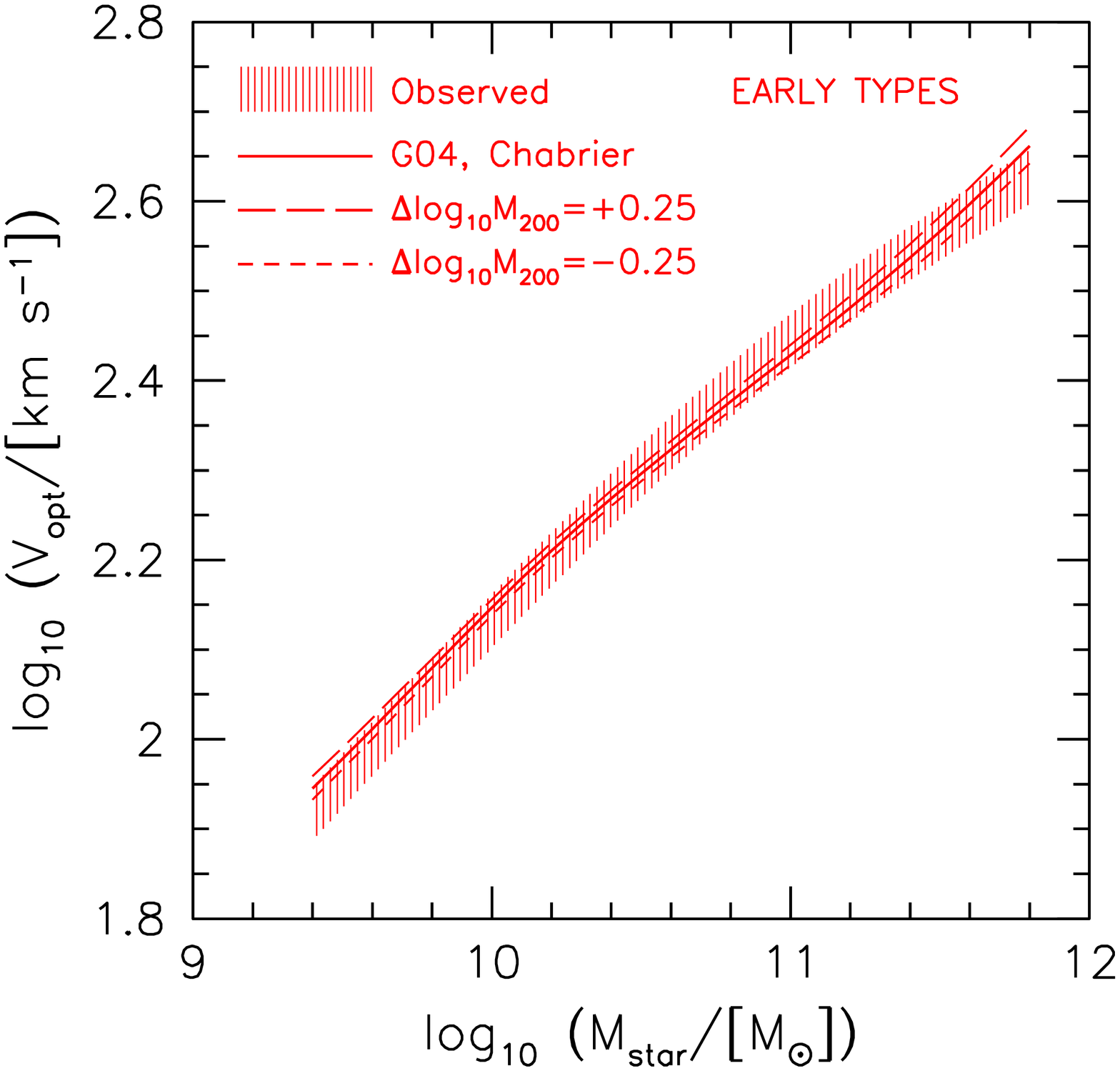,width=0.46\textwidth}}
\caption{TF and FJ relations for models with Gnedin \etal (2004, G04)
  halo contraction and a Chabrier (2003) IMF. The observations, with
  $2\sigma$ uncertainties, are given by shaded regions.  The dashed
  lines show the effect on the model by changing the halo masses by
  $\pm 0.25$ dex, which corresponds to the $2\sigma$ systematic
  uncertainty. The model reproduces the slopes of both relations, and
  the zero point of the FJ relation, but not the zero point of the TF
  relation.}
\label{fig:vm_type}
\end{figure*}

We now construct mass models for early-type and late-type galaxies, using
various assumptions about the stellar IMF, and adiabatic contraction,
and compare these to the observed optical circular velocity - stellar
mass (VM) relations.  We start with a model with standard (Gnedin
\etal 2004, G04) halo contraction and a standard (Chabrier 2003) IMF.
The predicted VM relations for early-type and late-types are shown in
Fig.~\ref{fig:vm_type}.  The shaded regions show the systematic
uncertainty in the observed relations. The solid lines show the model
relations, with the fainter lines corresponding to the systematic
uncertainty on the halo masses.  Note that for low stellar masses
($\Mstar \lta 10^{10.0}\Msun$ for early-types and $\Mstar \lta
10^{9.5}$ for late-types) we have extrapolated the halo mass - stellar
mass relation, and thus our results in these regions should be treated
with more caution.

The model nicely reproduces the slopes of the VM relations, and the
zero point of the VM relation for early-types, but it does not
reproduce the zero point for late-types.  At fixed stellar mass,
$M_{\rm star}$, the optical circular velocity, $V_{\rm opt}$, is
over-predicted by the model. This problem is known as the TF zero
point problem (e.g., Dutton \etal 2008).
As discussed in previous papers (Dutton \etal 2007; Dutton \& van den
Bosch 2009) there are three principle solutions to this problem of
simultaneously matching the rotation velocity - stellar mass
(i.e., Tully-Fisher) relation, disk size - stellar mass relation and
halo mass - stellar mass relation in the context of $\LCDM$.

\begin{itemize}
\item {Reduce the stellar mass}. 

\item {Reduce {\it pristine} halo concentration.} 

\item {Reverse or prevent halo contraction}. 

\end{itemize}

%% FIGURE  13

%% FIGURE 14
\begin{figure*}
\centerline{
\psfig{figure=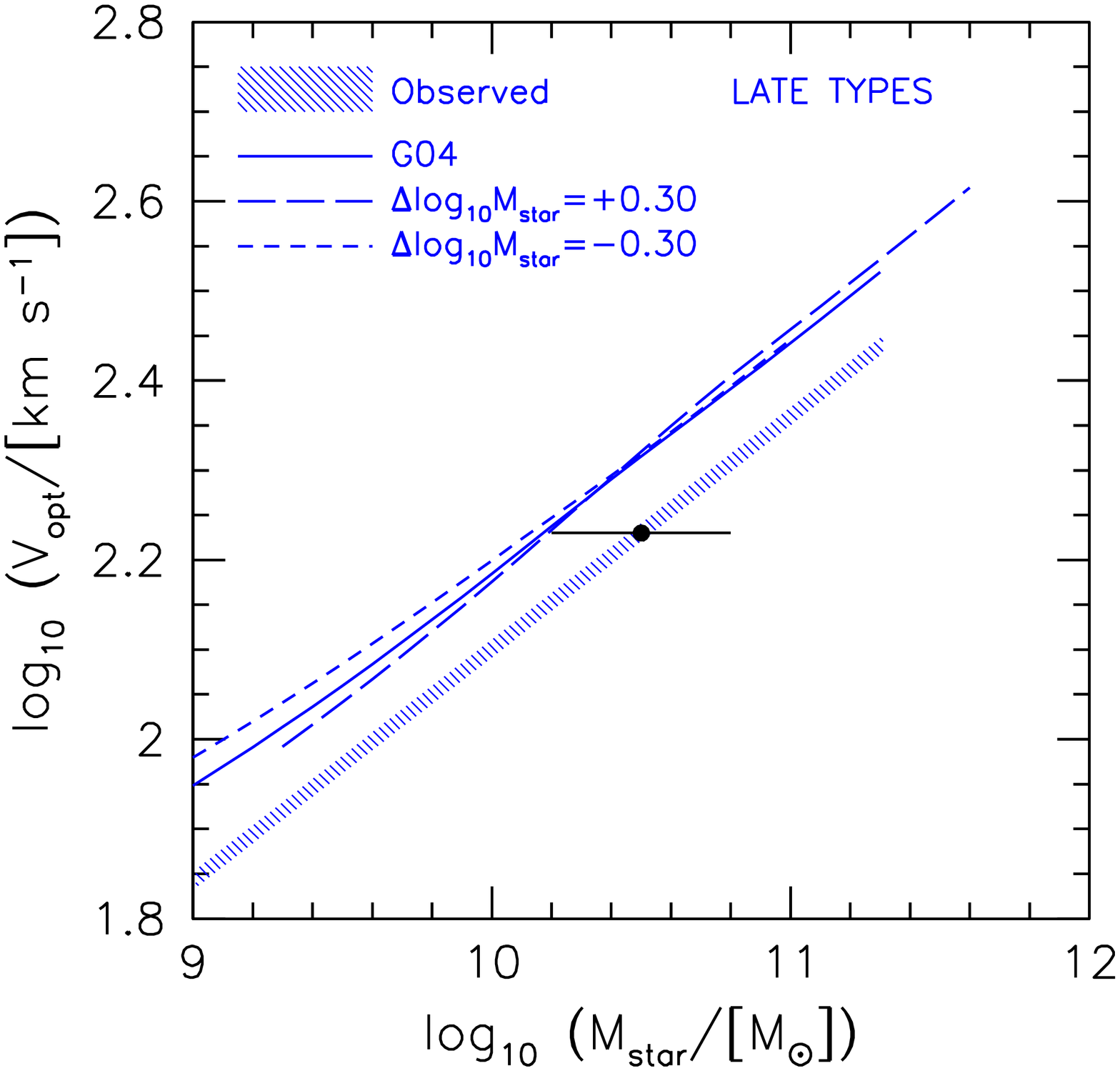,width=0.46\textwidth}
\psfig{figure=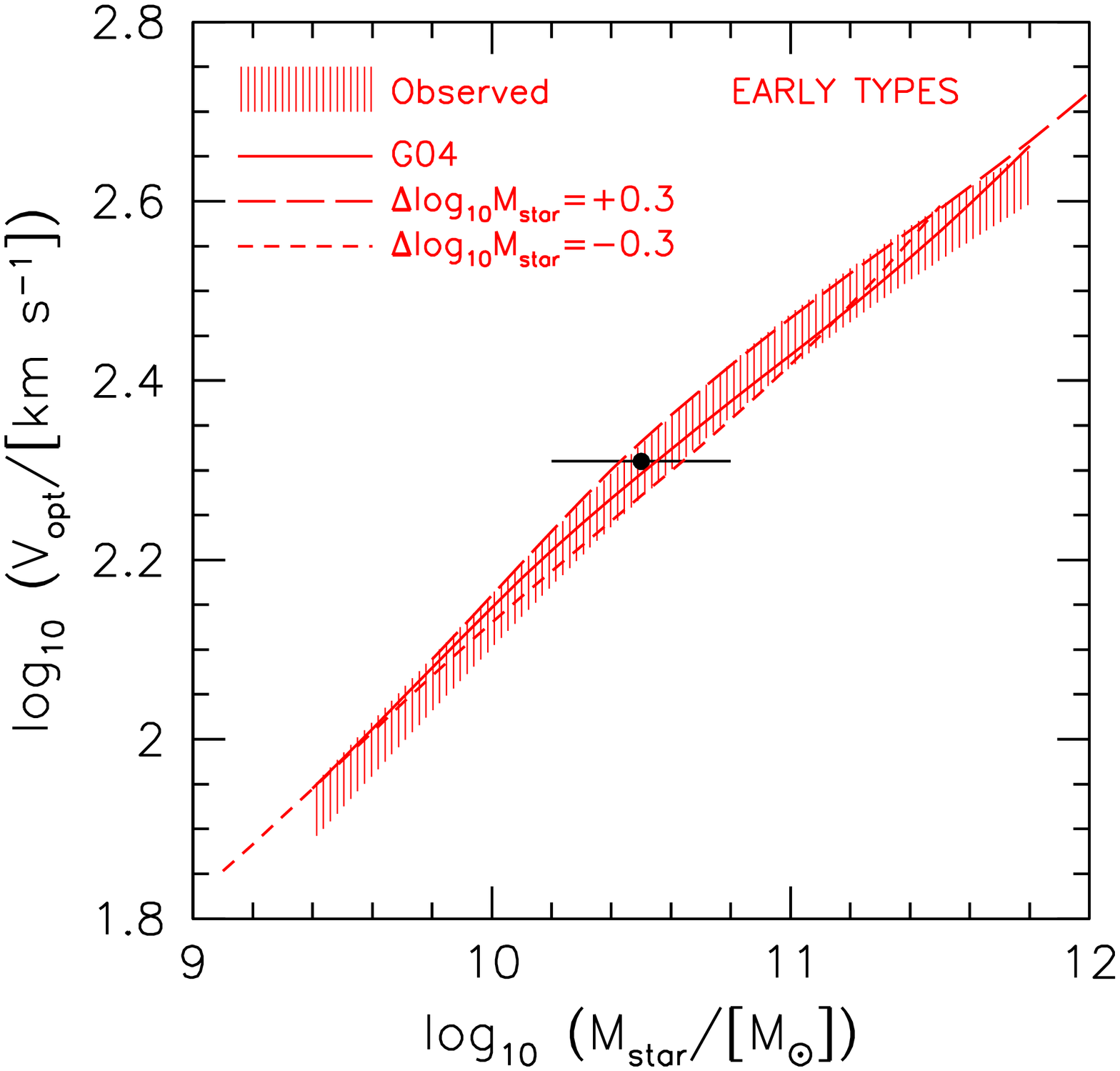,width=0.46\textwidth}}
\caption{Effect of stellar mass normalization on $V_{\rm opt}-\Mstar$
  relations for early-type (right panel) and late-type (left panel)
  galaxies.  The model relations are approximately independent to
  changes in stellar mass normalization of $\pm 0.3$ dex. This is
  because models with higher stellar masses (at fixed sizes and halo
  masses) also have higher $\Vopt$. By contrast the observed relations
  are strongly dependent on the stellar mass normalization (black
  horizontal bar). Thus reconciling the standard halo contraction
  model with observations requires late-type galaxies have stellar
  masses lower by factor of $\sim 2$. }
\label{fig:vm_type_imf}
\end{figure*}

%% FIGURE 15
\begin{figure*}
\centerline{
\psfig{figure=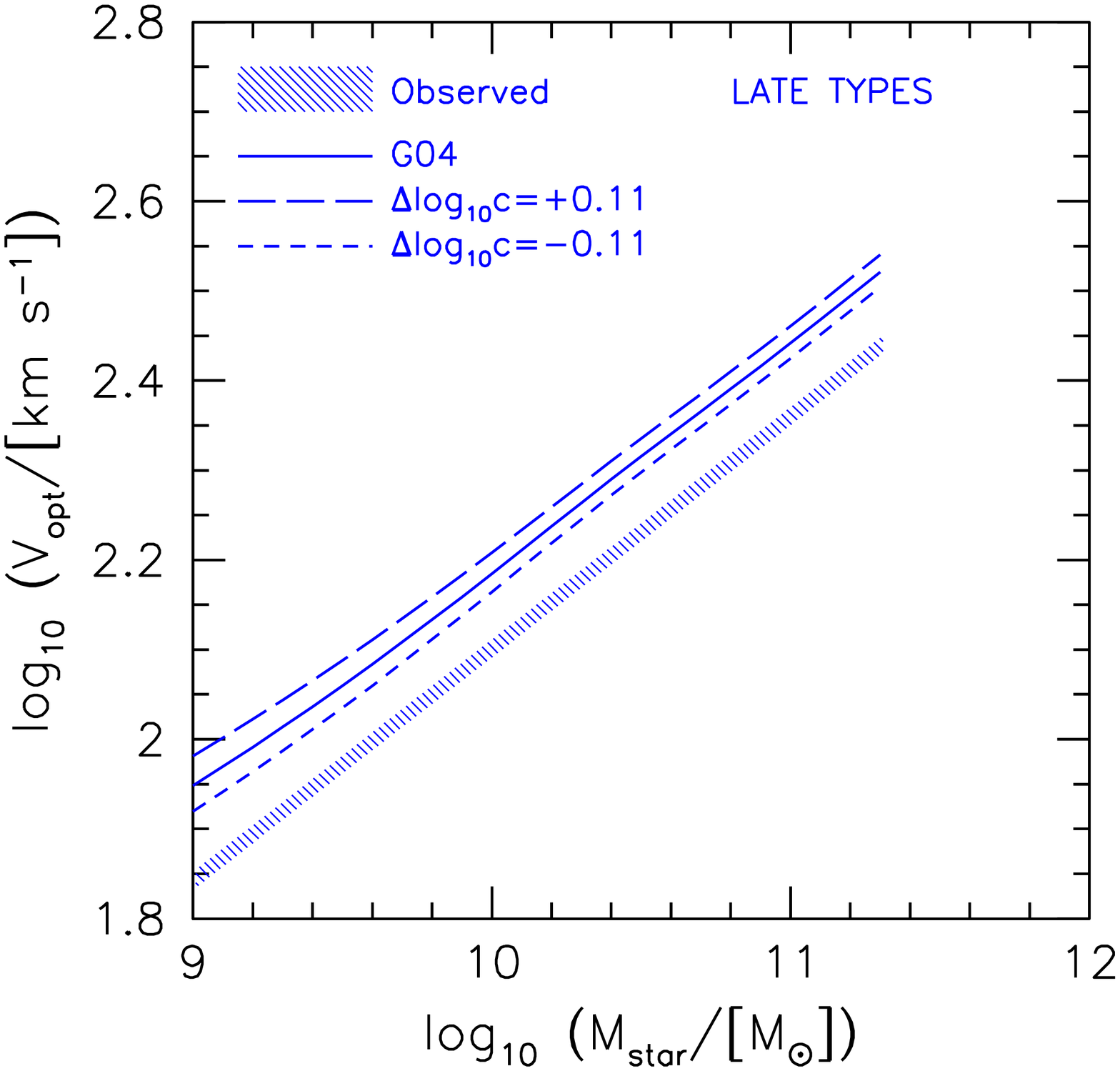,width=0.46\textwidth}
\psfig{figure=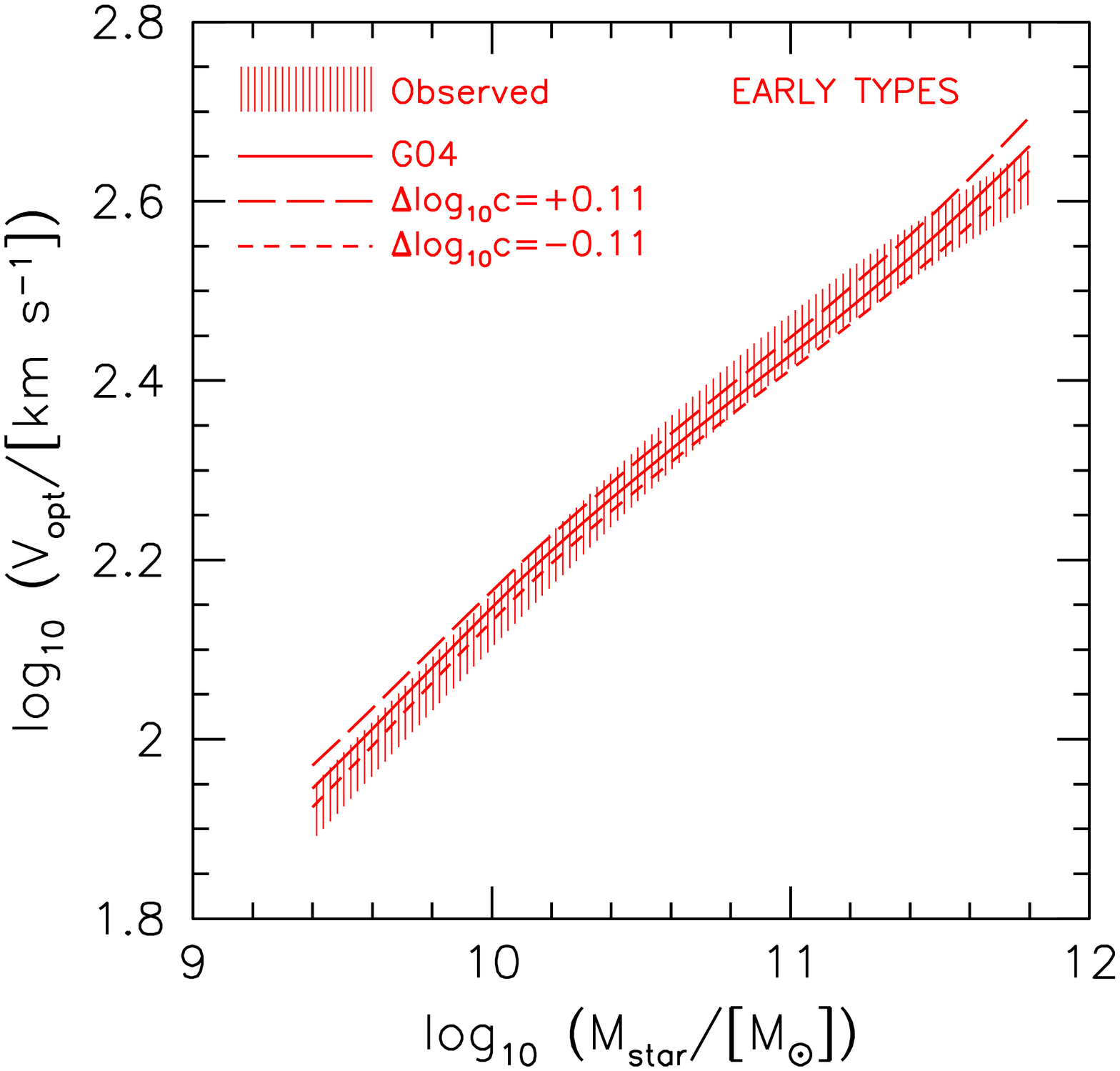,width=0.46\textwidth}}
\caption{Effect of halo concentration on $V_{\rm opt}-\Mstar$ relation
  for early-type (right panel) and late-type (left panel)
  galaxies. Higher halo concentrations result in higher optical
  circular velocities. The scatter in halo concentrations is $\simeq
  0.11$ dex (Macci\`o \etal 2008), and thus concentrations biases in
  both early-type and late-type galaxies are likely to be small.}
\label{fig:vm_type_c}
\end{figure*}

A change in stellar mass normalization can result from either a
systematic error in inferring stellar masses from observed photometry,
or due to a change in the stellar IMF.  A Salpeter IMF results in
stellar masses $\simeq 0.24$ dex higher than a Chabrier (2003) IMF
(which we adopt here as our standard IMF). Reducing stellar masses
would require an IMF with fewer low mass stars, or conversely, more
high mass stars.  The latter results in higher luminosities, and hence
lower mass-to-light ratios. However, there is a limit to how much the
stellar $M/L$ can be reduced, especially for old or moderately old
stellar populations, because stellar populations with bottom light
IMFs become dominated by stellar remnants at late times (van Dokkum
2008).

Fig.~\ref{fig:vm_type_imf} shows the effects on the VM relation of
changing the stellar mass normalization by $\pm 0.3$ dex. The
horizontal bar shows the changes this causes in the observed relation.
Interestingly, the model relations are largely unaffected by these
changes in stellar mass. This is because increasing the stellar mass
(at fixed galaxy size and halo mass) results in higher optical
circular velocity, and vice versa for lower stellar masses. Thus
changes in the stellar mass normalization move model galaxies along
the VM relation. Fig.~\ref{fig:vm_type_imf} shows that to reconcile a
model with standard halo contraction with the observed VM relation of
late-type galaxies requires stellar masses lower by a factor of $\sim
2$. However, lowering the stellar masses of early-type galaxies by the
same factor would remove the agreement. This suggests that early-type
and late-type galaxies cannot share the same IMF and halo response to
galaxy formation.

%% FIGURE 16
\begin{figure*}
\centerline{
\psfig{figure=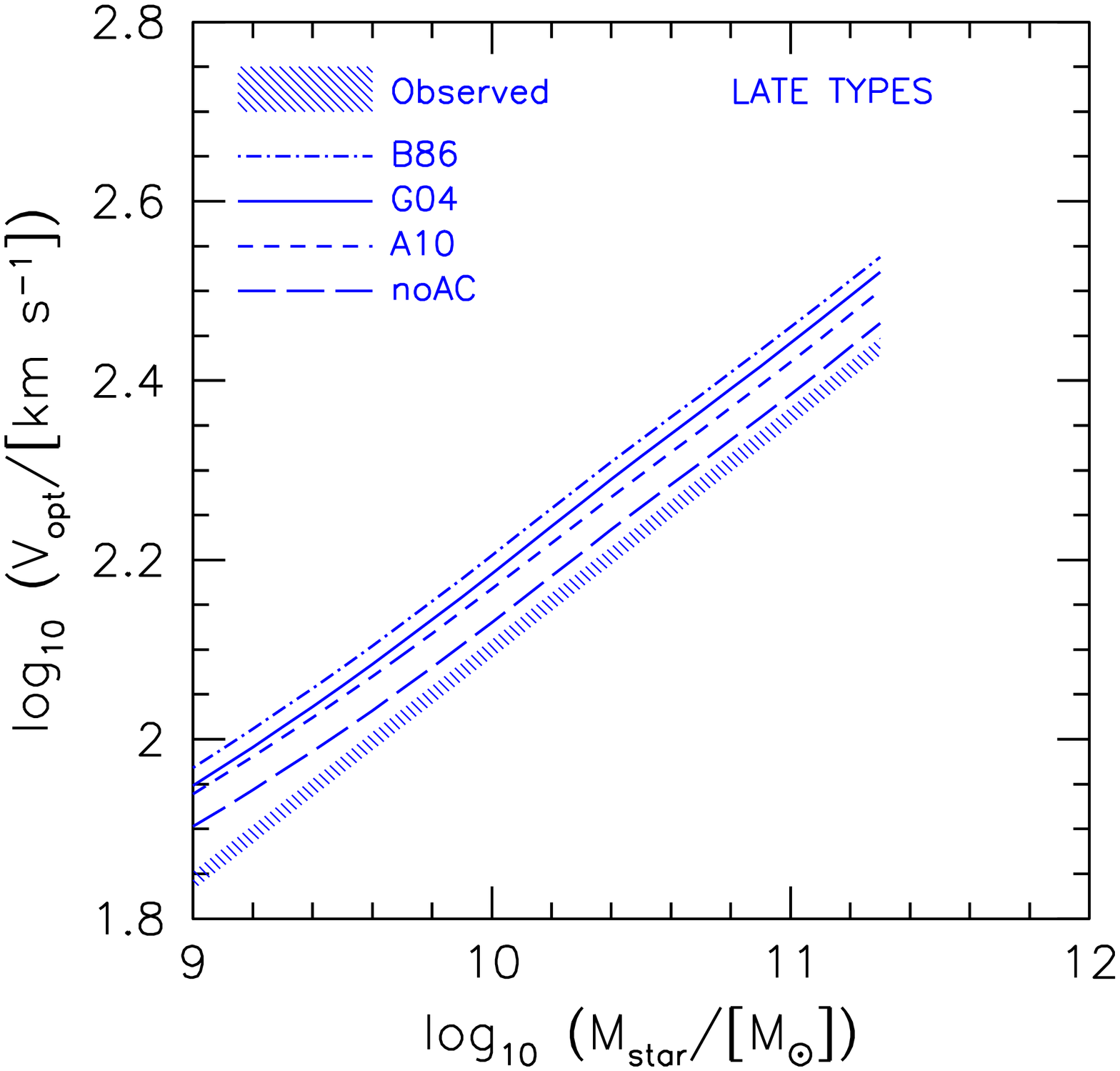,width=0.46\textwidth}
\psfig{figure=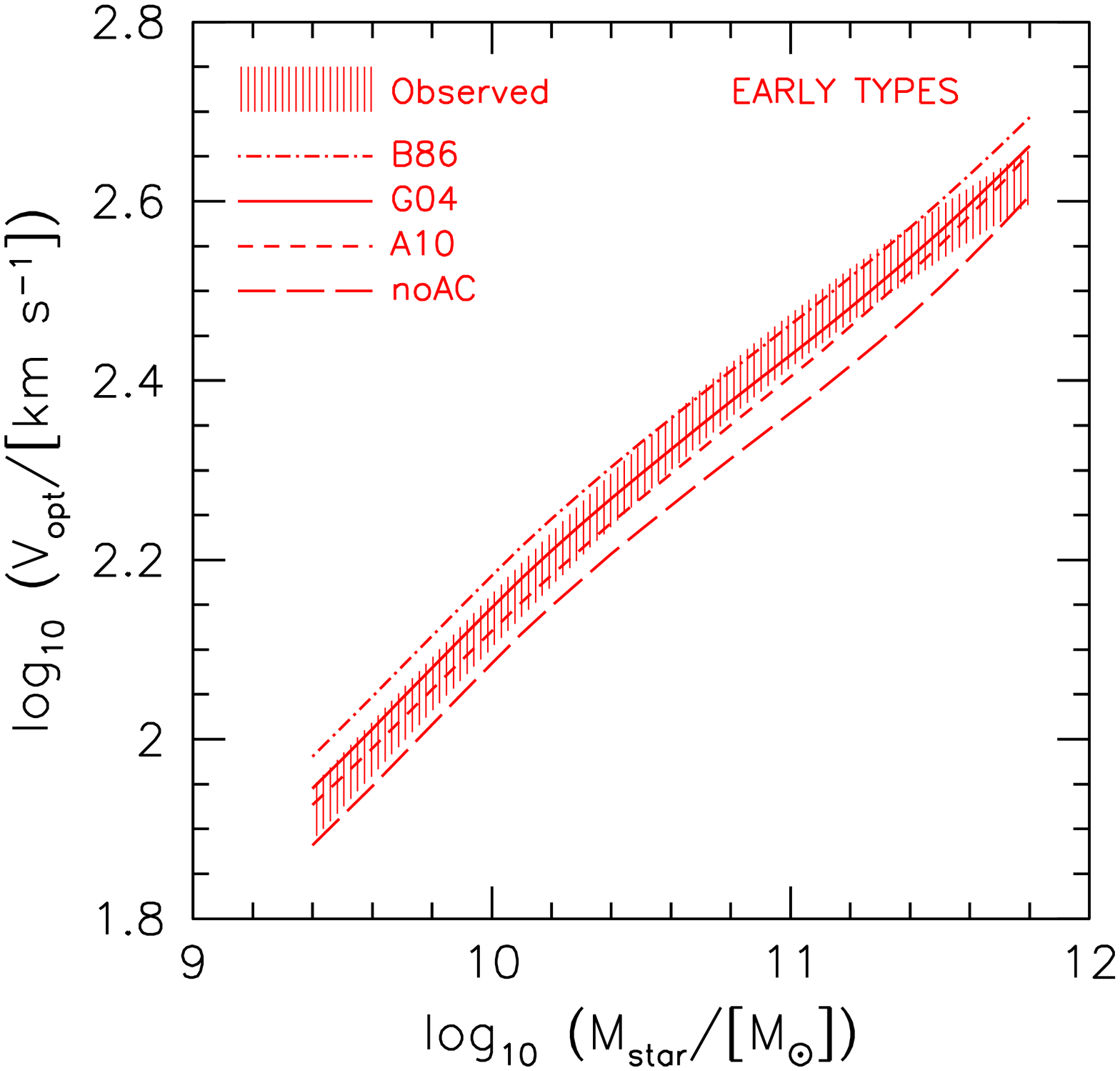,width=0.46\textwidth}}
\caption{Effect of adiabatic contraction on $V_{\rm opt}-\Mstar$
  relation for early-type (right panel) and late-type (left panel)
  galaxies. Adiabatic contraction results in higher optical circular
  velocities. Compared to a model with no halo contraction (long
  dashed lines), the increase in optical circular velocities is
  $\simeq 0.1$ dex for the Blumenthal \etal (1986, B86) model, $\simeq
  0.08$ dex for the Gnedin \etal (2004, G04) model, and $\simeq 0.05$
  dex for the Abadi \etal (2010) model. For these models (with a
  Chabrier IMF and standard halo concentrations) early-type galaxies
  favor models with strong halo contraction, whereas late-type
  galaxies favor models with halo expansion.}
\label{fig:vm_type_ac}
\end{figure*}

A change in the {\it ``pristine''} halo concentration could be
possible by either a formation bias, since halo concentration is
strongly correlated with halo formation time (Wechsler \etal 2002), or
a change in cosmology. The magnitude of any formation bias is
constrained by the scatter in halo concentrations. For relaxed haloes,
the scatter in halo concentrations, at fixed halo mass, is $\simeq
0.11$ dex (e.g., Jing 2000; Wechsler \etal 2002; Macci{\'o} \etal
2007).  At low stellar masses ($\Mstar < 10^{10}\Msun$) the majority
of galaxies are late-types, and thus a formation bias is only expected
to effect early-type galaxies. At high stellar masses $\Mstar \gta
10^{11} \Msun$) the majority of galaxies are early-types, and thus a
formation bias is only expected to effect late-types.

An alternative way to reduce the initial halo concentrations is to
reduce the amplitude of the power spectrum on galaxy scales. The most
effective way to achieve this is through a change in $\sigma_8$ (the
amplitude of the linear power spectrum today on scales of $8\hMpc$).  The
concentration scales roughly linearly with $\sigma_8$, so that a 0.11
dex change in halo concentrations (equivalent to the $1\sigma$
intrinsic scatter) requires a change in $\sigma_8$ of a factor of
$\simeq 1.3$. This is much larger than the reported uncertainties in
$\sigma_8$ of 0.014 dex (Komatsu \etal 2009), and thus the
uncertainties in halo concentrations from uncertainties in cosmological
parameters are likely to be small. 

Fig.~\ref{fig:vm_type_c} shows the effect of changing the halo
concentration (by $\pm 0.11$ dex, i.e., the $1\sigma$ intrinsic
scatter) on the VM relations of the standard
model. Increasing/decreasing the halo concentration results in
higher/lower $V_{\rm opt}$. The changes are relatively small. For high
mass late-type galaxies a formation bias resulting in 0.1 dex lower
halo concentrations is the most that is feasible, but this has very
little effect on the zero point of the VM relation. Thus changes in
halo concentrations are not likely to resolve the TF zero point
problem.

As discussed in \S~\ref{sec:ac}, recent cosmological simulations
suggest that the Gnedin \etal (2004) halo contraction formalism
over-predicts the amount of halo response (Abadi \etal 2010; Tissera
\etal 2010; Pedrosa \etal 2010).  Fig.~\ref{fig:vm_type_ac} shows the
effect of the halo contraction model on the VM relations adopting a
Chabrier IMF. The differences in $\Vopt$ between a model with no
contraction and the Blumenthal \etal (1986, B86) model are $\simeq
0.1$ dex, for both early-type and late-types. The Gnedin \etal (2004, G04)
model results in only slightly less contraction than the B86
model. The Abadi \etal (2010, A10) model results in $\simeq 0.05$ dex
increase in $\Vopt$, so that the VM relation lies half way between the
B86 model and no contraction. The model without halo contraction is
only marginally consistent with the observed VM relation for
late-types, and thus a model with halo expansion would fit the data
better. For early-types the Gnedin \etal (2004) halo contraction model
provides the best fit to the data, assuming a Chabrier IMF. The B86
and A10 halo contraction also provide acceptable fits given
uncertainties in halo masses.

%% FIGURE 17
\begin{figure}
\centerline{
\psfig{figure=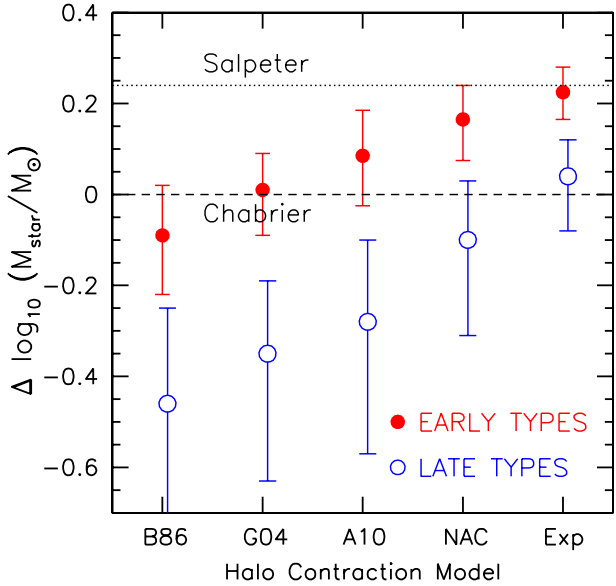,width=0.47\textwidth}}
\caption{Offset in stellar masses required to match the zero point of
  the VM relations as a function of halo response model, calculated at
  $\log_{10} (\Vopt/\kms)=2.30$, for early-type (red filled symbols)
  and late-type (blue open symbols) galaxies. The models correspond
  to: B86 (Blumenthal \etal 1986); G04 (Gnedin \etal 2004); A10 (Abadi
  \etal 2010); NAC no halo contraction; Exp (halo expansion with
  $\nu=-0.5$ in Eq.~\ref{eq:nu}. The error bars show the effects of
  $2\sigma$ systematic errors on the zero points of the VM and
  $M_{200}$-$\Mstar$ relations. For fixed IMF (i.e., horizontal lines)
  early-type galaxies require stronger contraction than late-type
  galaxies, while for fixed halo response (vertical direction)
  early-type galaxies require heavier IMFs than late-type galaxies.}
\label{fig:zp_ac}
\end{figure}

\subsection{Implications for a universal IMF and halo response }
We have shown that a model with standard halo contraction and a
Chabrier stellar IMF reproduces the slopes of the VM relations of
early-type and late-types, and the zero point of the VM relation for
early-types, but not the zero point for late-types.
Fig.~\ref{fig:zp_ac} shows the offsets in stellar mass required to
match the zero point of the VM relations. Early-types are shown with
red filled symbols and late-types with blue open symbols. The error
bars correspond to the $2\sigma$ uncertainty in the observed VM zero
point and halo masses. This shows that for any given halo response
model, a simultaneous match of the VM relations for early-type and
late-types requires different stellar mass normalizations.
Alternatively, for models with a universal IMF, early-types
require stronger halo contraction than late-types.

More specifically, for models with halo contraction (B86, G04, A10)
early-type galaxies require stellar masses higher by a factor of
$\simeq 2$ than late-types. For early-type galaxies halo contraction
models are consistent with a Chabrier IMF. For late-types halo
contraction models require lighter IMFs than Chabrier (at least 0.1
dex for A10, 0.2 dex for G04 and 0.25 dex for B86). For models with
uncontracted NFW haloes early-types favor an IMF $\simeq 0.17$ dex
heavier than Chabrier, while late-types favor an IMF $\simeq 0.1$ dex
lighter than Chabrier.

%% FIGURE 18
\begin{figure*}
\centerline{
\psfig{figure=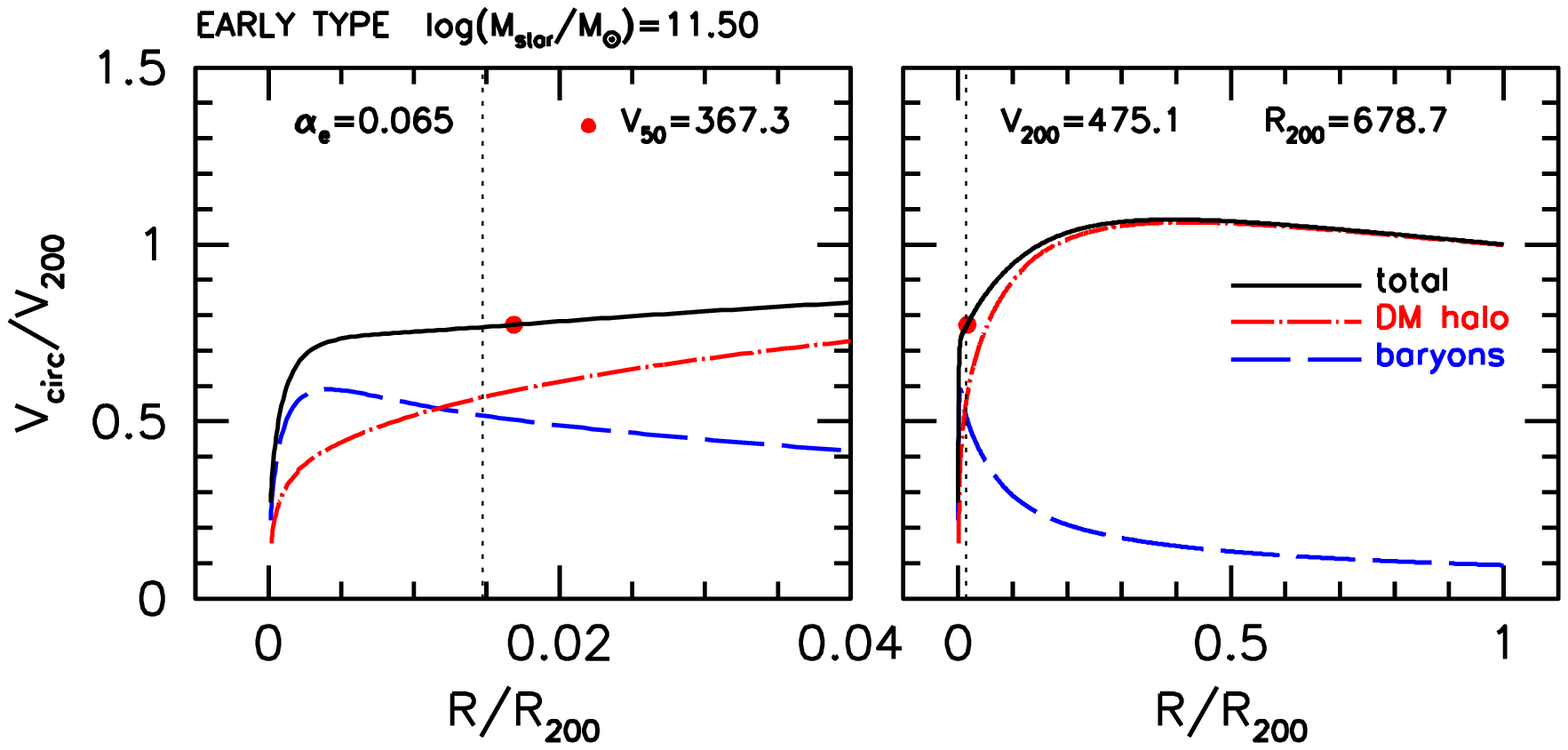,width=0.49\textwidth}
\psfig{figure=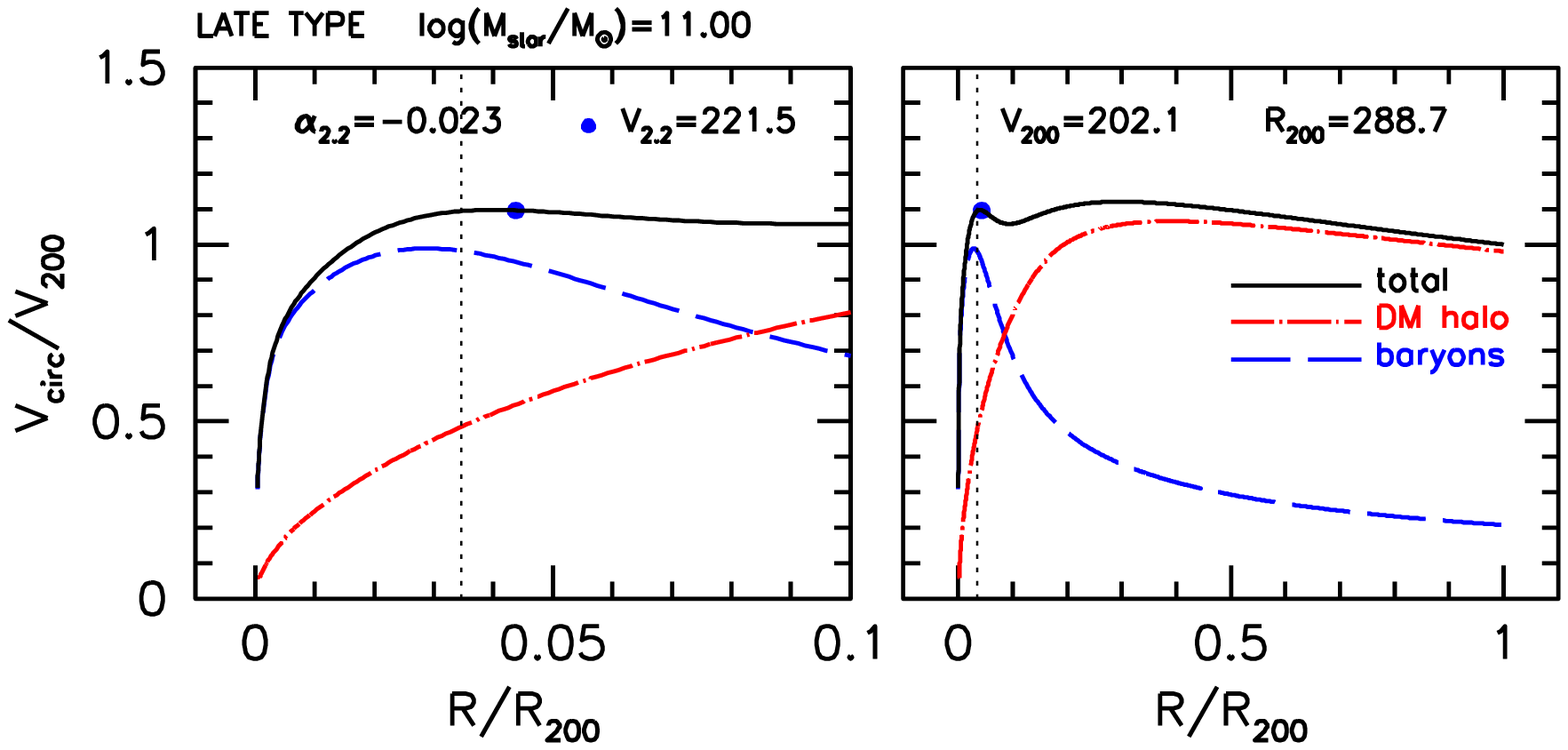,width=0.49\textwidth}
}

\centerline{
\psfig{figure=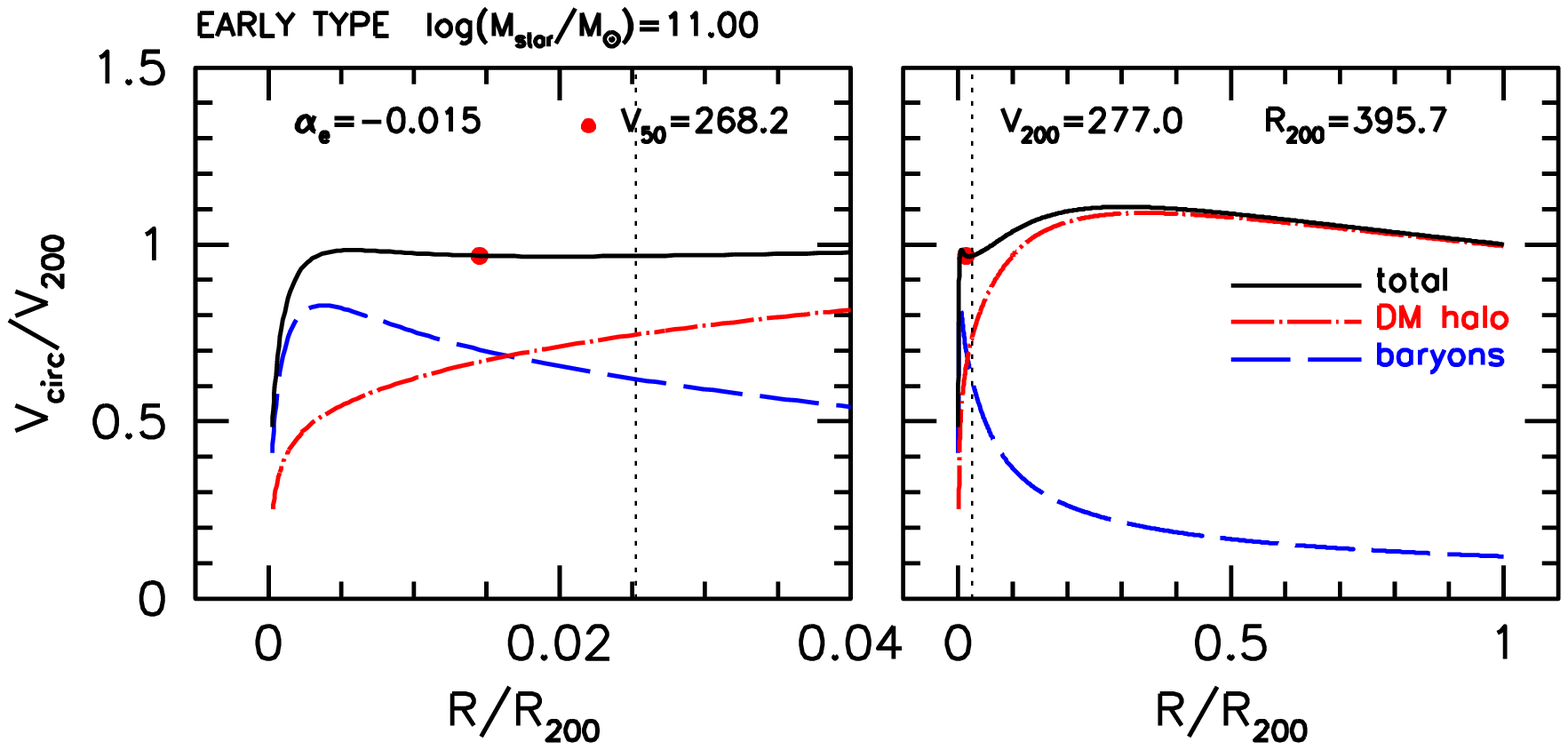,width=0.49\textwidth}
\psfig{figure=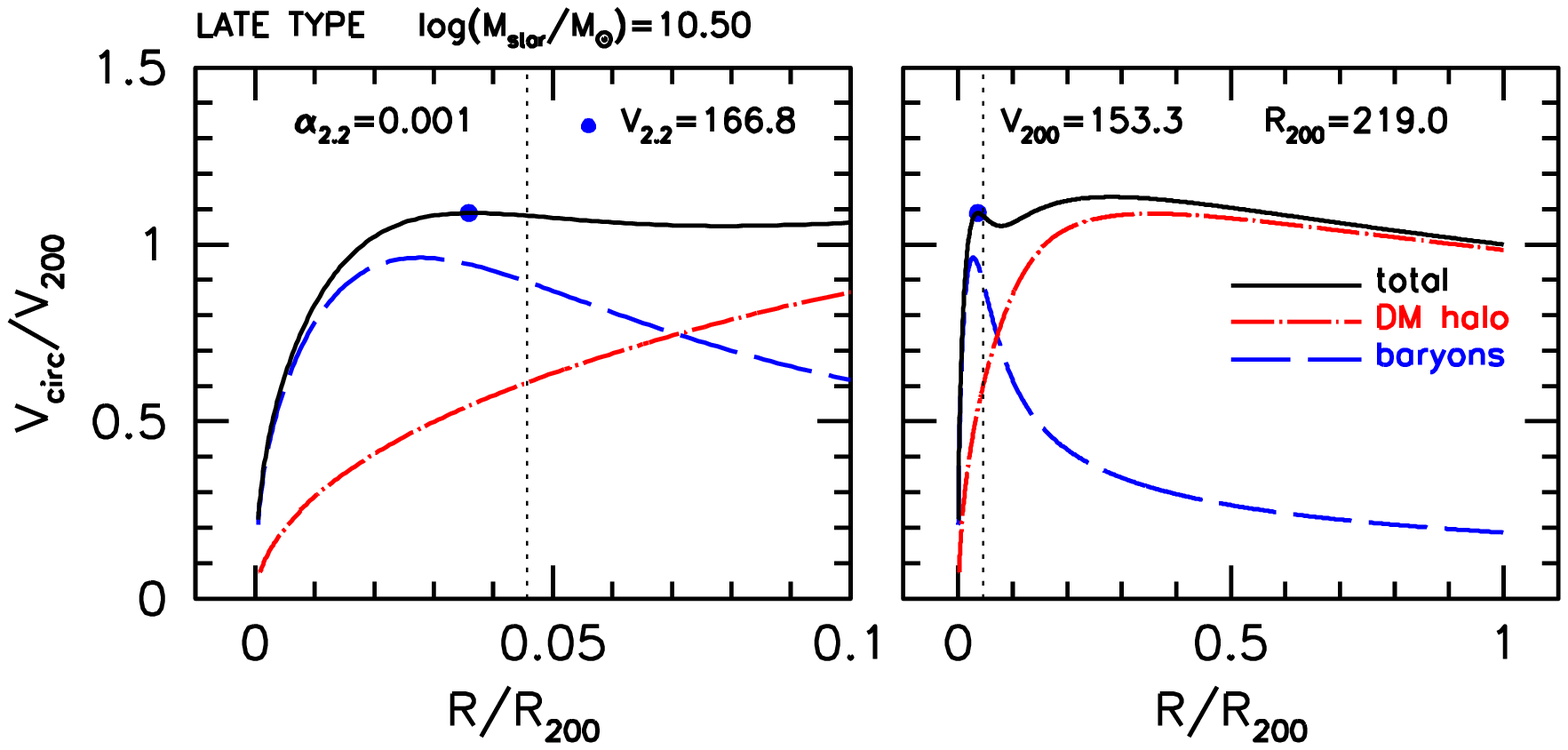,width=0.49\textwidth}
}

\centerline{
\psfig{figure=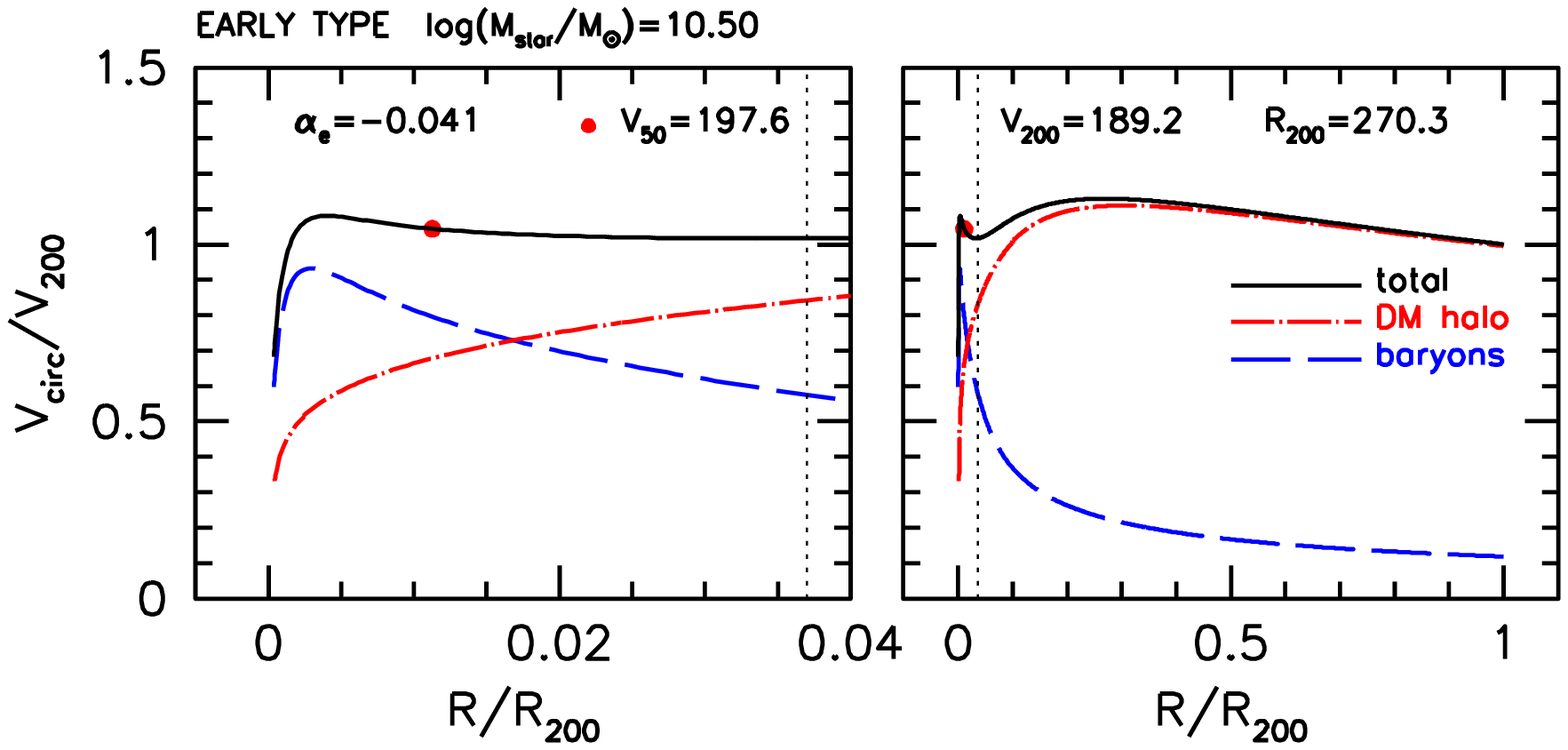,width=0.49\textwidth}
\psfig{figure=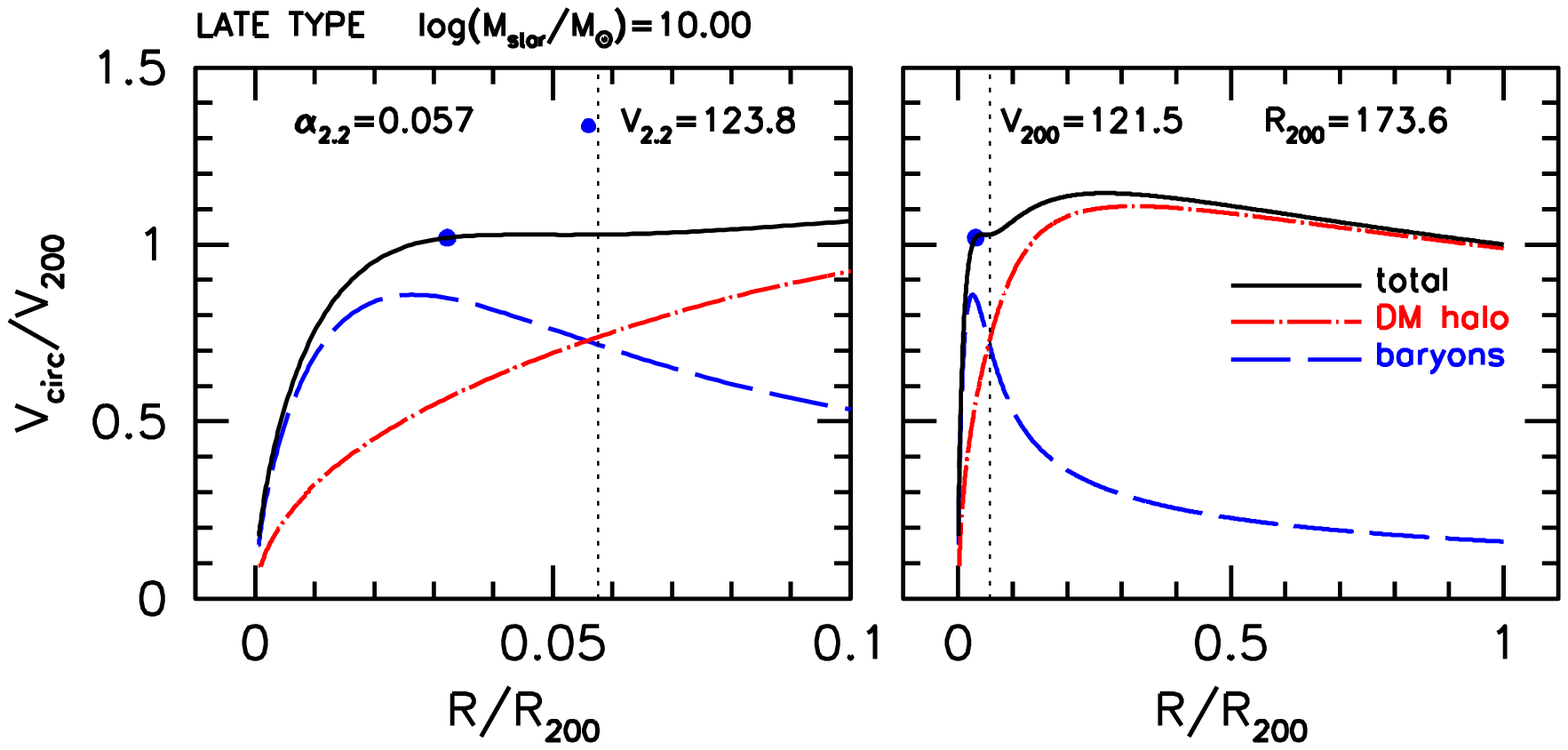,width=0.49\textwidth}
}

\centerline{
\psfig{figure=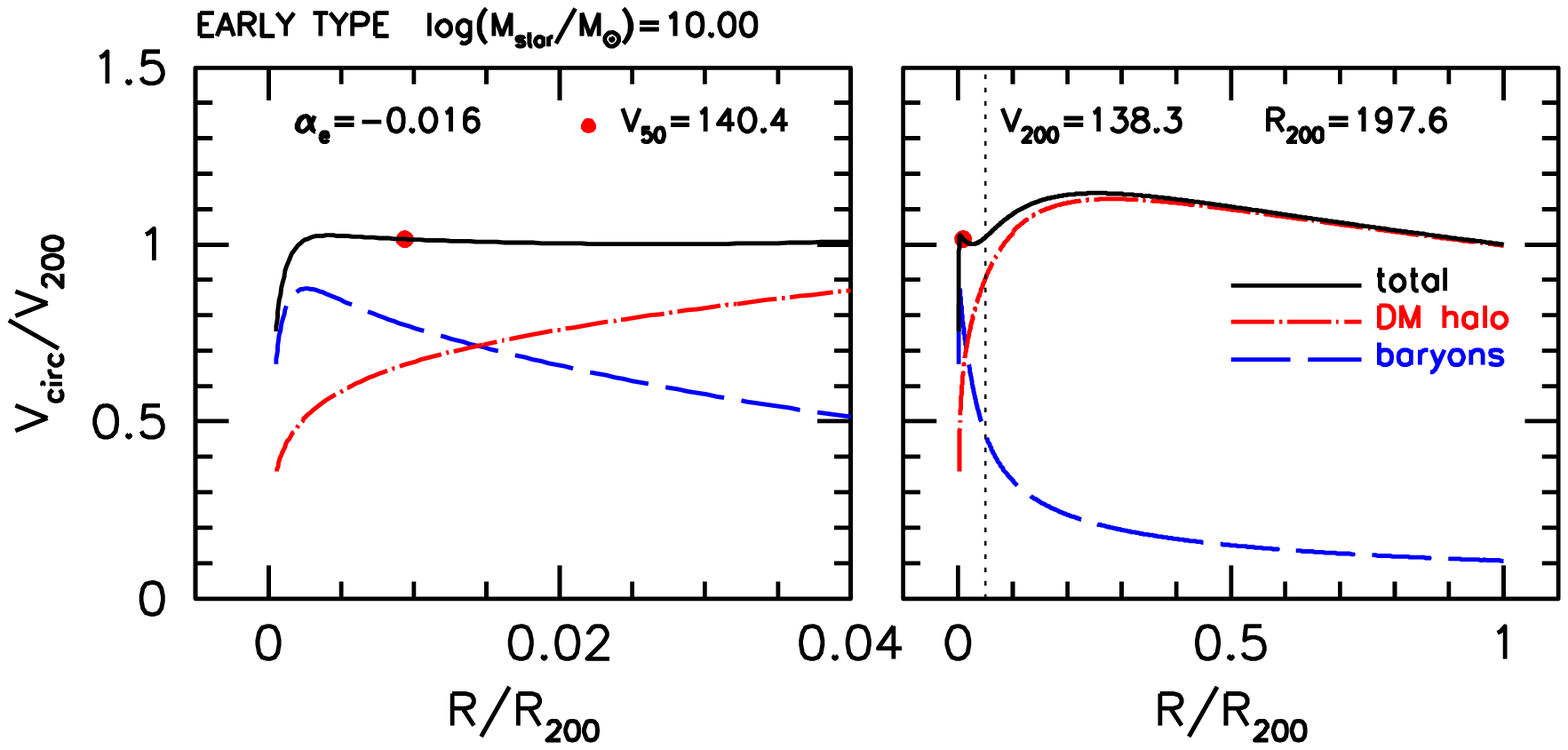,width=0.49\textwidth}
\psfig{figure=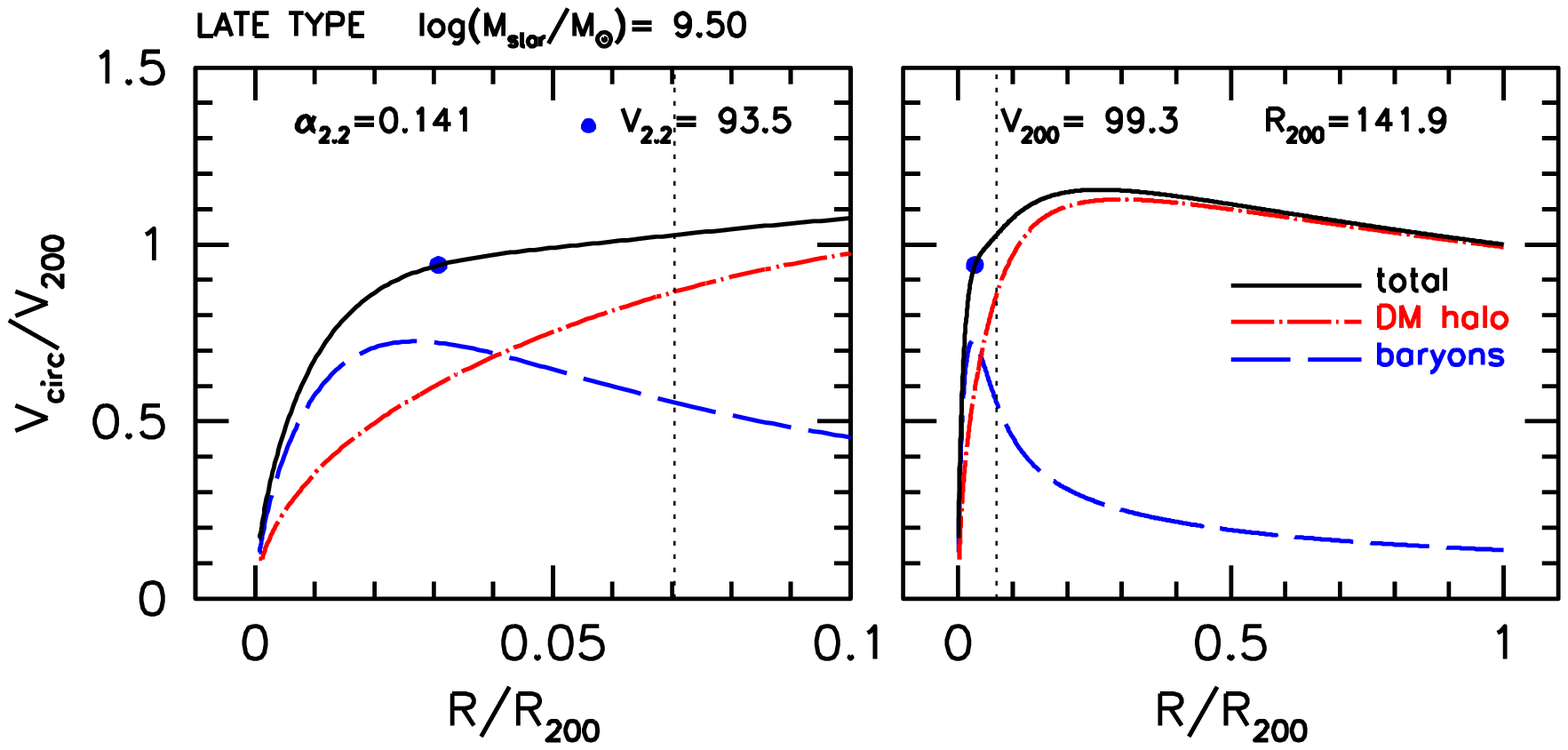,width=0.49\textwidth}
}
\caption{Circular velocity profiles for model early-type galaxies
  (left panels) and late-type galaxies (right panels) that reproduce
  the FJ and TF relations, respectively. All models have a Chabrier
  (2003) stellar IMF. The early-type galaxy models have halo
  contraction according to Gnedin \etal (2004), while the late-type
  galaxies have halo expansion with $\nu=-0.5$ in Eq.~\ref{eq:nu}.
  For each model galaxy the two panels show the circular velocity
  profiles out to the virial radius, and the inner 4\%/10\% of the
  virial radius, or early-type/late-type galaxies.  To give a physical
  reference point, in all panels the vertical dotted line corresponds
  to 10 kpc.  The stellar masses (in $\Msun$), virial circular
  velocities (in $\kms$) and virial radii (in kpc) decrease from top
  to bottom as indicated.  For early-type galaxies the red dot
  corresponds to the circular velocity at the effective radius, which
  occurs at $\sim 1\%$ of the virial radius. The logarithmic slope of
  the circular velocity profile at the effective radius is given by
  $\alpha_{\rm e}$ and is close to zero for all galaxies. Likewise,
  for late-type galaxies the blue dot corresponds to the circular
  velocity at 2.2 optical disk scale lengths radius, which occurs at
  $\sim 4\%$ of the virial radius. The logarithmic slope of the
  circular velocity profile at 2.2 scale lengths is given by
  $\alpha_{\rm 2.2}$ which is also close to zero for all galaxies.}
\label{fig:rc}
\end{figure*}

%% FIGURE 19
\begin{figure*}
\centerline{
\psfig{figure=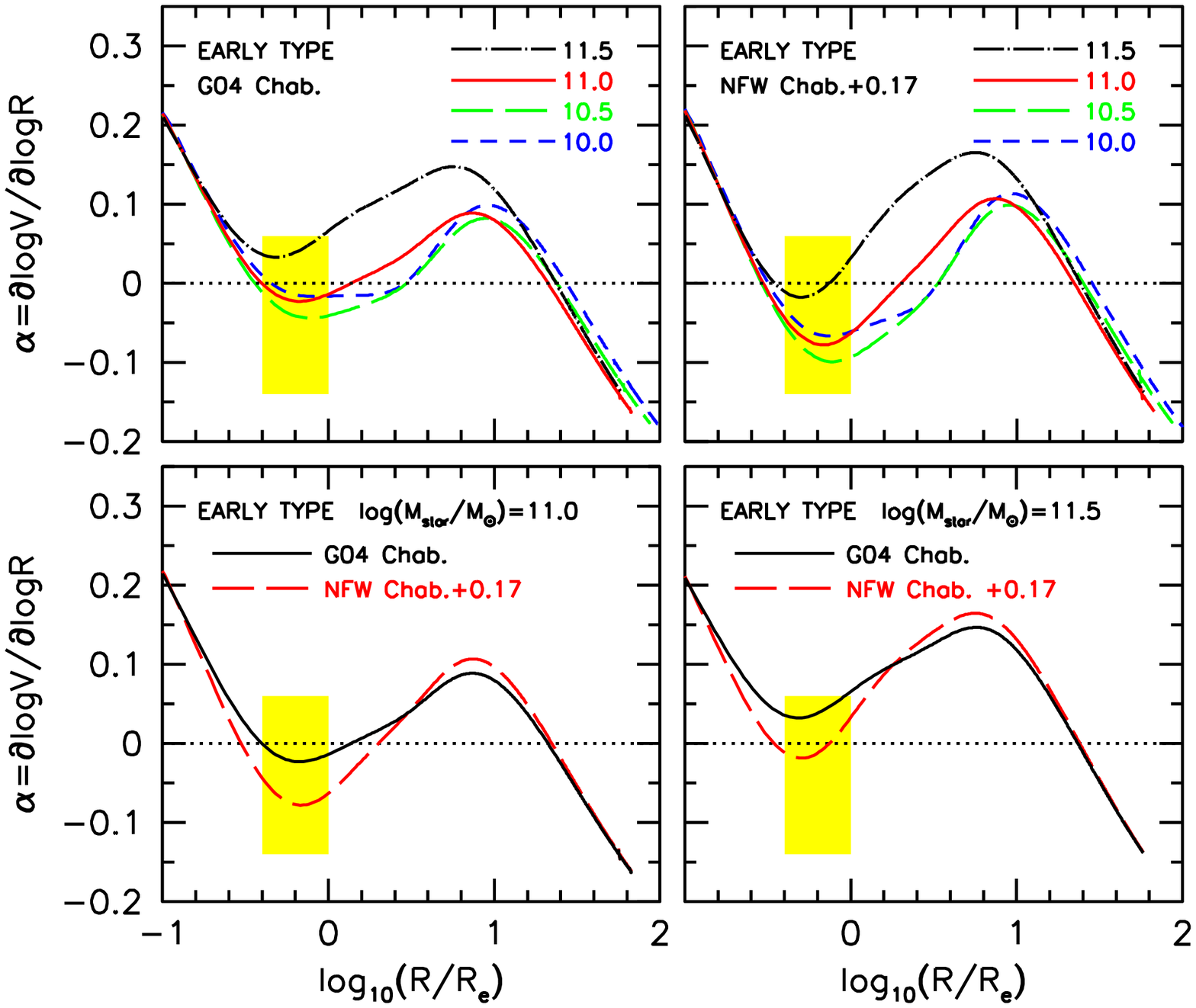,width=0.80\textwidth}
}
\caption{Logarithmic slope of the circular velocity profile for early
  type galaxies. The upper panels show the dependence on stellar mass,
  with $\log_{10}(\Mstar/\Msun)$ as indicated (normalized to a
  Chabrier IMF). The upper left panel shows models with Gnedin \etal
  (2004; G04) halo contraction and a Chabrier IMF, while the upper
  right panel shows models with uncontracted NFW haloes and stellar
  masses 0.17 dex higher than a Chabrier IMF. Both of these models
  reproduces the FJ relation.  All models have close to isothermal
  ($\alpha=0$) circular velocity profiles over a wide range of radii,
  and in particular in the region of overlap between baryonic and dark
  matter. The yellow shaded region shows the observed value (with
  $2\sigma$ systematic uncertainties) for massive $[11 \lta
  \log_{10}(\Mstar/\Msun) \lta 11.6]$ early type galaxies from
  Koopmans \etal (2009). Both sets of models are consistent with these
  data.  The lower panels show the difference between models with and
  without halo contraction for $\log_{10}(\Mstar/\Msun)=11.0$ (lower
  left) and $\log_{10}(\Mstar/\Msun)=11.5$ (lower
  right). Distinguishing between these models requires measuring
  $\alpha$ to better than 5\%.}
\label{fig:alpha}
\end{figure*}

\subsection{Circular velocity profiles and the bulge-disk-halo conspiracy}
\label{sec:alpha}
So far our analysis has considered only the optical circular velocity
at a single radius. But there is potentially useful information to be
obtained by including circular velocity measurements at additional
radii. In addition, the close to constant circular velocity profiles
in the optical regions of early and late-type galaxies has often been
termed a ``conspiracy'' between the baryons and the dark matter
(e.g., van Albada \& Sancisi 1986; Koopmans \etal 2009).

Fig.~\ref{fig:rc} shows example model circular velocity profiles for
early-type and late-type galaxies at a range of stellar masses. These
models have stellar masses according to Chabrier IMF and halo response
chosen to reproduce the FJ and TF relations. For early-type galaxies
this requires Gnedin \etal (2004) halo contraction, while for
late-type galaxies this requires halo expansion with $\nu=-0.5$ in
Eq.~\ref{eq:nu}.  The axes are scaled to the virial radii and circular
velocities. For each galaxy model, the right panel shows the model out
to the virial radius, $R_{200}$, while the left panel shows the model
in the inner 4\% of $R_{200}$ for early-types and the inner 10\% of
$R_{200}$ for late-types. To give a common reference point to the
scales in the various panels, the dotted vertical lines corresponds to
10 kpc.

It is immediately apparent that these model galaxies have roughly
constant circular velocity profiles, especially within a few optical
half-light radii.  As previously shown in Dutton \etal (2010b) we find
that both early-types and late-types have $V_{\rm opt}/V_{200}\simeq 1$,
where $\Vopt = V_{2.2}$ for late-types and $\Vopt=V_{\rm circ}(R_{\rm
  e})$ for early-types.  We do note however, that the mass profiles
are not exactly isothermal between the optical half-light radius and
the virial radius. Outside the inner $\sim 10\%$ of the virial radius
the circular velocity profiles rise, reaching a maximum at $\sim 30\%$
of the virial radius.

The shape of the circular velocity profiles can be quantified by the
logarithmic slope at a given radius:
\begin{equation}
\alpha(r) = \partial \log V_{\rm circ}(r) / \partial \log r.
\end{equation}
For a spherical mass density with $\rho(r)\propto r^{-\gamma}$, the
slope of the density profile is related to the slope of the circular
velocity profile via $\gamma = 2-2\alpha$.

By combining strong gravitational lensing with stellar dynamics
Koopmans \etal (2009) found that for massive early-type galaxies the
mass density slope within the effective radius
$\gamma=2.085^{+0.025}_{-0.018}$, assuming isotropic velocity
dispersions, with a systematic uncertainty of $\simeq 0.1$.  Other
authors find similar mean values of $\gamma$ (e.g., Dobke \& King
2006; Auger \etal 2010; Barnabe \etal 2011).  Thus the results from
Koopmans \etal (2009) are consistent with $1.88 \lta \gamma \lta 2.29$
(at the $2\sigma$ level), which corresponds to $ -0.14 \lta \alpha
\lta 0.06$.  Our models for massive early-types are within this range
and thus are consistent with current observational constraints
(Fig.~\ref{fig:alpha}).

Models with different halo response and IMFs can give different values
of $\alpha$. For example, an early-type galaxy with a stellar mass
of $\Mstar=10^{11}\Msun$ assuming a Chabrier IMF and Blumenthal \etal
(1986) halo contraction results in $\alpha(R_e)\simeq -0.01$, while a
model with 0.17 dex higher stellar mass and no halo contraction
results in $\alpha(R_e)\simeq -0.06$. Both of these models reproduce
the observed circular velocity at $R_e$, but the model without halo
contraction has higher circular velocities at smaller radii because of
the more massive baryonic component. So in principle, measuring the
slope of the total mass profile within the effective radius can help
to discriminate between models with different halo response and IMF.
However, the current systematic uncertainties in $\alpha$ are $\sim
0.05$, which is similar to the differences we wish to measure.

The fact that $\alpha \simeq 0$ is sometimes referred to as the
bulge-halo {\it conspiracy} for early-types and disk-halo {\it
  conspiracy} for late-types (e.g., van Albada \& Sancisi 1986;
Koopmans \etal 2009).  A conspiracy implies that there is an
underlying physical mechanism that forces galaxies to have $\alpha
\simeq 0$, despite numerous ways in which galaxies would otherwise
have $\alpha \ne 0$. Sometimes halo contraction is invoked as such a
mechanism (e.g., Blumenthal \etal 1986; Klypin \etal 2002).  However,
we find that models with different halo responses (contraction, no
contraction and expansion) and different IMFs all result in $-0.2 \lta
\alpha \lta 0.2$ for a wide range of galacto-centric radii $0.1 \lta
R/R_{\rm e} \lta 100$. Thus the observations fact that $\alpha\simeq
0$ cannot be used to discriminate between different models for halo
response. Rather, our results suggest that $\alpha \simeq 0$, over a
wide range of radii, is a natural consequence for galaxies embedded in
\LCDM haloes.

Using cosmological hydrodynamical simulations Duffy \etal (2010) find
a tension between the need for strong feedback (in order to reproduce
the observed low baryon fractions) and the need for strong dissipation
(in order to match the isothermal density profiles inferred from
lensing/dynamics). This is in conflict with our results which show
that early-type galaxies embedded in \LCDM haloes with the correct
baryon fractions reproduce isothermal density profiles within an
effective radius. Thus we do not support the idea that the lensing
observations are biased towards galaxies with steeper density
profiles. Rather, the tension found by Duffy \etal (2010) may simply
be the result of systematic effects in the comparison between
simulations and observations. The main cause for concern is that the
simulated galaxies are at $z=2$ whereas the observed galaxies are at
$z\sim 0$. This makes it hard to know the appropriate radii within
which to measure $\gamma$. Since $\gamma$ depends on radius, it is
critical to compare observations and simulations at the same radii.
 
\section{Discussion}
We have established that either the stellar initial mass function, or
the halo response to galaxy formation cannot be the same for
early-type galaxies (ETGs) and late-type galaxies (LTGs). We now
discuss physical mechanisms that could cause halo response to be
different from the standard adiabatic contraction hypothesis, and
others which could cause the IMF to vary.  We discuss how these
mechanisms might vary with galaxy type, and hence if they can explain,
qualitatively our results.

Two defining differences between ETGs and LTGs that may have
consequences of the IMF and halo response, are the concentration of
the stars and the star formation histories. ETGs have higher
concentrations (associated with a spheroidal component) and they
formed the bulk of stars at earlier times than LTGs.

\subsection{What could cause halo contraction to vary with galaxy type?}

There are a number of processes that occur during galaxy formation
which can alter the structure of the dark matter halo. Some of these
can also alter the morphological type of a galaxy. We outline these
processes below.

\begin{itemize} 

\item Smooth accretion: Smooth accretion of gas onto the central
  galaxy is the quintessential process that is expected to result in
  (adiabatic) halo contraction (Blumenthal \etal 1986). This process is
  expected to result in disk growth and star formation, and thus LTGs,
  but it could also occur during the history of ETGs.

\item Dissipative major mergers: Major mergers which involve
  significant amounts of gas a.k.a. ``wet mergers'' are expected to
  result in halo contraction due to the mass that accumulates at the
  center of the galaxy. This process turns LTGs into ETGs.

\item Non-dissipative major mergers: Major mergers which do not
  involve gas a.k.a. ``dry mergers'' are expected to result in
  effective halo contraction due to mixing of stars and dark matter
  through violent relaxation. This contraction effect partially undoes
  the segregation of baryons and dark matter produced by dissipation.
  Note that this is a different physical process than the standard
  adiabatic halo contraction.  This process only occurs for ETGs.

\item Clumpy accretion/minor mergers: Clumpy accretion of stars or gas
  in the form of minor mergers can cause halo expansion due to
  dynamical friction (e.g., El-Zant \etal 2001, Elmegreen \etal 2008;
  Romano-Diaz \etal 2008; Johansson \etal 2009).  The clumps need to
  be baryon dominated, or else dark matter will be brought in to
  replace the dark matter that is removed by dynamical friction. The
  clumps also need to be dense, or else they will be tidally
  disrupted before they can alter the center of the halo. Clumpy cold
  accretion is more likely to occur at high redshifts (Dekel \etal
  2009). This process can occur in both ETGs and LTGs.

\item Feedback: The energy and momentum feedback from supernova,
  stellar winds and AGN can cause halo expansion under the following
  conditions: (1) By removing large amounts of baryons on a timescale
  much faster than in which they were accumulated (e.g., Navarro \etal
  1996b; Gnedin \& Zhao 2002; Read \& Gilmore 2005; Governato \etal
  2010); or (2) By inducing large scale bulk motions (Mashchenko \etal
  2006; 2008).  These processes are expected to be more effective in
  galaxies with the following properties: (1) lower mass and lower
  bulge fraction (i.e., LTGs), due to the shallower potential wells;
  (2) higher gas fractions, because a large fraction of the baryons
  need to be expelled for this process to be effective, and stellar
  mass cannot be expelled by feedback; (3) higher redshifts, due the
  order of magnitude higher specific star formation rates.

\item Bars: Galactic bars can cause halo expansion due to dynamical
  friction on the bar from the dark matter halo (Weinberg \& Katz
  2002). Since gas can get driven to the center of the galaxy, bar
  formation could also result in halo contraction. Bars require disks
  in which to form, and since bulges help to stabilize disks, bars are
  more frequent in late-type galaxies. However, bars are known to
  exist in S0 galaxies, which are considered to be ETGs, and thus the
  effects of bars on dark matter haloes are not necessarily restricted
  to LTGs.

\end{itemize}

Major mergers have long been thought to be the key process that
determines the morphological type of a galaxy (Toomre \& Toomre 1972),
with major mergers destroying disks, and producing spheroidal
galaxies. Galaxy types are observed to vary strongly with stellar
mass: low mass galaxies are predominately late-types (disk dominated,
gas rich, star forming), while high mass galaxies are predominantly
early-types (bulge dominated, gas poor, non star forming). This basic
trend can be understood as a consequence of the mass dependence of the
frequency of major mergers and the mass dependence of gas fractions of
the progenitor galaxies (Maller 2008; Hopkins \etal 2009a,b). The mass
dependence of the effects of dissipative major mergers can also
qualitatively explain the tilt of the Fundamental Plane (Dekel \& Cox
2006).

Thus the key physical process that occurs during early-type galaxy
formation, which does not occur during late-type galaxy formation, is
a major merger. Whether the merger is dissipative or non-dissipative
we expect the haloes to contract, but for different reasons.  This
expectation needs to be tested, and quantified, with numerical
simulations of galaxy/halo mergers.  For late-type galaxies, a
combination of clumpy cold accretion, and feedback during the early
phases of galaxy formation could plausibly result in net halo
expansion (e.g., Mo \& Mao 2004).

Under this scenario, galaxies with a higher fraction of their stars in
a spheroid (i.e., a classical bulge) should have experienced more halo
contraction. This is qualitatively consistent with our result that for
a fixed IMF early-type galaxies have more halo contraction than
late-type galaxies (e.g., Fig.~\ref{fig:zp_ac}).  To test this further
for early-types and late-types separately, would require the velocity
- stellar mass, halo mass - stellar mass, and structural scaling
relations to be measured for galaxies with different bulge
fractions. In the meantime we note that earlier type spirals have
higher rotation velocities at fixed K-band luminosity and stellar mass
than spiral galaxies in general (Noordermeer \& Verheijen 2007;
Williams \etal 2010). Such a trend would be expected if earlier type
spirals experienced more halo contraction than later type spirals. But
there could be other explanations, such as more compact baryons, so it
is too early to say if this supports our simple scenario.

\subsection{What could cause the IMF to vary with galaxy type?}
Observations in the Galactic disk suggest that the IMF has a power-law
shape at masses above $1 \Msun$, and that it turns over at lower
masses (Kroupa 2001; Chabrier 2003). This turnover can be modeled by a
log-normal distribution with a characteristic turnover mass $m_{\rm
  c}$ (Chabrier 2003). The value of $m_{\rm c}$ is $\sim 0.1 \Msun$ in
the disk of the Milky Way.

Larson (1998, 2005) has argued that the characteristic turnover mass
may largely be determined by the thermal Jeans mass, which strongly
depends on the temperature of the ISM. An increased ISM temperature at
higher redshifts is robustly expected based on the temperature of the
cosmic microwave background, and also plausibly from higher star
formation rates (SFR) which result in more supernova heat input, and
lower metallicities, which result in less efficient cooling.

An evolving IMF, in which the characteristic mass increases with
increasing redshift, provides an explanation for a number of
discrepancies: The difference in evolution of dynamical $M/L$ ratios
and colors of early-type galaxies (van Dokkum 2008); The difference in
evolution of the galaxy SFR - stellar mass relation between \LCDM
galaxy formation models and observations (Dav\'e 2008); The difference
between the observed stellar mass density of the universe and the
implied stellar mass density from integrating the cosmic star
formation history (Larson 2005; Hopkins \& Beacom 2006; Fardal \etal
2007). 

As shown by van Dokkum (2008), for IMFs with $m_{\rm c} < 0.08$
(i.e., more bottom heavy than a Chabrier IMF), the stellar $M/L$ ratios
increase for decreasing $m_{\rm c}$. For $m_{\rm c} > 0.08$ (i.e., more
bottom light than a Chabrier IMF) the stellar M/L ratio decreases.
However, the relation between stellar $M/L$ and $m_{\rm c}$ is not
monotonic.  As $m_{\rm c}$ increases beyond $\sim 0.3 \Msun$ the
stellar $M/L$ can actually {\it increase}. This is because the mass
function becomes dominated by stellar remnants. For old enough stellar
populations (Age $\sim 5-10$ Gyr) with $m_{\rm c} \sim 1$ the stellar
$M/L$ can equal or even exceed that of a Salpeter IMF.
 
Since ETGs form their stars at higher redshifts than LTGs, the
evolving IMF as proposed by Van Dokkum (2008) and Dav\'e (2008) would
cause ETGs to have higher present day stellar $M/L$ ratios than
LTGs. The normalizations of the $M/L$ ratios are expected to be
Salpeter like for ETGs, and not lower than 0.1 dex below Chabrier for
LTGs (van Dokkum 2008).  In the context of our results, stellar $M/L$
ratios close to Salpeter for ETGs are inconsistent with halo
contraction. If the IMF for LTGs is close to Chabrier, then this would
also favor models with halo expansion or no halo contraction. Thus
this evolving IMF requires that the haloes of both ETGs and LTGs do
not contract in response to galaxy formation.

Recently van Dokkum \& Conroy (2010) derived constraints on the
stellar IMF in the cores of massive elliptical galaxies using stellar
absorption lines in the near-IR. They find strong evidence for an IMF
with a steeper low mass slope than a Salpeter IMF, i.e., a
bottom-heavy IMF.  If the IMF is bottom-heavy throughout massive
elliptical galaxies, and not just in their centers, then this is the
opposite result to what is expected from the evolving IMF models of
van Dokkum (2008) and Dav\'e (2008).

This bottom heavy IMF results in stellar $M/L$ ratios a factor of
$\sim 1.4$ higher than a regular Salpeter IMF. As shown in
Fig.~\ref{fig:fdm} the dark matter fractions within the effective
radii for the most massive early-type galaxies are $\sim 0.3$ for a
Salpeter IMF. Thus, given the current uncertainties in $V_{\rm
  c}/\sigma_{\rm e}$, the IMF from van Dokkum \& Conroy (2010) is
permitted to apply globally in massive early-type galaxies, and not
just in their centers.  However, this IMF would strongly over-predict
the total masses within the effective radii of intermediate mass
($\Mstar \sim 10^{10}\Msun$ ) early-type galaxies.  Thus based on our
mass models we do not expect the bottom-heavy IMF of van Dokkum \&
Conroy (2010) to be universal across early-type galaxies of different
masses.

\subsection{Comparison with previous studies}
There are several recent studies that have addressed dark halo
contraction and the stellar initial mass function of galaxies (Treu
\etal 2010; Schulz \etal 2010; Trujillo-Gomez \etal 2010; Auger \etal
2010a; Napolitano \etal 2010). Most of these have focused on
massive early-type galaxies.  Although the individual conclusions
vary, all of them are consistent with the following: ETGs with a
Chabrier IMF plus un-contracted NFW haloes with standard halo
concentrations do not have enough mass within the effective radius.

Schulz \etal (2010) and Trujillo-Gomez \etal (2010) advocate models
with halo contraction (Gnedin \etal 2004, and Blumenthal \etal 1986,
respectively) to provide this additional mass, whereas Treu \etal
(2010) and Auger \etal (2010a) advocate a Salpeter IMF.  Schulz \etal
(2010) argue against a Salpeter IMF based on the results of Cappellari
\etal (2006). However, there are some caveats to this line of
reasoning. Firstly the dynamical masses from Cappellari \etal (2006)
are consistent with a Salpeter IMF for the most massive
galaxies. Secondly the dynamical masses from Cappellari \etal (2006)
assume mass follows light, which is expected to result in an
underestimate of the dynamical masses if the dark matter fractions
within the effective radii are significant. In \S \ref{sec:vsigma} we
showed that the Cappellari \etal (2006) dynamical masses imply that,
on average, $V_{\rm c}/\sigma_{\rm e} = 1.44 \pm 0.01$.  In
Fig.~\ref{fig:fdm} we showed that if $V_{\rm c}/\sigma_{\rm e} \gta
1.6$ then a Salpeter IMF is consistent for ETGs of all masses.  Thus a
Salpeter IMF is allowed for the most massive ETGs, and it is not yet
robustly ruled out for intermediate mass ETGs (where the dark matter
fractions are expected to be the lowest).

Auger \etal (2010a) combined strong lensing, weak lensing and stellar
dynamics for a sample of 53 massive elliptical galaxies to place
constraints on the stellar IMF. They conclude that, given their model
assumptions, the data strongly prefer a Salpeter like IMF over a
lighter IMF such as Chabrier of Kroupa. While we agree that a model
with a Salpeter like IMF can reproduce the observations, we find that
our data cannot distinguish between models with Salpeter and Chabrier
IMFs.  Below we discuss two areas that could contribute to these
differences: anisotropy and halo masses.

The use of strong lensing and stellar dynamics can constrain the slope
of the total mass profile within an effective radius (e.g., Koopmans
\etal 2006). As we show in \S~\ref{sec:alpha} this information can
help distinguish between models with different IMFs.  However, a
major source of systematic uncertainty is the anisotropy of the
stellar orbits (Koopmans \etal 2009). Auger \etal (2010a) assumed
isotropic orbits, $\beta=0$, and thus may have inadvertently favoured
a particular IMF.

In order to constrain the halo contraction model the halo mass needs
to be accurately determined. This is for 2 reasons. Firstly, the halo
mass is needed to predict the typical pristine halo concentration
using cosmological N-body simulations. Secondly, to provide an
accurate normalization of the halo mass profile. For a fixed IMF, the
relation between stellar mass and halo mass has been determined using
halo abundance matching, weak lensing and satellite kinematics. These
techniques yield consistent results, and in particular for massive
early-type galaxies (Dutton \etal 2010b; More \etal 2011), and thus
provide a consistency check on the models of Auger \etal (2010a).
Taking a model with a Chabrier IMF and Gnedin \etal (2004) halo
contraction, Auger \etal (2010a) find halo masses a factor of $\sim
0.5$ dex lower than obtained by Moster \etal (2010).  The relation
from Moster \etal (2010) is in good agreement with the relations we
use in this paper for massive early-type galaxies (see Fig. 1 in
Dutton \etal 2010b). Thus Auger \etal (2010a) is inferring abnormally
low halo masses at fixed stellar mass, which may be biasing their
results (as previously discussed by Tortora \etal 2010).

Schulz \etal (2010) argue that the halo contraction models of Abadi
\etal (2010) and Blumenthal \etal (1986) are inconsistent
with the data. We disagree with this conclusion, as we show that for a
Chabrier IMF, all three halo contraction models are consistent with
the data, given reasonable systematic uncertainties.  Furthermore,
distinguishing between the Gnedin \etal (2004) and Abadi \etal (2010)
models requires stellar masses to be measured to an accuracy of 0.1
dex. Such accuracy may in principle be achievable for early-type
galaxies (Gallazzi \& Bell 2009), but the current limiting factors are
uncertainties in stellar population synthesis models (Conroy, Gunn, \&
White 2009).

The degeneracy between halo contraction and stellar IMF for
early-types was also discussed by Napolitano \etal (2010), with
similar qualitative conclusions as we find here.  However, a
limitation of this study was the treatment of the stellar mass to halo
mass ratio as a free parameter. In our analysis the halo masses are
constrained through results from weak lensing and satellite
kinematics, which enables us to make more quantitative conclusions
regarding the nature of dark halo response for a given IMF.  Finally,
we note that the conclusion of Napolitano \etal (2010) that the
relation between the central dark matter density and effective radius
provides evidence for cuspy dark matter haloes is in fact degenerate
with the IMF.

Our conclusions for late-type galaxies are in agreement with those of
Dutton \etal (2007), namely that for a Chabrier IMF, halo expansion is
required to match the zero point of the TF relation (as well as galaxy
sizes).  Trujillo-Gomez \etal (2010) claim their results are
incompatible with the conclusions of Dutton \etal (2007). However,
their figures show that a model with Blumenthal \etal (1986) adiabatic
halo contraction is consistent with the velocity-luminosity (VL)
relation of early-types, but it does not match the VL relation of
late-types. Their model without halo contraction provides a better fit
to the VL relation of late-types.  Their results are thus consistent
with our findings.

\subsection{Future prospects} 

There are a number of techniques that are capable of constraining the
IMF and/or dark matter fractions in galaxies. Here we give a brief
outline of these.

Upper limits to stellar $M/L$ ratios are obtainable from both strong
lensing and dynamical models.  The strongest constraints are expected
for intermediate mass early-type galaxies, as these are expected to
have the highest baryon fractions within the effective radius
(Fig.~\ref{fig:fdm}). The ATLAS3D project (Cappellari \etal 2011)
contains $\sim 10$ times more galaxies than studied by Cappellari
\etal (2006), and thus promises to provide stronger constraints on
$V_{\rm circ}(R_{\rm e })/\sigma_e$ over a wider range of galaxy
masses than previous dynamical studies. Strong lensing has the
potential to provide stronger constraints than dynamical models, but
it is currently limited by the sparsity of known strong lenses with
$\Mstar\sim 10^{10}\Msun$. Furthermore, low mass early-type galaxies
tend to be satellites, which adds an extra complication to inferring
total masses from strong lensing. Strong lensing will be able measure
the total projected mass accurately, but the problem will be
disentangling the mass of the satellite from that of its host.

There is a well known disk-halo degeneracy that plagues the
decomposition of galaxy rotation curves into baryonic and dark matter
components (e.g., van Albada \& Sancisi 1986; van den Bosch \& Swaters
2001; Dutton \etal 2005). Strong gravitational lensing of high
inclination disk-dominated galaxies can provide a complementary
information to that obtainable from kinematics. Specifically, strong
lensing measures projected mass, and ellipticity of projected
mass. Both of which depend on the disk mass fraction, and thus a
combined strong lensing and dynamics analysis can place constraints on
the stellar $M/L$ ratio (Dutton \etal 2011, in prep).  This technique
has not been fully exploited due to the lack of known disk dominated
strong lenses. However, searches for spiral galaxy strong lenses are
underway (F{\'e}ron \etal 2009; Sygnet \etal 2010; Treu \etal 2011, in
prep), and thus the primary limitation of this method will be soon
overcome.

An absolute constraint on disk masses can be obtained by using the
fact that the disk surface mass density is a function of the vertical
velocity dispersion and the disk scale height (Bottema 1993). This
method is being applied by the Disk Mass Survey (Verheijen \etal 2007;
Bershady \etal 2010). They are measuring the disk mass density profile
from vertical velocity dispersions and a statistical measurement of
disk scale heights. By subtracting off the observed gas mass density
this gives the stellar mass density profile. This method is limited to
regions of galaxies where the disk dominates the baryons, i.e., it
does not apply to elliptical galaxies or the bulges of spiral
galaxies. It is also a statistical method, since it requires knowledge
of two parameters that cannot be measured simultaneously.

Constraints on dark matter fractions can be obtained from the scatter
in the velocity-mass (VM) and size-mass (RM) relations. The basic idea
is that the strength of correlation between residuals of the VM and RM
relations depends on the dark matter fraction. This method has been
applied to late-type galaxies (Courteau \& Rix 1999; Dutton \etal
2007; Gnedin \etal 2007), but the interpretation in terms of dark
matter fractions are not always unique.  We plan to apply this method
to early-type galaxies in a future paper (Dutton \etal 2011, in prep.)

The low-mass end of the IMF can be constrained with stellar absorption
lines (van Dokkum \& Conroy 2010). The lines are weak and at $\sim
900$ nm, so that this method is only applicable to non-star forming
galaxies.  The galaxy redshifts are currently limited to be very low
by detector technology. This method has only currently been applied to
the centers of massive early-type galaxies, with evidence for an IMF
more bottom heavy than Salpeter. It would be very interesting to see
this method applied radially and in lower mass early-types.

%%%%%%%%%%%%%%%%%%%%%%%%%%%%%%%%%%%%%%%%%%%%%%%%%%%%%%%%%%%%%%%%%%%%%%
%% SECTION 6: SUMMARY
%%%%%%%%%%%%%%%%%%%%%%%%%%%%%%%%%%%%%%%%%%%%%%%%%%%%%%%%%%%%%%%%%%%%%%

\section{Summary}
\label{sec:sum}

We use structural and dynamical scaling relations of early-type
galaxies (ETGs) and late-type galaxies (LTGs) to place constraints on
the stellar initial mass function (IMF) and dark halo response to
galaxy formation, which is commonly modeled as adiabatic contraction
(AC). We build bulge-disk-halo models that by construction reproduce
the observed structural scaling relations of galaxies: optical size vs
stellar mass; bulge fraction vs stellar mass; gas mass vs stellar
mass; and gas size vs stellar mass.  The dark matter haloes are
constrained to reproduce the observed halo mass vs stellar mass
relation from satellite kinematics and weak lensing (Dutton \etal
2010b), and the concentration - halo mass relation from cosmological
N-body simulations (Macci\`o \etal 2008). Lastly, the Tully-Fisher
(TF) and Faber-Jackson (FJ) relations provide a constraint on the
total mass within 2.2 disk scale lengths for LTGs and the half-light
radius, $R_{\rm e}$, for ETGs.

A key uncertainty in the constraint from the FJ relation is the
conversion between the observed stellar velocity dispersion within the
half-light radius, $\sigma(<R_{\rm e})\equiv \sigma_{\rm e}$, and the
circular velocity within the half-light radius, $V_{\rm circ}(R_{\rm
  e})\equiv V_{\rm c}$. Using results from the SLACS survey we show
that for massive early-type galaxies ($\Mstar > 10^{11} \Msun$), on
average $V_{\rm c}/\sigma_{\rm e}=1.54\pm0.02$. We show that the
SAURON results from Cappellari \etal (2006), which make use of a
smaller sample, but cover a much larger range in stellar mass $3\times
10^{9} - 4\times 10^{11} \Msun$, imply that on average $V_{\rm
  c}/\sigma_{\rm e}=1.44\pm0.01$. The inconsistency of these two
results implies that there are systematic effects which bias either or
both of these results.

Based on the observed scaling relations alone, we calculate the
spherical dark matter fraction within the half-light radius for ETGs
and 2.2 disk scale lengths for LTGs.  For LTGs the dark matter
fraction increases with increasing stellar mass (in agreement with
previous studies), while for ETGs the dark matter fraction reaches a
minimum for a stellar mass of $\Mstar \sim 10^{10}\Msun$ (Assuming a
Chabrier IMF), and velocity dispersions of $\sigma_{\rm e} \sim 100
\kms$.  The dark matter fraction increases towards both lower and
higher masses.

High mass ETGs are consistent with a Salpeter IMF (i.e., the stellar
mass fraction implied by the structural scaling relations is less than
the dynamically inferred mass.)  However, a Salpeter IMF is ruled out
for galaxies with velocity dispersions $\sigma_{\rm e} \sim 100 \kms$,
unless $V_{\rm c}/\sigma_{\rm e} \gta 1.6$. Thus improved constraints
on $V_{\rm c}/\sigma_{\rm e}$ for intermediate mass ETGs from strong
lensing or dynamical modeling would provide the strongest upper limits
to the IMF.

Our bulge-disk-halo models reproduce the slopes of the FJ and TF
relations, for ETGs and LTGs respectively. However, models with a
universal IMF and universal halo response to galaxy formation are
unable to {\it simultaneously} match the zero points of the FJ and TF
relations. For a given AC model, ETGs require higher stellar mass
normalizations (i.e., IMFs with higher stellar mass-to-light ratios)
than LTGs. For a given IMF, ETGs require stronger halo contraction.

For early-type galaxies, models with a Chabrier IMF and adiabatic
contraction according to Gnedin \etal (2004) provide good fits to the
FJ relation. Models with adiabatic contraction according to Blumenthal
\etal (1986) and Abadi \etal (2010) also provide good fits, within the
systematic uncertainties in dark halo masses and $V_{\rm c}/\sigma_{\rm
  e}$.  Distinguishing between the models of Gnedin \etal (2004) and
Abadi \etal (2010), and hence constraining the stability of AC to
bombardment from major and minor mergers, requires measuring stellar
masses to within 0.1 dex, which seems beyond the reach of current SPS
models.  Models without adiabatic contraction favor an IMF $\simeq
0.17$ dex heavier than Chabrier.

For late-type galaxies, models with adiabatic contraction require
lighter IMFs (i.e., lower stellar mass-to-light ratios) than Chabrier
(at least 0.10 dex for Abadi \etal 2010, 0.20 dex for Gnedin \etal
2004, and 0.25 dex for Blumenthal \etal 1986). Matching the TF
relation with a Chabrier IMF requires mild halo expansion, in
agreement with Dutton \etal (2007).  Producing stellar masses lower by
$\sim 0.3$ dex is not possible from plausible variations in the IMF,
and thus lower stellar masses would need to arise from systematic
uncertainties in stellar population synthesis models or the
application of these models in deriving stellar masses.

Evolution of the IMF as proposed by van Dokkum (2008) and Dav\'e
(2008) would cause ETGs to have higher present day stellar $M/L$ than
LTGs.  The normalization for ETGs would be significantly higher than
Chabrier, and thus disfavoring dark halo contraction.

Alternatively, if the IMF is universal, we envision the following
scenario which results in halo expansion in late-types and halo
contraction in early-types.  We suppose that the distinguishing
feature between ETGs and LTGs is that ETGs assemble a significant
fraction of their stellar mass in major merger events. For galaxies
without a major mergers, the dark haloes expand due to a combination
of dynamical friction from baryonic clumps during the early phases of
galaxy formation (El-Zant \etal 2001; Mashchenko \etal 2006; Elmegreen
\etal 2008) and/or supernova/stellar wind driven mass outflows
(Navarro \etal 1996b; Read \& Gilmore 2005; Governato \etal
2010). These galaxies will become late-types (star forming and disk
dominated). When galaxies experience a major dissipative merger, large
amounts of gas accumulate at the center of the remnant, resulting in
standard halo contraction (i.e., Gnedin \etal 2004). Subsequent
bombardment from non-dissipative mergers may reduce the amount of halo
contraction (Johansson \etal 2009; Abadi \etal 2010).  A prediction of
this scenario would be that bulge dominated LTGs should have more
contraction than disk dominated LTGs of the same overall mass.

Finally we show that our models naturally reproduce flat and
featureless circular velocity profiles within the optical regions of
both early and late-type galaxies for a wide range of halo responses
(including contraction and expansion) and IMFs.

\section*{Acknowledgments} 
A.A.D. acknowledges financial support from the Canadian Institute for
Theoretical Astrophysics (CITA) National Fellows program.
S.C. acknowledges the support of a Discovery grant by the Natural
Science and Engineering Research Council of Canada. The work of
A.D. was partly supported by ISF grant 6/08, by GIF grant
G-1052-104.7/2009, by a DIP grant, by the Einstein Center at HU, and
by NSF grant 1010033 at UCSC.  This research has made use of NASA's
Astrophysics Data System Bibliographic Services.

Funding for the  Sloan Digital Sky Survey (SDSS)  has been provided by
the Alfred  P. Sloan  Foundation, the Participating  Institutions, the
National  Aeronautics and Space  Administration, the  National Science
Foundation,   the   U.S.    Department   of   Energy,   the   Japanese
Monbukagakusho,  and the  Max Planck  Society.  The SDSS  Web site  is
http://www.sdss.org/.

The SDSS is managed by the Astrophysical Research Consortium (ARC) for
the Participating Institutions. The Participating Institutions are The
University of Chicago, Fermilab, the Institute for Advanced Study, the
Japan Participation  Group, The  Johns Hopkins University,  Los Alamos
National  Laboratory, the  Max-Planck-Institute for  Astronomy (MPIA),
the  Max-Planck-Institute  for Astrophysics  (MPA),  New Mexico  State
University, University of Pittsburgh, Princeton University, the United
States Naval Observatory, and the University of Washington.

%%%%%%%%%%%%%%%%%%%%%%%%%%%%%%%%%%%%%%%%%%%%%%%%%%%%%%%%%%%%%%%%%%%%%%
%%  REFERENCES
%%%%%%%%%%%%%%%%%%%%%%%%%%%%%%%%%%%%%%%%%%%%%%%%%%%%%%%%%%%%%%%%%%%%%% 

\label{lastpage}
\end{document}